\documentclass[a4paper,useAMS,usenatbib]{mnras}
\usepackage{graphicx}
\usepackage{color,ulem}
\usepackage[usenames,dvipsnames,svgnames,table]{xcolor}
\usepackage{array}
\newcolumntype{P}[1]{>{\centering\arraybackslash}p{#1}}

\newcommand{\mgta}{}

\newcommand{\gtrsim} {\ga}
\newcommand{\lesssim} {\la}

\def\aap{A\& A}
\def\aapr{A\& A Rv}

\def\aj{{AJ}}
\def\araa{ARA\&A}

\def\apj{ApJ}
\def\apjl{ApJL}
\def\apjs{ApJS}

\def\mnras{MNRAS}
\def\nar{NewAR}

\def\nat{Nature}
\def\pasj{{PASJ}}

\def\'#1{\ifx#1i{\accent"13\i}\else{\accent"13#1}\fi}

\newcommand{\av} {\alpha_{\rm vir}}
\newcommand{\call} {{\cal L}}
\newcommand{\cs} {c_{\rm s}}

\newcommand{\eff} {\epsilon_{\rm ff}}
\newcommand{\Eg} {E_{\rm g}}
\newcommand{\Ek} {E_{\rm k}}
\newcommand{\Ektot} {E_{\rm k,tot}}
\newcommand{\Etot} {E_{\rm tot}}

\newcommand{\hii}{{\sc Hii}}
\newcommand{\kj}{k_{\rm J}}
\newcommand{\kms}{{\rm ~km~s}^{-1}}

\newcommand{\LJ} {L_{\rm J}}

\newcommand{\LR} {{\cal L}}
\newcommand{\LS} {${\cal L}$-$\Sigma$}
\newcommand{\Mo} {M_0}
\newcommand{\Mf} {M}
\newcommand{\Mgas}{M_{\rm gas}}
\newcommand{\MJ} {M_{\rm J}}
\newcommand{\MJrms} {M_{\rm J,rms}}
\newcommand{\MJo} {M_{\rm J,0}}

\newcommand{\Ms} {{\cal M}_{\rm s}}
\newcommand{\Mstar}{M_{*}}
\newcommand{\Msun}{M_{\odot}}
\newcommand{\Mtot}{M_{\rm tot}}
\newcommand{\navg}{\langle n \rangle}
\newcommand{\NJ} {N_{\rm J}}
\newcommand{\nmj} {n_{\rm MJ}}
\newcommand{\nsf} {n_{\rm SF}}

\newcommand{\pcc}{{\rm ~cm}^{-3}}
\newcommand{\psc}{{\rm ~cm}^{-2}}
\newcommand{\rhomeanf} {\bar \rho_{\rm f}}
\newcommand{\Rof} {R_{\rm 0}}
\newcommand{\Rf} {R}

\newcommand{\rhoo}{\rho_0}
\newcommand{\rhosf}{\rho_{\rm SF}}

\newcommand{\rrms}{\rho_{{\rm rms}}}
\newcommand{\rs}{r_{\rm s}}
\newcommand{\sfeff}{{\rm SFE_{\rm ff}}}
\newcommand{\SSFR}{\Sigma_{\rm SFR}}
\newcommand{\savg}{\langle s \rangle}
\newcommand{\sturb}{\sigma_{\rm turb}}
\newcommand{\sv}{\sigma_v}
\newcommand{\tauff} {\tau_{\rm ff}}
\newcommand{\tauffo} {\tau_{{\rm ff},0}}

\newcommand{\taumu} {\tau_{\rm M,u}}

\newcommand{\tff} {t_{\rm ff}}
\newcommand{\tffo} {t_{{\rm ff},0}}

\newcommand{\tmu} {t_{\rm M,u}}
\newcommand{\vff} {v_{\rm ff}}
\newcommand{\vg} {v_{\rm g}}

\newcommand{\vl} {v_{\parallel}}
\newcommand{\vvir} {v_{\rm vir}}

\newcommand{\VS}{V\'azquez-Semadeni}
\newcommand{\Myr} {{\rm Myr}}

\newcommand{\etal}{et al.}
\newcommand{\bfg}{\begin{figure}}
\newcommand{\efg}{\end{figure}}
\newcommand{\bfgw}{\begin{figure*}}  
\newcommand{\efgw}{\end{figure*}}    
\newcommand{\beq}{\begin{equation}}
\newcommand{\eeq}{\end{equation}}
\newcommand{\bce}{\begin{center}}
\newcommand{\ece}{\end{center}}
\newcommand{\ben}{\begin{enumerate}}
\newcommand{\een}{\end{enumerate}}

\title[Global Hierarchical Collapse. The scenario] {Global
Hierarchical Collapse In Molecular Clouds. Towards a Comprehensive Scenario} 

\author[V\'azquez-Semadeni et al.] {Enrique
V\'azquez-Semadeni$^1$\thanks{E-mail: e.vazquez@irya.unam.mx}, 
Aina Palau$^1$, Javier Ballesteros-Paredes$^1$,  \newauthor
Gilberto C.\ G\'omez$^1$, and Manuel Zamora-Avil\'es$^{2,1}$
\\
$^1$Instituto de Radioastronom\'\i a y Astrof\'\i sica,
Universidad Nacional Aut\'onoma de M\'exico, Apdo. Postal 3-72, Morelia,
Michoac\'an, 58089, M\'exico\\
$^2$CONACYT-Instituto Nacional de Astrof{\'i}sica, \'Optica y
Electr{\'o}nica, Luis E. Erro 1, 72840 Tonantzintla, Puebla,
M{\'e}xico}

\begin{document}
\label{firstpage}
\pagerange{\pageref{firstpage}--\pageref{lastpage}}

\maketitle

\begin{abstract}

We present a unified description of the scenario of global hierarchical
collapse (GHC). GHC constitutes a flow regime of (non-homologous)
collapses within collapses, in which all scales accrete from their
parent structures, and small, dense regions begin to contract at later
times, but on shorter timescales than large, diffuse ones. The different
timescales allow for most of the clouds' mass to be dispersed by the
feedback from the first massive stars, maintaining the cloud-scale star
formation rate (SFR) low. MCs, clumps, and cores are not in equilibrium,
but rather are either undergoing contraction or dispersal. The main
features of GHC are: 1) The gravitational contraction is initially very
slow, and begins when the cloud still consists of mostly atomic gas. 2)
Star-forming MCs are in an essentially pressureless regime, causing
filamentary accretion flows from the cloud to the core scale to arise
spontaneously. 3) Accreting objects have longer lifetimes than their own
free-fall time, due to the continuous replenishment of material. 4) The
clouds' total mass and its molecular and dense mass fractions increase
over time. 5) The clouds' masses stop growing when feedback becomes
important. 6) The first stars appear several megayears after global
contraction began, and are of low mass; massive stars appear a few
megayears later, in massive hubs. 7) The minimum fragment mass may well
extend into the brown-dwarf regime. 8) Bondi-Hoyle-Lyttleton-like
accretion occurs at both the protostellar and the core scales,
accounting for an IMF with slope $dN/dM \propto M^{-2}$. 9) The extreme
anisotropy of the filamentary network explains the difficulty in
detecting large-scale infall signatures. 10) The balance between
inertial and gravitationally-driven motions in clumps evolves during the
contraction, explaining the approach to apparent virial equilibrium,
from supervirial states in low-column density clumps and from subvirial
states in dense cores. 11) Prestellar cores adopt Bonnor-Ebert-like
profiles, but are contracting ever since when they may appear to be
unbound. 12) Stellar clusters develop radial age and mass segregation
gradients. We also discuss the incompatibility between supersonic
turbulence and the observed scalings in the molecular hierarchy. Since
gravitationally-formed filaments do not develop shocks at their axes, we
suggest that a diagnostic for the GHC scenario should be the absence of
strong shocks in them. Finally, we critically discuss some recent
objections to the GHC mechanism.

\end{abstract}



\section{Introduction} \label{sec:intro}


Molecular clouds (MCs) have long been known to exhibit supersonic
linewidths \citep{Wilson+70} and to possess masses much larger than
their thermal Jeans mass \citep[e.g.,] [] {Blitz93}. One of the first
interpretations of the supersonic linewidths observed in MCs
\citep{Wilson+70} was that the clouds are in global
gravitational collapse \citep{GK74}, but this suggestion was quickly
rejected by \citet{ZP74} and \citet{ZE74} on the basis of two main
arguments: One, that, if the clouds were in global collapse, then the
star formation rate (SFR) would be much larger than the actual observed
value. Second, radial, cloud-scale motions should cause absorption lines
produced by the gas surrounding star-forming regions to be
systematically redshifted from emission lines produced at the region
itself, since the surrounding gas should be infalling into the
star-forming region. Since this systematic shift had not been observed
\citep[but see] [for a  possible recent detection] {Barnes+18},
\citet{ZP74} ruled out global collapse, as well as any cloud-scale
systematic, radial motions. As a plausible alternative, \citet{ZE74}
suggested that the observed supersonic linewidths in MCs correspond
instead to {\it small-scale} turbulence,  so that it can provide a
roughly isotropic pressure gradient. This view has prevailed until
today.

In the present-day scenario \citep[see, e.g., the reviews by][]{MK04,
BP+07, MO07, BT07, HF12}, the clumps and cores within MCs are thought to be
produced by the supersonic compressions caused by the turbulence in
the cloud, and the turbulence is expected to cascade from the large
scales where it is injected all the way down to the smallest (dense
core, $\lesssim 0.1$ pc) scales, so that the clumps at intermediate
scales are themselves internally turbulent. In turn, the clumps are
believed to be supported against their self-gravity by their internal
turbulence, and to collapse once the latter has been dissipated, so
that the turbulent support is lost \citep[e.g.,] [] {Goodman+98,
BT07}, or when the compression is strong enough that the local Jeans
mass
becomes smaller than the clump's own mass \citep[e.g.,][] {Sasao73,
Padoan95, VS+03, Gomez+07, GO+09}.

However, over the last decade, a number of recent observational studies
have suggested that star-forming MCs may be in a state of global
gravitational contraction after all. On the basis of comparisons between
millimeter interferometric observations and numerical simulations,
\citet{Peretto+07} suggested that the elongated clump NGC 2264-C may
be in the process of collapsing and fragmenting along its long axis.
Several molecular-line kinematic studies of clouds and their internal
filaments suggest that these systems are undergoing global multi-scale
and multi-epoch collapse \citep[e.g.,] [] {Galvan+09, Peretto+13,
Beuther+15, Liu+15, Liu+16b, Friesen+16, Jin+16, Hacar+17, Csengeri+17,
Yuan+18, Jackson+19, Barnes+19}, with the filaments in particular
exhibiting longitudinal flow \citep[e.g.,] [] {Sugitani+11, Kirk+13,
FL+14, Motte+14, Peretto+14, Tackenberg+14, JS+14, Hajigholi+16,
Wyrowski+16, Juarez+17, Rayner+17, Lu+18, Baug+18, Gong+18,
Ryabukhina+18, Dutta+18, Chen+19} that feeds central ``hubs'', the sites
where the filaments converge \citep{Myers09}.

On the numerical side, self-consistent simulations of giant molecular
cloud (GMCs) formation by converging motions in the warm diffuse
atomic medium, as suggested by \citet{BP+99}, \citet{HP99}, and
\citet{HBB01}, have shown that
\begin{enumerate}
\item The clouds are dynamical entities that accrete from their diffuse
environment and therefore grow in mass \citep[e.g.,] [] {BP+99a, AH05,
Banerjee+09}.

\item The turbulence generated by inherent instabilities in the
compressed material \citep[e.g.,] [] {KI02, AH05, Heitsch+05,
Heitsch+06, VS+06, Wareing+19} is only {\it moderately} supersonic, with
typical sonic Mach numbers $\Ms \sim 3$. This value corresponds to that
observed at the scale of individual ``molecular clumps'' ($\lesssim 1$
pc), but falls significantly short of the typical values for large
clouds and GMCs \citep[see, e.g.,] [] {HB04}.

\item Because of this relatively low Mach number, the turbulence is not
enough to support a GMC, and so the cloud begins to undergo global
collapse shortly after its mass becomes larger than its thermal Jeans
mass \citep[e.g., ] [] {VS+07,VS+10, Heitsch+08b, Carroll+14}. 

\item  The collapse process, however, is extremely nonlinear, being
initially very slow, and violently accelerating towards its final stages
\citep{Girichidis+14, ZV14}.

\item Nevertheless, no star formation occurs during the first several
  {megayears} of the global collapse, both in the non-magnetic \citep[e,g,] []
{VS+07, HH08, Carroll+14} and the magnetized case \citep[e.g.] []
{VS+11, Fogerty+16, ZA+18},  due to the slowliness of the
initial stages of the collapse.

\item As shown by \citet{Camacho+16}, \citet{Ibanez+16}, and
\citet{BP+18}, the clouds and clumps defined in the simulations follow
the scaling relation observed for molecular clouds,
\beq
\frac{\sv}{R^{1/2}} \approx \sqrt{G \Sigma},
\label{eq:vir}
\eeq
\citep[e.g.,] [] {Field+11, Heyer+09, Leroy+15}, where $\sv$ is the
cloud velocity dispersion, $R$ is its approximate radius, and $\Sigma$
is its column density. On the other hand, \citet{Ibanez+16} have shown
that this scaling is {\it not} reproduced in simulations of the ISM with
supernova-driven turbulence when the self-gravity is not included.
Instead, purely supernova-driven turbulence in the ISM generates a
scaling of the form $\sv \propto R^{1/2}$ \citep[i.e., $\sv R^{-1/2} =$
cst., as corresponds to strongly supersonic turbulence; e.g.,] [] {Padoan+16},
which corresponds mainly to the scaling observed in GMCs in the outer
parts of galaxies, but is not representative of the majority of
star-forming MCs \citep[e.g.,] [] {Heyer+09, BP+11a, Leroy+15,
Traficante+18a}. 

\end{enumerate}

{Item (vi) above suggests that self-gravity may be the main driver of the motions at all scales in a cloud, and not
necessarily implying virialization as it is frequently interpreted, but rather infall, as will be
discussed in Sec.\ \ref{sec:energy_balance}.}

In addition, \citet{BH04} investigated the collapse of
thin, finite sheets of cold gas, formed by converging flows in the warm
atomic gas. On the basis of those results, \citet{HB07} suggested
that the morphology of the Orion A cloud, and the location of the Orion
Nebula Cluster within it, may be explained simply in terms of a global
gravitational contraction of the cloud. Recently, evidence for this
cloud-scale contraction in Orion has been presented by
\citet{Hacar+17}.

In the light of this mounting evidence of global MC gravitational
contraction, in \citet[] [see also V\'azquez-Semadeni et al.\ 2017]
{VS+09}, we suggested that MCs in general might be undergoing a
process of {\it  global, hierarchical collapse}
(GHC), in which the clouds develop multi-scale
gravitational contraction, similar to the hierarchical fragmentation
scenario proposed by \citet{Hoyle53}, but seeded with nonlinear
density fluctuations produced by the turbulence. This mechanism
constitutes a {\it mass cascade}, somewhat analogous to the energy
cascade of turbulence, although with some important differences,
already discussed by \citet[] [see also Field et al.\ 2008] {Hoyle53},
most important of which is the fact that the mass cascade implies that
all scales accrete from their parent structures.

In view of the above, the time is ripe for a 
re-analysis of the evidence and arguments leading to the 
extended notion that MCs require support, and for a comprehensive
formulation of the GHC scenario, which has been presented in scattered
form in various papers \citep{VS+07, VS+09, VS+10, VS+17, HH08,
Heitsch+09, BP+11a, BP+15, BP+18, ZA+12, ZV14, Kuznetsova+15,
Kuznetsova+17, Kuznetsova+18a, Kuznetsova+18b, Naranjo+15, Camacho+16},
including some new calculations, concerning the timescales involved in
the process.

The plan of the paper is as follows: In Sec.\ \ref{sec:no_support} we
start with a critical discussion of the arguments that originally led to
the notion that MCs require support (Sec.\ \ref{sec:deconstr}), and then
continue with a discussion of some problems with the notion of
turbulent support (Sec.\ \ref{sec:probs_turb}). Next, in Sec.\
\ref{sec:nonlin_Hoyle} we revisit the scenario of hierarchical
gravitational fragmentation in an isothermal medium \citep{Hoyle53},
and discuss its extension to the case of nonlinear density
fluctuations produced by turbulence, in particular the timescales
involved in the sequential onset of collapse of different scales. In
Sec.\ \ref{sec:nature_of_collapse} we then revisit some particular features and
misconceptions about the mechanism of non-homologous gravitational
collapse that explain various structural features of observed cores.
Next, in Sec.\ \ref{sec:model}, we give a comprehensive
description of the GHC model, covering
the development of filamentary structure, the increase of the SFR, the
formation of massive star-forming regions, the termination of the local
SF episodes and the self-regulation of star formation (SF) by feedback,
and finally the resulting stellar cluster structure. In Sec.\
\ref{sec:disc} we discuss implications and caveats of the model, and in Sec.\
\ref{sec:concl} we conclude with a summary and some final
remarks.

\section{Do molecular clouds need to be supported?}
\label{sec:no_support}

In this section we revisit the arguments that originally led to the
established notion that MCs need to be supported against gravity,
showing that they do not hold in the light of current observational data
of the structure and dynamics of MCs, and then we summarize some
difficulties with the hypothetical turbulence that would be able to
provide support for the clouds.

\subsection{Deconstructing the {standard} objections against global cloud collapse}
\label{sec:deconstr}

As mentioned in Sec.\ \ref{sec:intro}, one of the first interpretations
of the supersonic linewidths observed in MCs
\citep{Wilson+70} was that the clouds are in global
gravitational collapse \citep{GK74, Liszt+74}. However, this proposal
was shortly afterwards rejected by \citet[] [hereafter ZP74] {ZP74}
and \citet[] [hereafter ZE74] {ZE74}, on
the basis of two main arguments:
\begin{enumerate}
\item {\it The SFR conundrum:} If the clouds were in global collapse,
then the star formation rate (SFR) would be much larger than the
actual observed value.

\item {\it The line-shift-absence conundrum:} Absorption lines
produced by the gas surrounding star-forming regions should be
systematically redshifted from emission lines produced at the region
itself, since the surrounding gas should be infalling into the
star-forming region. Such systematic redshifts, however, were not
observed.

\end{enumerate}

In view of these arguments, ZE74 argued that
the supersonic linewidths observed in MCs should correspond to {\it
small-scale} supersonic turbulent motions. Note that the requirment that
the motions be small-scale is essential, since it is necessary to avoid
cloud-scale systematic motions that would produce the non-detected line
shifts between envelope and core regions.

However, the notion of small-scale turbulence (the only one that can
resolve the ZE74 conondrums) faces several problems of its own. First,
the SFR conundrum has had {an alternative}  solution that dates back to the same
times in which it originated, namely that the clouds may be destroyed
early in their life cycle by their first stellar products, so that only
a small fraction of their mass manages to be converted to stars during
the collapse \citep[e.g.,] [] {Field70, Whitworth79, Elm83, Cox83,
Franco+94}. That is, rather than the clouds being in equilibrium and
maintaining a low SFR over their entire lifetimes, the alternative is
for SF to constitute a highly intermittent process in space and time,
with short and intense bursts of SF that destroy their parent clouds
before most of their mass can be converted to stars.

Indeed, recent numerical simulations have shown that the ionizing
radiation from massive stars is generally capable of eroding away MCs
of masses up to $\sim 10^5 \Msun$ \citep[e.g.,] [] {Dale+12, Colin+13,
Wareing+17a, Wareing+17b, Haid+19}, although the destruction and
dispersal of MCs more massive than that may require additional driving
mechanisms, and remains a topic of debate.
{For example, using a one-dimensional, spherically symmetric model,
\citet{Murray+10} proposed that radiation pressure is an essential
ingredient for dispersing GMCs in all environments, while protostellar
jets may be important only during early evolutionary stages of the
clouds, and ionising radiation is effective in Milky-Way-type galaxies,
but not in starbursts. Numerical simulations of initially spherical
clouds with decaying turbulent velocity fields confirm that
photoionising radiation appears incapable of destroying clouds more
massive than $10^5 \Msun$ \citep[e.g.,] [] {Dale+12, Dale+13}, although
simulations also suggested that radiation pressure would be less
effective than simple 1D models predicted, because of the development of
Rayleigh-Taylor instabilities that limit the coupling between radiation
and matter \citep[e.g.,] [] {KT12, KT13, Reissl+18}. On the other
hand,} the effectiveness of supernova (SN) explosions in destroying
clouds depends strongly on the relative locations of the SNe and the
dense gas, although when the SNe are self-consistently positioned, cloud
dispersal is generally accomplished {for low- to intermediate-mass clouds}
\citep[e.g.,] [] {IH15, IH17, Koertgen+16, Seifried+18}. {Apparently, the
destruction of massive clouds requires the combined action of winds and
the thermal and radiation pressures \citep{Rahner+17}. }

In any case, it is well
established observationally that stellar associations and clusters rid
themselves of their gas within 5-10 Myr \citep[e.g.,] [] {Leisawitz+89,
LL03, Shimoikura+18}.

On the other hand, the line-shift-absence conundrum of ZP74 is easily
explained through geometrical considerations. Essentially, the
arguments leading to this conundrum assume that the collapse is roughly
spherically symmetric and monolithic, so that the infall motions are
coherent, and directed towards a single collapse center, at the
geometrical center of the cloud. This assumption is inconsistent with
our current understanding of the structure of MCs, which are known to be
far from spherically symmetric, and instead consist of an intrincate
and inhomogeneous network of filaments and clumps within them
\citep[e.g.,] [] {Bally+87, Feitzinger+87, Gutermuth+08, Myers09,
Juvela+09, Andre+10, Henning+10, Menshchikov+10, Molinari+10,
Arzoumanian+11, Busquet+13}. The central clumps (``hubs'') appear to
accrete from the filaments, while in turn the filaments seem to accrete
radially from their surroundings \citep{Schneider+10, Kirk+13,
Peretto+14, Lu+18, Williams+18, Shimajiri+19, Gong+18}. Thus, the
geometry is far from being spherically symmetric, and therefore the
accreting gas is not isotropically distributed around the collapse
centers (the hubs).  In addition, the velocity field is highly complex
and chaotic \citep[e.g.,] [] {GV14, ZA+17, Gomez+18}, so there is no
reason to expect a {\it systematic} redshift of the absorption lines
produced in the gas surrounding the hubs. Instead, the accretion flow is
most directly observed as velocity-centroid gradients along the
filaments, directed towards the hubs.  Indeed, synthetic CO observations
of simulations of the regime often show only marginal or no evidence for
infall profiles, due to the chaotic motions and perhaps velocity
crowding effects, although the line profiles do look similar to observed
ones \citep[e.g.,] [] {Heitsch+09, Heiner+15, Clarke+18}. Nevertheless,
{recent dedicated searches for evidence of infall signatures in CO
lines from GMCs have met with success. For example, \citet{Schneider+15}
have found the classical combination of self-absorbed and blue-skewed
optically thick lines ($^{12}$CO ($3
\rightarrow 2$)) together with centrally-peaked optically thin ($^{13}$CO ($1
\rightarrow 0$)) lines, indicating collapse in the molecular gas
surrounding IRDC G28.37+0.07, while \citet{Barnes+18} have measured
shifts between the lines of $^{12}$CO (tracing gas in the outer parts of
the cloud) and $^{13}$CO (tracing gas deeper into the cloud) in the
pc-scale, massive clumps of the CHaMP survey, finding systematic
velocity differentials between the two lines that imply an average mass
accretion timescale of $\sim 16$ Myr, consistent with the timescales we
discuss in this paper (cf.\ Sec.\ \ref{sec:summary} and Fig.\
\ref{fig:timeline}).}

\subsection{Difficulties with turbulent support} \label{sec:probs_turb}

It is also very important to note that the ZE74 line-shift-absence
conundrum would not only rule out global collapse of clouds, but also MC
turbulence in general, as we presently understand it. Indeed, 
real-world turbulence
is characterized by an energy spectrum {\it containing most of the
energy at large scales}. This implies that the average velocity
difference between points in the flow is larger for larger separations
between the points, with this trend continuing up to the scale at which
the energy is injected \citep[the ``energy injection'', or ``energy
containing'' scale; see, e.g.,][]{Frisch95}. 

It is important to emphasize that the larger velocities at large scale
refer to {\it coherently-moving structures}. That is, they do not only
imply a larger velocity dispersion for larger regions; rather, they
mainly imply large-scale coherent motions, up to the scale of the clouds
themselves.  Therefore, the large-scale turbulent motions must
shred, compress, or gyrate the clouds, rather than keeping them in a
quasi-static state.

Indeed, studies aimed at determining the energy-containing scale in
MCs conclude that it corresponds to the largest scales in the clouds
\citep[e.g.,] [] {Brunt+09}, and so the dominant motions in the clouds
are those at their largest scales. This implies that the objection to global
gravitational contraction by ZE74 would also apply to the large-scale
motions that dominate the turbulence in MCs. One concludes that either
both global gravitational collapse {\it and} large-scale-driven 
compressible turbulence are ruled out by the ZE74 argument, or else
neither one is.


In addition, plain turbulence as we understand it today has problems
explaining some observed properties of MCs: 
\begin{enumerate}
\item Clouds both with and without SF have similar nonthermal velocity
  dispersions \citep[e.g.,][]{WB98} 
\item Clouds appear to be close to equipartition between their kinetic
and gravitational energies, except at low column densities, where
large kinetic energy excesses are often found \citep[e.g.,] []
{Larson81, Heyer+09, Kauffmann+13, Leroy+15}. This fact has
traditionally been interpreted as virial equilibrium, although
\citet{BP+11a} have argued that it rather corresponds to the signature
of gravitational collapse at all scales.\footnote{In addition, as
discussed by \citet{BP06}, virial equilibrium, as it is generally
discussed in the literature, assumes that (a) the time derivatives are
zero (while in reality, they cannot be measured); (b) clouds have fixed
masses and there is neither mass nor energy flux through their boundaries
\citep[while in reality, there should be substantial exchange of these
quantities between the clouds and their surroundings, {first as
accretion during cloud growth, and then as return of material during the
cloud disruption;} see also] [] {BP+99a, KH10, VS+10, Ibanez+17}; (c) the
gravitational potential arises only from the cloud itself and is
approximately that of a uniform sphere \citep[while in reality the
clouds are far from homogeneous and the environment may dominate the
gravitational potential; see also] [] {BP+18}; (d) turbulence {is
continuously driven and strong enough that it} can
prevent collapse on large scales and promote collapse at smaller scales
(while, as we argue in this paper, simulations {with self-consistent
turbulence driving by the formation of the clouds and by stellar
feedback} do not support this
assumption. {Instead, the turbulence generated by the cloud assembly
itself cannot support the clouds, while the feedback destroys the clouds
rather than supporting them.})}
\end{enumerate}

Indeed, the first property rules out driving from internal stellar
sources, because this driving cannot explain the turbulence in
starless clouds.  Conversely, if the driving is external, then strong
fine-tuning would be required in order to satisfy the second property,
because there is no {\it a priori} reason why the turbulence induced
in the cloud should acquire precisely an equipartition level. The
induced motion might just as well be either too strong, with the
effect of dispersing the cloud, or too weak, being insufficient to
support the cloud and thus allowing subsequent collapse.

Another view of external driving is that presented by
\citet{Padoan+16}, who have suggested that MCs in general do {\it not}
exhibit relations corresponding to near equipartition (of the form of
eq.\ [\ref{eq:vir}]). {Instead, they suggest that the clouds}
exhibit a plain turbulence-scaling relation $\LR \equiv \sv/R^{1/2} =$
cst, {where $\LR$ is hereinafter referred to as} the {\it Larson
ratio}.\footnote{{This quantity is sometimes referred to as the {\it
velocity scaling} or the {\it size-linewidth coefficient.}}}

\citet{Padoan+16} report the scaling $\LR \approx $ cst.\ 
for their simulations of the SN-driven ISM, and noted that this is the
scaling observed for clouds in the Outer Galaxy Survey
\citep{Heyer+01}. They thus argued that this should be considered as the
general scaling, because the Outer Galaxy Survey is the largest MC
survey available. However, concerns exist about this
interpretation. First, the simulation by \citet{Padoan+16} is confined
to a (250 pc)$^3$ box, and so the thermal and kinetic energy injected
by the SNe cannot escape to high altitudes above the Galactic plane
and drive a Galactic fountain, as it should, and so the simulation is
most probably overdriven. Second, the Outer Galaxy clouds are likely
to have lower column densities and thus to be less influenced by
self-gravity than their inner-Galaxy counterparts. This is consistent
with the fact that data sets from several nearby galaxies
\citep{Leroy+15} exhibit in general a Heyer-like scaling $\LR \propto
\Sigma^{1/2}$, except for the lowest column clouds, which tend to be
dominated by turbulence, and have $\LR \sim$ cst., although with a
large scatter. This trend is reproduced in simulations of less intense
turbulence in the ISM \citep{Camacho+16, Ibanez+16}.

Note that the scenario of turbulence driven by internal feedback from
stellar sources has been assumed by many analytical or semi-analytical
models, which have suggested that the SFR can self-regulate and maintain
near-virial balance in the clouds \citep[e.g.,][] {NS80, FC83, McKee89,
KMM06, Goldbaum+11}, and by numerical simulations of the effect of
feedback from stellar outflows on clump-sized ($\sim 1$ pc) structures
\citep[e.g.,] [see also the reviews by \VS\ 2011 and Krumholz et
al.\ 2014] {NL05, NL07, LN06, Carroll+09}. However, the full numerical
simulations of MCs mentioned above \citep{Dale+12, Dale+13, KT12,
KT13, Colin+13, IH15, IH17, Koertgen+16, Wareing+17a, Wareing+17b,
Seifried+18}, which have included various forms of feedback, generally
suggest that the result is to destroy the cloud by a combination of
``evaporation'' to an ionized phase and dispersal of the remaining
cold, dense fragments, rather than maintaining a near-equilibrium
state in the cloud. An analytical model that has considered ionizing
feedback in non-spherical clouds, with the resulting cloud dispersal,
has been presented by \citet[] [see also V\"olschow et al.\ 2017]
{ZA+12}.

On the other hand, the possibility of external driving has been favored by
numerical simulations of cloud formation by converging flows in the warm
neutral medium (WNM)
\citep[e.g.,] [] {VS+07, VS+10, HH08, KH10, Micic+13}. However, the
turbulence produced in the clouds during their assembly is generally
found to be too weak to support the cloud, and global collapse ensues
nevertheless. This is essentially a consequence of the facts that a) the
turbulence injected by the assembly process remains at a constant level,
while the mass of the cloud continues to increase \citep[see also] []
{FW06}, and b) this constant level of the velocity dispersion is only
moderately supersonic \citep[sonic Mach numbers $\Ms$ of a few; e.g.,]
[] {KI02, AH05, VS+06, VS+07, Banerjee+09}, significantly weaker than
the highly supersonic regime ($\Ms \gtrsim 10$) known to exist in GMCs
\citep[e.g.,] [] {HB04}.\footnote{Note that \citet{KH10} report that, in
their simulations, the internal velocity dispersion of {\it clumps}
produced in their colliding flow simulations agrees with a Larson
scaling, but it should be noted that the vast majority of the clumps
they examined had sizes smaller than 1 pc, so they did not probe the GMC
regime.}


Global hierarchical collapse, on the other hand, naturally explains the
properties of the observed motions in the clouds, because the motions are not
driven by stellar feedback, and so no stellar population needs to be
present in order to attain the observed motions, explaining property (1)
above. In addition, the motions necessarily correspond to equipartition
with the gravitational energy, since they amount essentially to
free-fall, and the free-fall speed is observationally indistinguishable
from the virial speed, given typical uncertainties in the measurements
\citep{BP+11a}, thus explaining property (2).

 It is worth remarking that the ISM and MCs are undeniably highly
turbulent, given their very large Reynolds numbers \citep[e.g.,] []
{ES04}. However, the conclusion from this section is that the
nonthermal motions observed in MCs are likely to consist of a
combination of infall and truly turbulent motions, resulting in
what \citet{BP+11a} refer to as ``chaotic infall''. The latter may tend to
oppose the collapse  at large scales (while producing small-scale
compressions), but they are likely not as strongly supersonic as they
are normally thought to be, since they may actually be seeded by the
converging flows that form the clouds, but only to a moderately
(rather than strongly) supersonic regime, and then fed by the collapse
itself \citep[e.g.,] [] {VS+98, RG12, MC15}, therefore being a
byproduct of the collapse rather than an external agent capable of
preventing it.

\section{Hoyle fragmentation revisited} \label{sec:nonlin_Hoyle}

\subsection{General considerations and assumptions} \label{sec:general_consid}

If we accept the hypothesis that the supersonic motions observed in the
clouds are {\it not} dominated by random turbulence \citep[except
perhaps those observed in the lowest-column-density clouds;] []
{Leroy+15, Camacho+16, Traficante+18a}, a simple alternative is to
return to the proposal of \citet{GK74}, that they are dominated by
infall. However, in view of the discussion in Sec.\
\ref{sec:deconstr} above, the infall should not be assumed to be
monolithic nor spherically symmetrical.  Instead, it can be safely 
considered to be {\it hierarchical},  as we will show in this
section. The fundamental underlying mechanism is essentially that
outlined by \citet{Hoyle53}, and later laid on a more quantitative
mathematical foundation by \citet{Hunter62, Hunter64}.  The mechanism is
based on the fact that, for a nearly {\it isothermal} medium \citep[and
more generally, for any polytropic flow with an effective polytropic
exponent $\gamma < 4/3$; e.g.,] [] {Chandra61, VPP96}, the Jeans mass
{\it decreases} with increasing density so that, as an isothermal cloud
of fixed mass $M$ contracts gravitationally, the number of Jeans masses
it contains increases with time, and so, new stages of collapse are
initiated as time proceeds, as sequentially smaller masses become
unstable. This trend continues until the flow ceases to behave
isothermally, as it eventually becomes optically thick, as pointed out
by \citet{Hoyle53} himself, who estimated that the final fragment masses
should be $\sim 0.1\, \Msun$, not far from the presently accepted lower
limit of the stellar initial mass function.

As appealing as it was, this mechanism later fell in disfavor when it
was realized that, for spherical, nearly uniform clouds {\it starting
with masses close to the Jeans mass and with linear (small-amplitude)
density fluctuations}, the largest growth rates correspond to the
largest size scales, and so the small-scale perturbations do not have
time to complete their collapse before the whole cloud does
\citep{Tohline80}. However, as mentioned above, the recent numerical
simulations of GMC formation including cooling and self-gravity 
show that fragmentation clearly does occur
\citep[e.g.,] [] {KI02, AH05, AH10, Heitsch+05, Heitsch+08a,
Heitsch+08b, VS+07, VS+10, VS+11, HH08, Hennebelle+08, Banerjee+09,
Colin+13, Micic+13, FK14, Carroll+14, Fogerty+16, Koertgen+16, Wareing+19}.

Two likely reasons why this {\it Tohline barrier}, as we will refer to it,
is overcome in the numerical simulations are that a) the clouds formed
by large-scale coverging flows that coherently trigger phase transitions
from the warm to the cold atomic phase quickly acquire many Jeans
masses, and b) the assembly mechanism drives moderately supersonic
turbulence ($\Ms \sim 3$) through the nonlinear thin shell instability
\citep[NTSI;] [] {Vishniac94}, which, aided by the thermal,
Kelvin-Helmholz and Rayleigh-Taylor instabilities, causes significantly
nonlinear (large-amplitude) density fluctuations \citep{KI02,
Heitsch+06, VS+06, Banerjee+09, Wareing+19}. Moreover, the geometry of the flow
collisions leads to the formation of flattened rather than spherical
clouds. Thus, the main conditions on which the {\it Tohline barrier} is
based, namely a spherical geometry, linear density fluctuations, and the
closeness to the Jeans mass, are violated in the mechanism of GMC
formation by converging flows in the warm neutral medium, and thus
Hoyle-type gravitational fragmentation appears feasible again.

\begin{figure*}
\includegraphics[width=1\hsize]{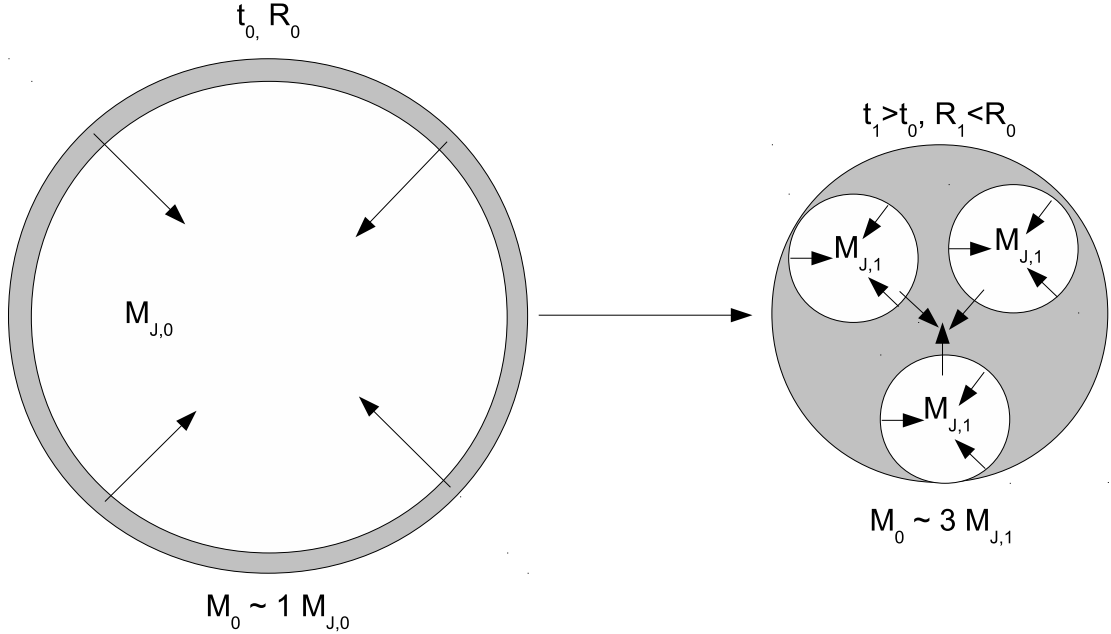}
\caption{Schematic representation of the first step of fragmentation. A
cloud of mass $M$ slightly greater than its initial Jeans mass $\MJo$
begins to contract gravitationally. As it contracts, its mean density
increases, and thus its Jeans mass decreases, so that at later times the
number of Jeans masses it contains is larger, and fragments again into
objects with masses of the order of the instantaneous Jeans mass. The
fragments begin to contract themselves, while participating of the
large-scale contraction flow as well.}
\label{fig:schematic}
\end{figure*}

In the remainder of this section, we first outline the sequential,
 global hierarchical fragmentation mechanism, and then compute
the timescale for the onset of the collapse of objects of different
masses. The mechanism proposed can be summarized as follows:

\begin{enumerate}

\item The  global, hierarchical gravitational collapse amounts to
a mass and energy cascade from large to small scales driven by
self-gravity \citep{Field+08}, so that at each moment in time each
structure is accreting from its parent structures and onto its own
child substructures.

\item The nonthermal motions in the gas consist of two main components:
a moderately supersonic turbulent background, with typical sonic Mach
number $\Ms \sim 3$, and a dominant infall flow.  

\item The global infall flow is {\it hierarchical}, and consists of
large-scale flows  (which are often filamentary; see Sec.\
\ref{sec:molec_fil_form}) directed toward the minimum of the
large-scale potential wells, on top of which ride flows directed
toward local potential minima, therefore causing Hoyle-like
fragmentation. This type of flow has been observed numerically
\citep{Smith+11, GV14, ZA+17, Gomez+18} and suggested observationally
\citep{Longmore+14}. The latter authors refer to this type of flow as a
``conveyor belt'' flow, and we will maintain this nomenclature.

\item The hierarchical collapse process consists of the sequential
destabilization of progressively smaller-mass density fluctuations, as
the global collapse causes an increase of the mean density within the
cloud, and thus causes the mean Jeans mass in the cloud to  decrease.
Therefore, progressively smaller-mass turbulent density fluctuations
become Jeans-unstable as time proceeds, and begin their own collapse
at sequentially later times.

\item As the cloud's mean density increases and the average Jeans mass
decreases, the cloud's mass becomes progressively larger than the mean
Jeans mass, and the large-scale collapse approaches a pressureless
regime. This causes the collapse motions to amplify anisotropies
\citep{Lin+65} and to produce sheets and filaments \citep{GV14}.

\item The collapse of each mass scale starts at a finite time during the
evolution of the cloud, with a finite initial radius, and from zero
local infall speed \citep{BP+18}. Figure \ref{fig:schematic}
schematically illustrates the first step of this hierarchical collapse
process.

\item While the small scales undergo their own collapse, they
simultaneously participate of the large-scale collapse; i.e., they fall
into the large-scale minimum of the gravitational potential, and so the
large-scale collapse amounts to the merger of the locally-collapsing
regions. 


\end{enumerate}

\subsection{Definitions and order-of-magnitude quantities}
\label{sec:seq_definitions}

In this section we will consider a Hoyle-like mechanism assuming
spherical geometry which, although higly idealized, illustrates the
essential aspects of GHC, and also serves as the basis for
further refinements, some of which we will consider in the subsequent
sections.

For the sake of calculation, we consider a spherical isothermal cloud
that starts out with a fixed mass $M$ just above its Jeans
mass. Although, as mentioned in Sec.\ \ref{sec:general_consid}, in
reality MCs are expected to quickly acquire many Jeans masses, the
calculation is made simpler by assuming our cloud contains just over
one Jeans mass but can fragment nevertheless and collapses
pressurelessly. This assumption also applies to later stages of
fragmentation. Generalization to a multi-Jeans mass scenario is
trivial. We also neglect the role of accretion onto each level
of the hierarchy.

The Jeans mass is in general defined as
\beq
\MJ = \frac{\pi^{5/2}}{6}\, \frac{\cs^3}{G^{3/2} \rho^{1/2}}.
\label{eq:MJz}
\eeq
That is, $\MJ$ is defined as the mass contained in a uniform sphere with
density $\rho$, isothermal sound speed $\cs$, and radius $\LJ/2$,
where 
\beq
\LJ = \left(\frac{\pi \cs^2} {G \rho} \right)^{1/2} 
\label{eq:LJ}
\eeq
is the Jeans length.

The cloud thus starts contracting gravitationally, but, as it contracts
and its mean density increases, its mean Jeans mass decreases, so the
number of Jeans masses it contains increases with time. 
Also, because fragments at all levels of the hierarchy are collapsing
on their own,  after some time (see below), they will all exhibit
internal contraction velocities of the order of their free-fall
velocity,
\beq
\vff = \sqrt{\frac{2 G \Mf} {\Rf}},
\label{eq:vff}
\eeq
where $\Mf$ and $\Rf$ are the mass and radius of a fragment,
respectively, {and we have neglected a geometrical coefficient of
order unity which, for a uniform-density sphere, is 3/5}. However, this velocity is an upper limit, because it
assumes that the fragment had an infinite radius when it began
contracting. In reality, as stated in item 6 of Sec.\
\ref{sec:general_consid}, the local contraction began with a certain
finite radius $\Rof$ equal to half the local Jeans length. Thus, the
actual infall speed can be calculated from energy conservation
\citep{BP+18}, writing $\Ek + \Eg = \Etot$, where $\Ek = 1/2 M
\vg^2$ {is the infall kinetic energy}, $\Eg = -GM^2/R$ {is the
instantaneous gravitational energy}, and $\Etot = -GM^2/\Rof$ is the
gravitational energy of the fragment (neglecting geometrical factors of
order unity) at the time {when it started its own contraction}. Thus,
the gravitationally-driven infall speed is given by
\beq
\vg = \sqrt{2G\Mf\left(\frac{1}{\Rf} - \frac{1}{\Rof} \right)}.
\label{eq:vg}
\eeq
Using the assumptions of sphericity and of a fixed mass $M$, this
expression can be written in terms of the column density $\Sigma
\equiv M/\pi R^2$ and the Larson ratio $\LR$ \citep[cf.\ Sec.\
\ref{sec:probs_turb};] [] {KM86, Heyer+09} as 
\beq
\LR_{\rm g} = \sqrt{2 \pi G \Sigma \left[ 1 - \left(\frac{\Sigma_0}
{\Sigma}\right)^{1/2} \right]},
\label{eq:LRg}
\eeq
where the subindex ``g'' denotes the Larson ratio corresponding to the
gravitational velocity given by eq.\ (\ref{eq:vg}), and $\Sigma_0$
denotes the initial column density of the object.

In addition, \citet{BP+18} also considered that the cloud may have
started with an initial inertial (i.e., not driven by the self-gravity
of the cloud) or ``turbulent'' velocity which, however, does not
necessarily provide support. Instead, it may constitute a large-scale
compression driven by a passing supernova shock front, or be a generic
compressive velocity field driven a larger-scale agent, such as the
gravitational potential of a stellar spiral arm into which the gas may
be entering. \citet{BP+18} further assumed that the inertial velocity
scales with the size of the clump as 
\begin{equation}
\sturb = v_0 \left(\frac{R}{R_0}\right)^\eta,
\label{eq:turb_scaling}
\end{equation}
where, for example, $\eta=1/2$ would correspond to a standard
supersonic turbulent scaling, while larger values would allow for the
possibility that the inertial motions dissipate in the objects as they
contract. As noted by \citet{BP+18}, the scaling given by eq.\
(\ref{eq:turb_scaling}) should not be confused with that followed by
the component of the velocity driven by self-gravity, which, as pointed out in
\citet{BP+11a}, is instead expected to follow a relation of the form
$\sv/R^{1/2} \propto \Sigma^{1/2}$ (cf.\ Sec.\ \ref{sec:probs_turb});
i.e., the gravitational velocity depends not only on size, as the
turbulent one does, but also on the column density. In this case,
adding the gravitational and the inertial components of the velocity
in quadrature, the resulting expression for the total Larson ratio is
\beq
\LR_{\rm tot} = \frac{\sigma_{\rm tot}} {R^{1/2}} = \sqrt{2 \pi
G\Sigma \left[1 - \left(\frac{\Sigma_0}{\Sigma}\right)^{1/2} \right] +
\frac{\sturb^2} {R}},
\label{eq:LRtot}
\eeq

\begin{figure*}
\includegraphics{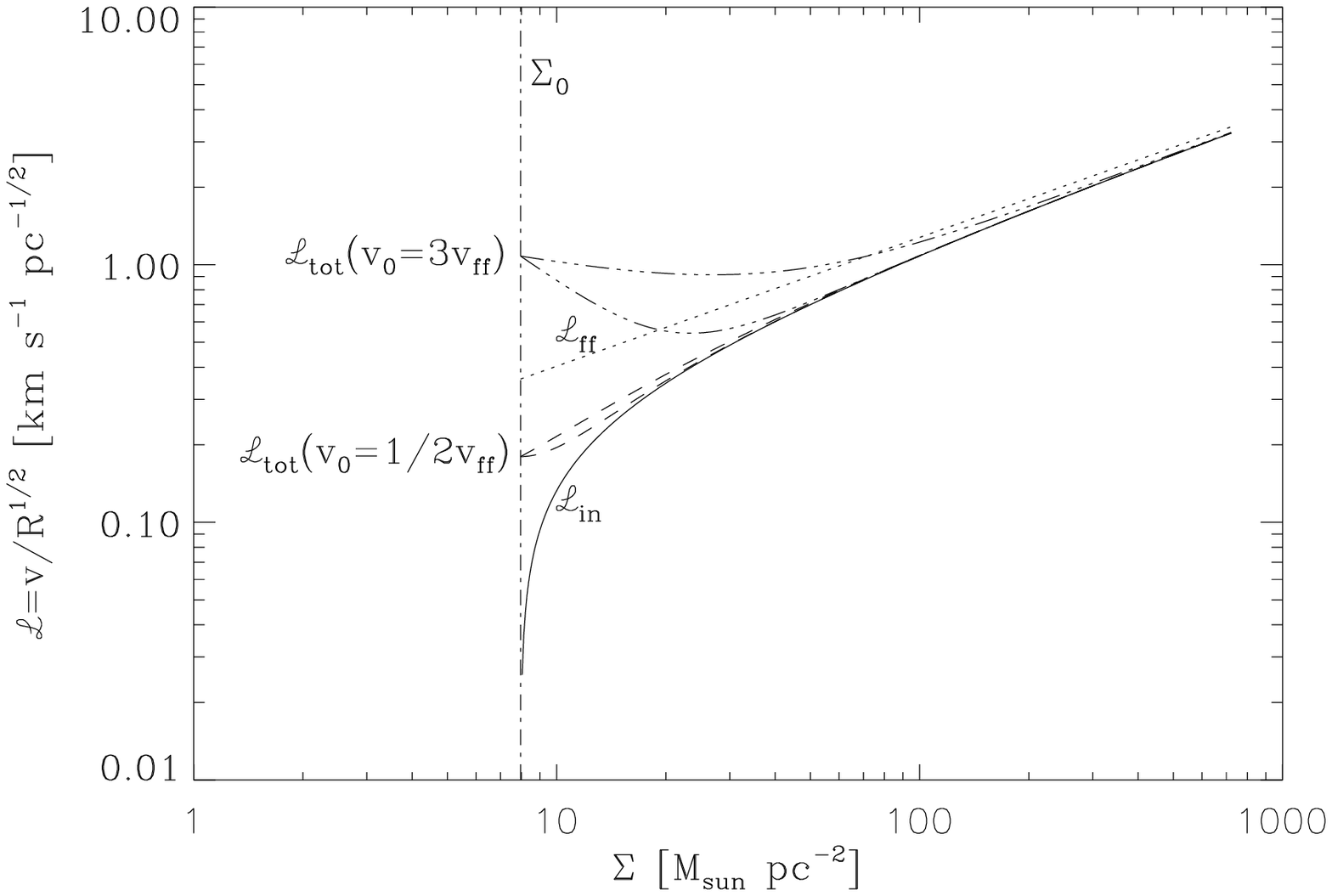}
\caption{Evolution in the $\LR$-$\Sigma$ plane of dense cores that
become Jeans unstable at time $t_0$, with initial radius $R_0 = 0.2$
pc and mass $M = 1\, \Msun$, implying an initial column density
$\Sigma_0$, shown by the vertical dashed-dotted line, for various
cases of initial inertial (i.e., non-self-gravitating) velocities,
given by eq.\ (\ref{eq:turb_scaling}), which however do not provide
support, but rather represent an external compression that initiates
the formation of the core. The {\it solid line} shows the locus of a
core with $v_0 = 0$. The {\it dotted line} shows the locus of a core
that contracts at the free-fall speed; i.e., assuming $\Rof =\infty$,
or, equivalently, $\Sigma_0 =0$. The {\it dashed lines} show the
evolutionary paths of 2 cores with a combined inertial+gravitational
velocity with $v_0 = 3 \vg$ (cf.\ eq.\ [\ref{eq:turb_scaling}]). The
{\it dash-triple dot lines} show the corresponding evolutionary paths
for 2 cores with $v_0 = 1/2 \vg$. For both sets of cores, the upper
curve corresponds to $\eta = 1/2$ in eq.\ (\ref{eq:turb_scaling}),
appropriate for a Burgers spectrum, and the lower one to $\eta = 2$,
loosely representing dissipation in dense objects.  Figure from
\citet{BP+18}.}
\label{fig:Lin_Lff}
\end{figure*}

Figure \ref{fig:Lin_Lff} shows the evolution of cores following eq.\
\ref{eq:LRtot}, for various cases of the inertial, or ``turbulent''
velocity in the $\LR$ {\it vs.} $\Sigma$ diagram \citep{KM86, Heyer+09},
which we refer to as the \LS\ diagram.
It is noteworthy that, in all cases, when the collapse is in an advanced
stage, the velocity approaches the limit given by the free-fall speed,
eq.\ (\ref{eq:vff}). 


{It is also convenient to write the virial paramenter, $\av \equiv 2
\Ektot/|\Eg|$ \citep{BM92}, in the case that the kinetic energy $\Ektot$
contains a contribution
from truly turbulent random motions, with velocity dispersion $\sturb$,
and another from gravitationally-driven infall motions, with
characteristic velocity $\vg$, given by eq.\ (\ref{eq:vg}). Noting
that, in the definition of $\av$, $\Eg = 1/2 M \vff^2$, and using eqs.\
(\ref{eq:vff}) and (\ref{eq:vg}), we obtain
\begin{eqnarray}
\av &=& \frac{2(\sturb^2 + \vg^2)} {\vff^2} \nonumber \\
&=& 2 \left[\frac{\sturb^2} {\vff^2} +  \left(1 - \frac{R} {R_0}
\right) \right].
\label{eq:alpha_vir_composite}
\end{eqnarray}
We will use this expression in our discussion of the inverse relation
between $\av$ and mass in Sec.\ \ref{sec:alpha-mass}.
}

To conclude this section, it is important to note that, in this basic
picture, we are neglecting all geometrical aspects, such as the fact
that, in realistic numerical simulations of MCs, the collapse is
systematically observed to develop filamentary structures
\citep{GV14}, and the collapse of filaments involves longer timescales
than the free-fall time for spherical symmetry \citep{Toala+12,
Pon+12}.

\subsection{The sequential onset of collapse of progressively lower-mass
objects}  \label{sec:seq_coll}  

We now calculate the timescales for sequential destabilization of
progressively smaller masses of the {\it typical} (rms) density
fluctuation. Note that below we also calculate a different timescale:
that of the {\it most extreme} density fluctuations, which are the first
to occur.

The calculation of the destabilization timescale for the typical density
fluctuation requires a specific model for the nature of the density
fluctuations, which, as stated in Sec.\ \ref{sec:general_consid}, need
to be either nonlinear, and/or contain many Jeans masses in order to
be able to continue fragmenting. Here we
perform a proof-of-concept calculation assuming that the fluctuations are
generated by moderately supersonic turbulence, which, as stated in
assumption 2 above, we generically describe by a typical Mach number
$\Ms \sim 3$, so that the typical turbulent density fluctuations
have an rms amplitude of order $\rrms/\rho_0 \sim \Ms^2 \sim 10$, where
$\rho_0$ is the mean density in the medium.

Let $M_0$ be the total mass of the cloud. The time for  an
arbitrary fluctuation of mass $M < \Mo$ to become unstable is the
time for the Jeans mass in the typical (rms) density fluctuation to
become equal to its actual mass $M$; that is, for $\MJrms(t) = M$. We
now proceed to calculate this time. The evolution of the radius as a
function of time during free-fall is a transcendental
equation. Therefore, in what follows, it will be convenient to use the
analytic fit to this evolution provided by \citet{Girichidis+14},
\beq
R(\tau) = R_0 (1 - \tau^2)^{a/3},
\label{eq:r_of_t}
\eeq
where $\tau \equiv t/\tffo$ is the time in units of the free-fall time
for the initial density ($\rho_0$) of the object, $\tffo$, $R_0$ is the
initial radius, and $a \approx 1.8614$ is a parameter for which the fit
remains within 0.5\% of the actual solution during the whole collapse
time.

A very important feature of the free-fall process is that it is extremely
nonlinear, {\it proceeding very slowly at the beginning, and 
accelerating enormously towards the end}. \citet{Girichidis+14} note 
that, after 50\% of the free-fall time, the radius has only decreased by
$\sim 16\%$, and that after 99\% of the collapse time, the radius has
only decreased by one order of magnitude.

Given expression (\ref{eq:r_of_t}) for the temporal evolution of the
radius, we can write an expression for the evolution of the mean density,
\beq
\rho(\tau) = \rho_0 (1 - \tau^2)^{-a},
\label{eq:rho_of_t}
\eeq
and, from this, an equation for the Jeans mass at the instantaneous
mean density; that is, the instantaneous mean Jeans mass in the cloud:
\beq
\MJ(\tau) = \MJ(\rho_0) (1 - \tau^2)^{a/2}.
\label{eq:MJ_of_t}
\eeq

Now, as stated above, the
typical rms density fluctuation is given by 
\beq
\rrms(\tau) \sim \Ms^2 \rho(\tau),
\label{eq:rhorms_of_tau}
\eeq 
and the Jeans mass in this fluctuation is
\beq
\MJrms(\tau) \approx \frac{\MJ(\rho_0)} {\Ms} \left(1 -
\tau^2\right)^{a/2}.
\label{eq:MJrms_of_tau}
\eeq

This equation can be then inverted to find the time required for a mass
$M = \MJrms$ to become unstable. We find for this time, in units of
the initial free-fall time in the cloud,
\beq
\taumu \equiv \frac{\tmu} {\tffo} \approx \left[1 - \mu^{2/a}
\right]^{1/2}, 
\label{eq:t_M_unst}
\eeq
where $\mu \equiv \Ms M/\MJ(\rho_0)$ is the fragment mass normalized to
the initial fluctuation Jeans mass in the cloud. 
Figure \ref{fig:ttot_of_mu} shows the dependence of this destabilization time
(in units of the initial free-fall time) for mass $M$ (in units of the
initial Jeans mass) for various values of the Mach number. We consider
only masses $\le \MJo/2$ assuming that fragmentation starts when there are
at least two Jeans masses in the cloud.
%
%
%
%
\bfg
\includegraphics[width=0.48\textwidth]{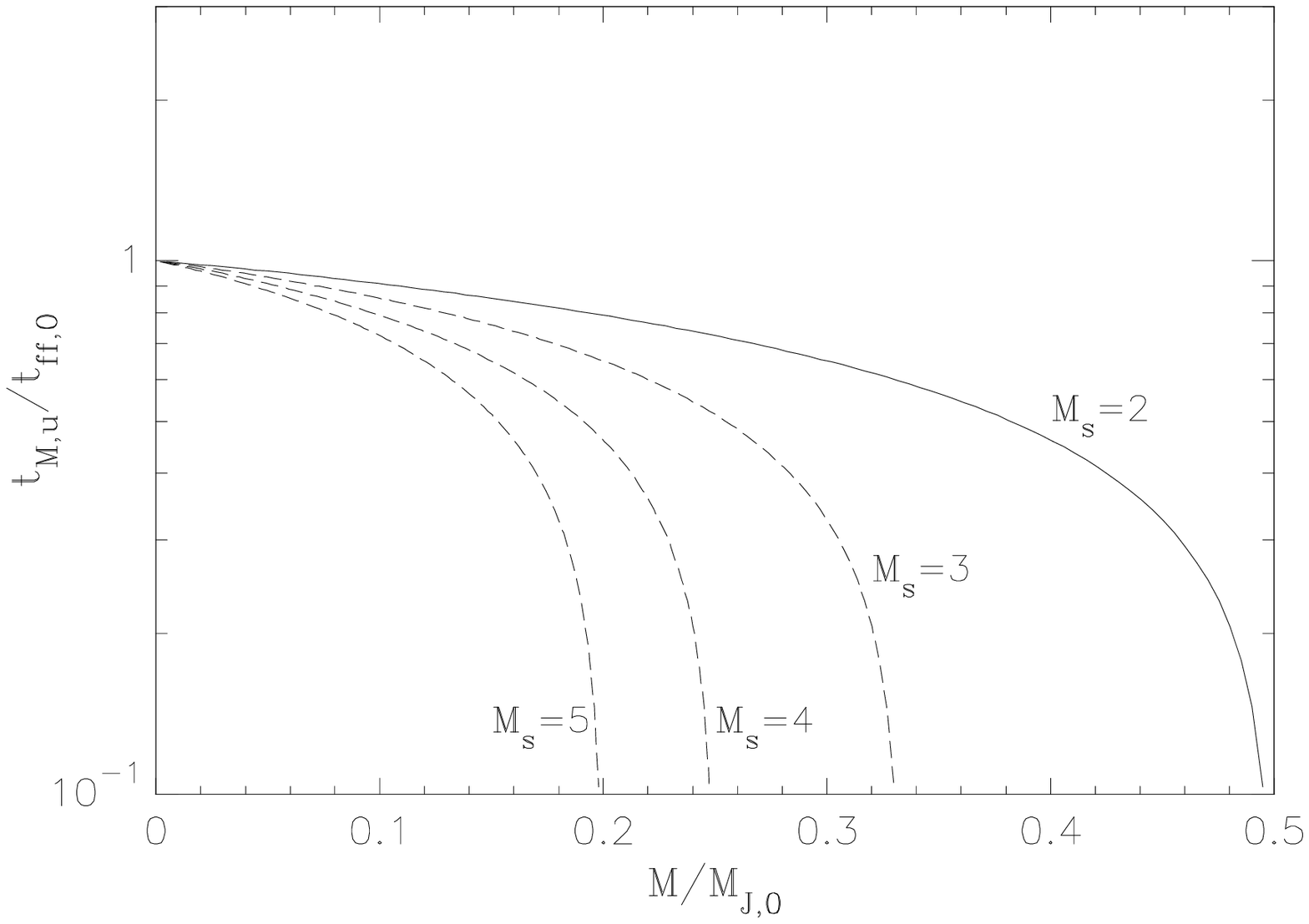}
\caption{Time for destabilization (i.e., onset of collapse) of a
fragment of mass $M$ for various values of the turbulent Mach number,
$\Ms$, as given by eq.\ (\ref{eq:t_M_unst}). The time and mass are
respectively normalized to the free-fall time and the Jeans mass at the
mean initial density.}
\label{fig:ttot_of_mu}
\efg
%
%
We can see that,  for larger Mach numbers, smaller masses can
begin to collapse during the collapse of the largest scale.

It is important to note that the above calculation is only an
approximation intended as a proof-of-concept for the sequential onset of
collapse of the hierarchy of mass scales in a cloud, and probably
provides only an upper limit to this time. A more precise calculation
should consider several additional factors that can shorten the time for
the onset of collapse of structures of mass $M$, such as:
\begin{itemize}

\item This calculation refers only to the {\it first} stage of
fragmentation for any given mass. We have neglected the subsequent
fragmentation of a fragment.

\item The collapse of a structure embedded in an also-contracting
background takes a {\it shorter} time than the standard collapse time
for isolated objects \citep{Toala+15}.

\item In reality, there is a distribution of density fluctuations, and
therefore fluctuations of a higher density are expected to exist (if
albeit rare), so that their free-fall times are shorter.

\end{itemize}

Nevertheless, the main point of the above calculation is to show the
fact that progressively smaller masses become unstable at progressively
later times.

\subsection{The time to reach the $n$-th level of fragmentation}
\label{sec:time_to_nth}

Another relevant quantity that can be computed for the fragmentation
process is the time to reach the $n$-th level of fragmentation. This can
be done by assuming that a new fragmentation stage begins every time an
object undergoing gravitational contraction in the previous stage
attains a number $\nmj$ of local Jeans masses as its density increases
and the local Jeans mass decreases.  This is consistent with the
empirical finding by \citet{Guszejnov+18} in their numerical simulations
that a structure must contain at least 3 Jeans masses in order for it to
fragment instead of collapsing monolithically. Thus, the Jeans masses
at stages $n$ and $n-1$ of the hierarchy are related by
\beq
M_{{\rm J},n} = \frac{M_{{\rm J},n-1}} {\nmj}.
\label{eq:mass_hier}
\eeq

As in the previous section, we
consider that the fragmentation hierarchy starts at the Jeans mass in
the rms turbulent density fluctuation which, according to eq.\
(\ref{eq:MJrms_of_tau}), at the onset of global collapse, is given by
$\MJrms(\tau = 0) = \MJo/\Ms$. We can then write the successive steps of
the hierarchy as

\noindent
Level 1:
\[
M_{\rm J,rms,1} = \frac{M_{\rm J,rms,0}} {\nmj} = \frac{1} {\nmj}
\frac{M_{\rm J,0}} {\Ms},
\label{eq:level1}
\]

\noindent
Level 2:
\begin{eqnarray*}
M_{\rm J,rms,2} &=& \frac{M_{\rm J,rms,1}} {\nmj} = \frac{1} {\nmj^2}
\frac{M_{\rm J,0}} {\Ms},\\
&\vdots& \nonumber
\label{eq:level2}
\end{eqnarray*}

\noindent
Level $n$:
\beq
M_{{\rm J,rms},n} = \frac{M_{{\rm J,rms},n-1}} {\nmj} = \frac{1} {\nmj^n}
\frac{M_{\rm J,0}} {\Ms}.
\label{eq:leveln}
\eeq

On the other hand, if $\tau_n$ is the time at which the $n$-th
fragmentation level is reached, then, from eq.\ (\ref{eq:MJrms_of_tau})
we can write
\beq
M_{\rm J, rms}(\tau_n) \equiv M_{{\rm J,rms},n} = \frac{M_{{\rm J},0}}
{\Ms} \left(1 - \tau_n^2 \right)^{a/2}.
\label{eq:MJrms_n}
\eeq

\begin{figure*}
\includegraphics[width=0.48\textwidth]{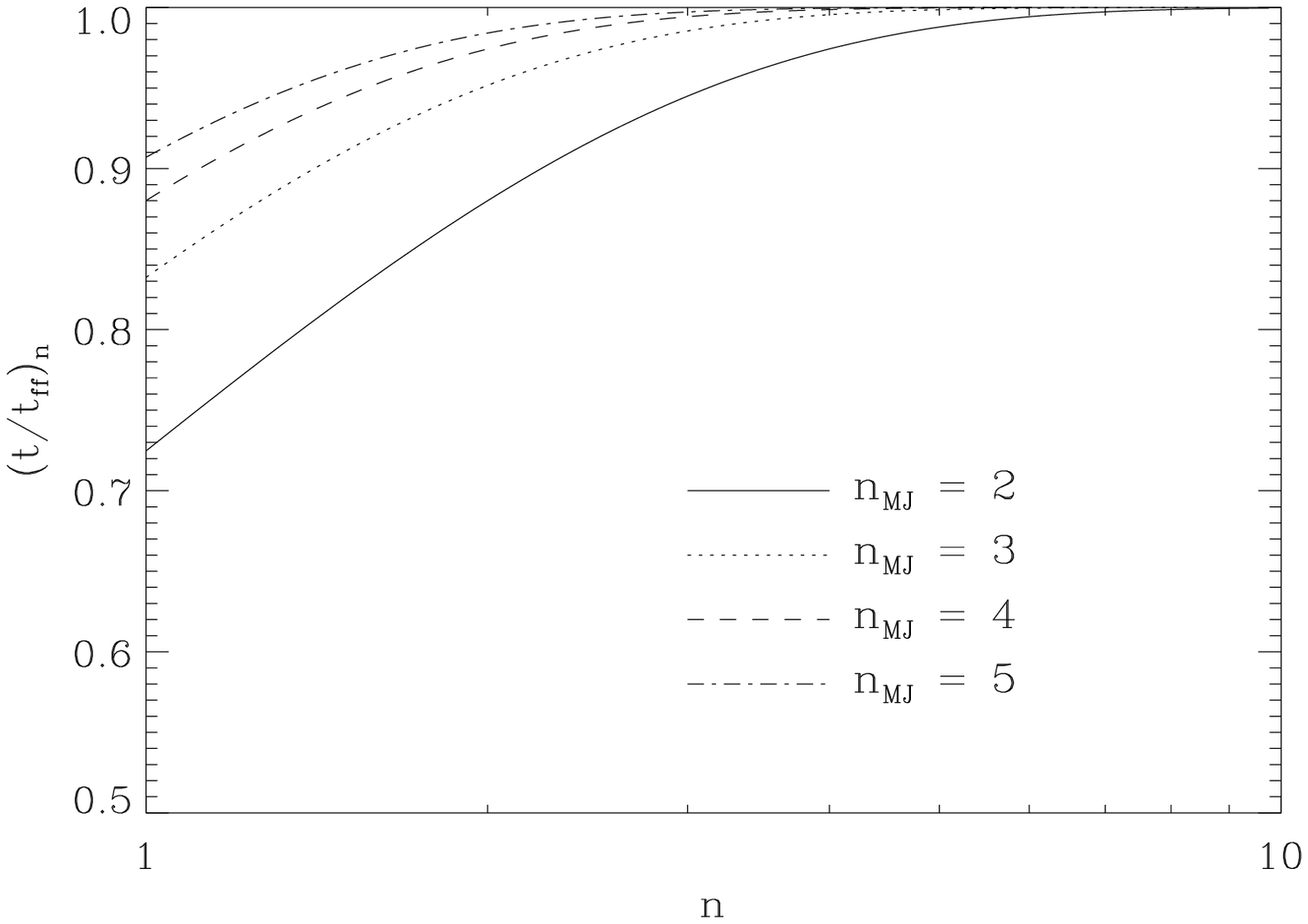}
\includegraphics[width=0.48\textwidth]{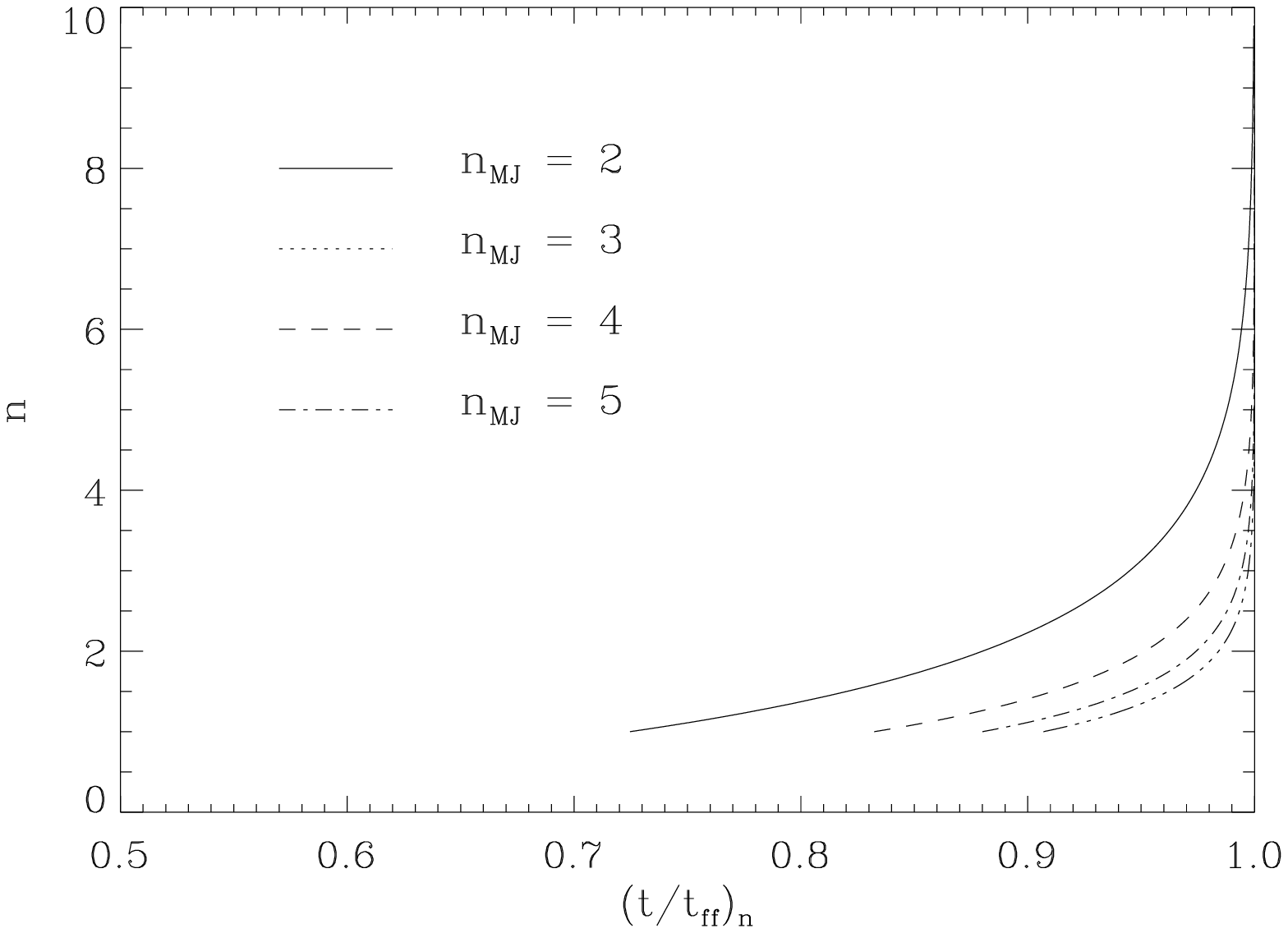}
\caption{{\it Left:} Time (in units of the initial free-fall time) to
reach the $n$-th level of fragmentation, as given by eq.\
(\ref{eq:tau_n}) for $\nmj =2$, 3, 4, and 5 (solid, dashed, dot-dashed,
and dash-dot-dotted lines, respectively). {\it Right:} Reverse plot,
showing the number of fragmentation levels reached at time $\tau$. The
line coding is the same as in the left panel.}
\label{fig:tau_n}
\end{figure*}

Therefore, solving for $\tau_n$ from equations (\ref{eq:leveln}) and
(\ref{eq:MJrms_n}), we find
\beq
\tau_n = \sqrt{1 - \left(\frac{1} {\nmj} \right)^{2n/a}},
\label{eq:tau_n}
\eeq
which gives the necessary time to reach the $n$-th level of
fragmentation. The {\it left} panel of Fig.\ \ref{fig:tau_n} shows plots
of this expression for $\nmj =2$, 3, 4, and 5 (solid, dashed,
dot-dashed, and dash-dot-dotted lines, respectively). The {\it right}
panel of this figure shows the reverse plot, of the number of
fragmentation levels reached at a given elapsed time.

It is noteworthy that this result does {\it not} depend
on the Mach number of the turbulence. We discuss insights and
implications of this result in Sec.\ \ref{sec:indep_t_Ms}.

\subsection{The time for the onset of the first fragment contraction}
\label{sec:1st_coll}

Until now, we have considered the times for the onset of gravitational
contraction {\it for the typical density fluctuations}, of amplitude
proportional to $\Ms^2$. However, the {\it first} fragments that will
become Jeans unstable are not those of typical amplitude, but rather the
most extreme ones, as they have the shortest free-fall times. We can
compute the time for the onset of the first collapses by noticing that they
can only occur when the total mass at the highest densities $M(n \ge
\nsf)$ is at least equal or higher than the Jeans mass at those
densities $\MJ(\nsf)$. That is, we write the condition for the
occurrence of the first collapse as
\beq
M(n \ge \nsf) \ge \MJ(\nsf),
\label{eq:cond_1st_coll}
\eeq
where we have denoted by $\nsf$ the density for which the free-fall time
is so much shorter than the average free-fall time in the cloud that it
can be considered instantaneous. That is, once condition
(\ref{eq:cond_1st_coll}) is met, the collapse of this region will occur
essentially immediately.

 We assume a lognormal  density PDF, appropriate for
supersonic, isothermal turbulence \citep{VS94, Padoan+97, PV98,
Federrath+08}, of the form
\beq 
P(s)=\frac{1}{\sqrt{2\pi \sigma_s^2}} ~ {\rm exp} \left[- \frac{(s -
\savg)^2}{2\sigma_s^2} \right], 
\label{eq:pdf}
\eeq
where $s \equiv \ln (\rho / \bar \rho)$, $b$ is the compressibility
parameter \citep{Federrath+08}, $\sigma_s^2= \ln (1+ b^2
\Ms^2)$, and $\savg =
-\sigma_s^2/2$. In this case, the mass $M(n \ge \nsf)$ is given by
\citep{KM05}
\begin{eqnarray}
M(n \ge \nsf) &=& \Mtot \int_{\rhosf}^\infty \rho \frac{dP} {d\rho}
d\rho \nonumber \\
&=& \frac{1}{2} \Mtot \left[1 - {\rm erf} \left( \frac{2 \,
s_{\rm SF}(t)-\sigma_s^2}{\sqrt{2} \sigma_s} \right) \right].
\label{eq:mass_frac}
\end{eqnarray}
Note that, in fact, this mass estimate is strictly a lower
limit, since the PDF contains no spatial information, and it is thus
possible that not all the mass above $\nsf$ lies in a single connected
object in the medium. However, since we are considering the global
density maximum of the flow, it is reasonable to assume that the next
densest fluid parcels belong to the same object. With this caveat in
mind, the {\it left panel} of Fig.\ \ref{fig:t_1st_coll} shows plots
of the ratio $M(n \ge \nsf)/\MJ(\nsf)$ at a fixed value of $\Mtot =
10^4 \Msun$, for various values of the rms Mach number, assuming
$\navg = 10^3\, \pcc$ and $\nsf = 10^6\, \pcc$, and the {\it right
panel} shows this ratio for various values of $\Mtot$ assuming $\Ms =
4$, and the same values of $\navg$ and $\nsf$. The horizontal line
shows the value of $\MJ(\nsf)$. The time of first collapse for these
various cases is given by the intersection of the curves and the
horizontal identity line. It can be seen that, in general, the time of
first collapse occurs relatively late in the collapse at the mean
density of the cloud,, at $t \sim 0.65$--$0.9\, \tffo$, where we
recall that $\tffo$ is the free-fall time at the initial mean
density. This is qualitatively consistent with the evolution of
numerical simulations, where the first collapsed objects appear
several {megayears} after the onset of the large-scale contraction, but a few
{megayears} before it terminates \citep[e.g.,] [] {VS+07, VS+10, HH08,
Carroll+14}.

\bfgw
\includegraphics[width=0.48\textwidth]{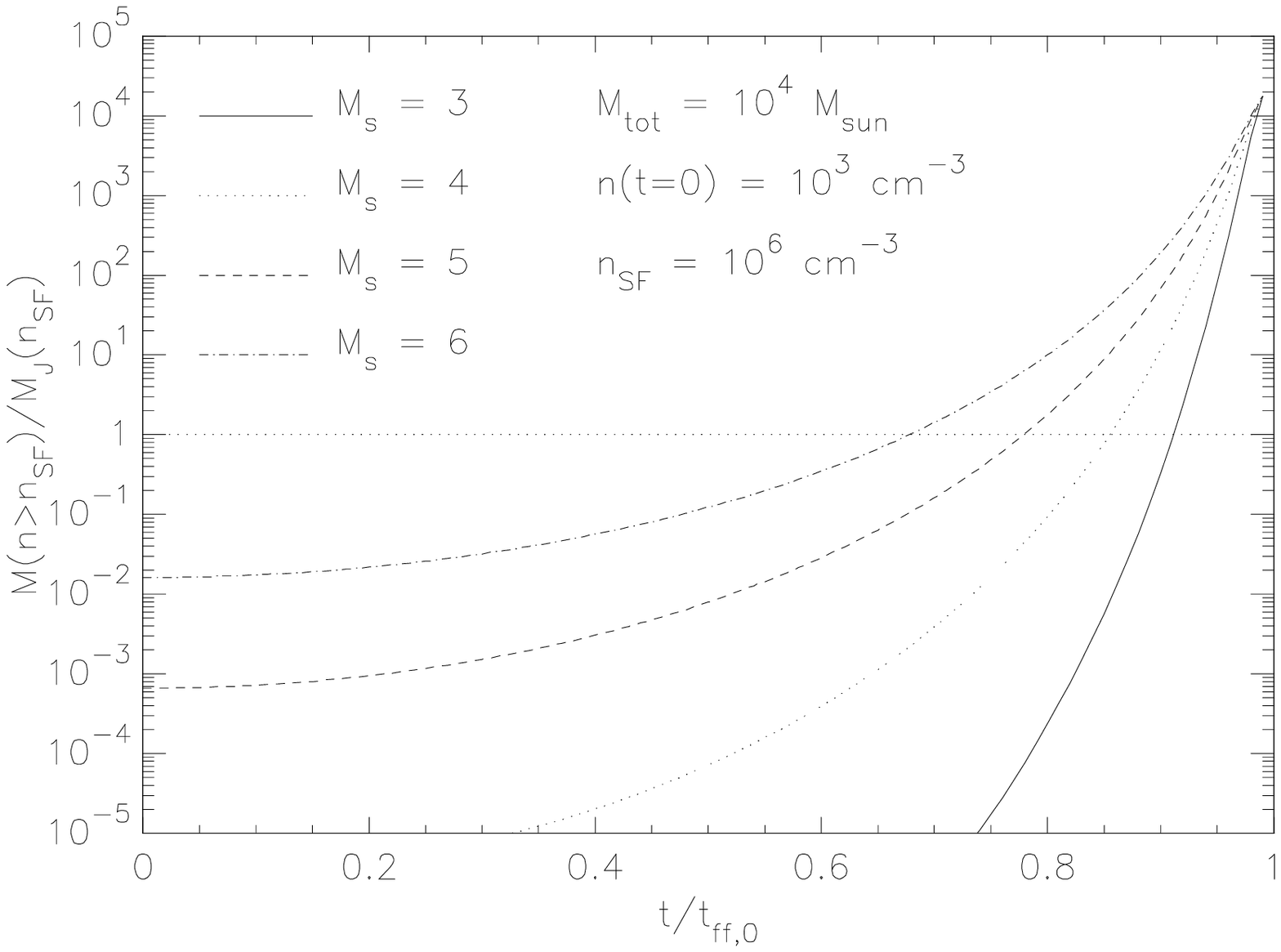} 
\includegraphics[width=0.48\textwidth]{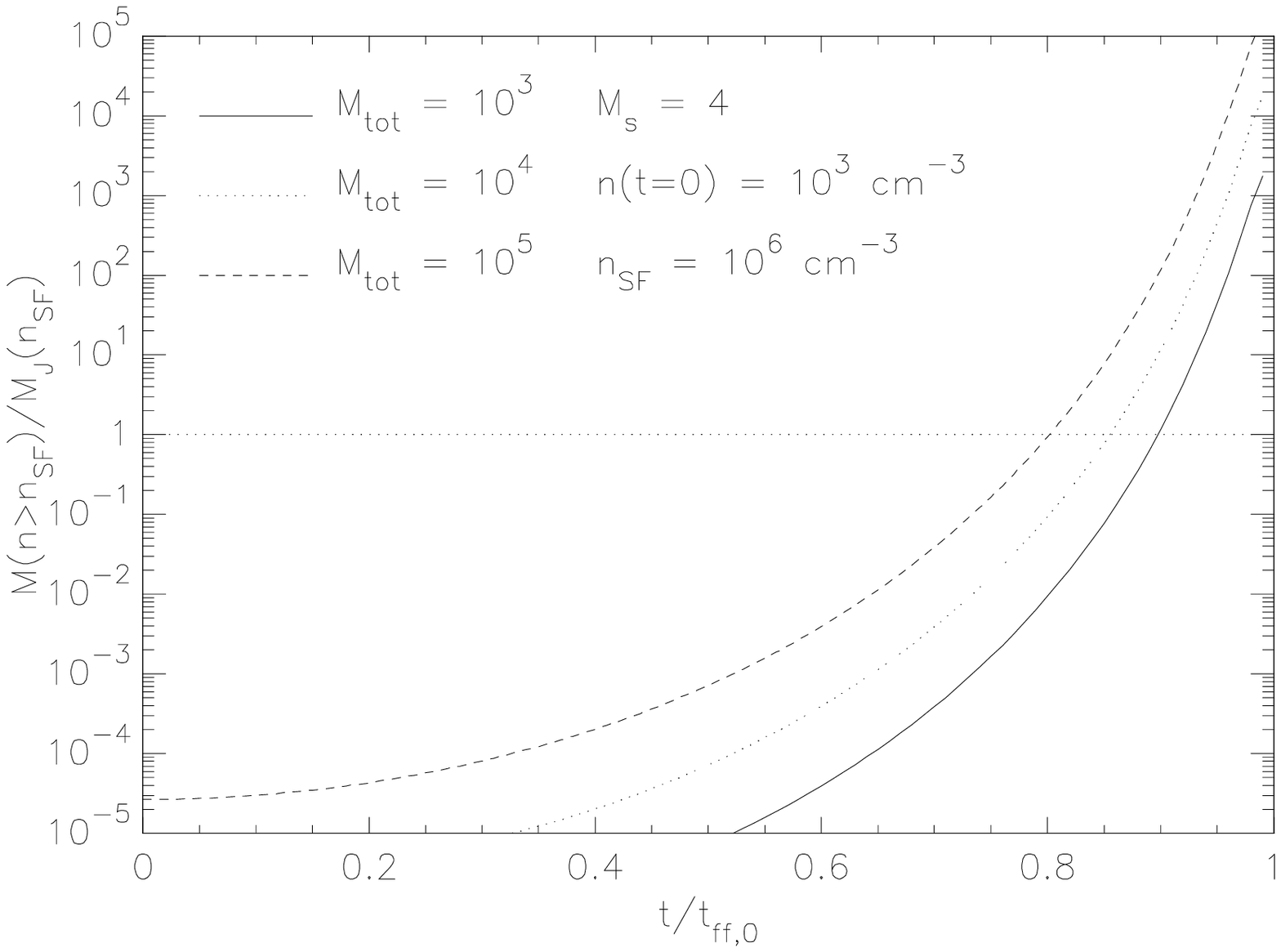}
\caption{Time for the onset of the first fragment contraction, computed as
the time at which the mass above a critical density for ``instantaneous
collapse'' ($\nsf$), denoted $M(n \ge \nsf)$ and shown by the various
curves, exceeds the Jeans mass at density $\nsf$ (horizontal dotted
line). This time is indicated by the point at which the curves intersect
the horizontal line. {\it Left panel:} Dependence of the time of first
collapse with the turbulent Mach number. {\it Right panel:} Dependence
of this time with the total initial mass of the cloud, $\Mtot$.} 
\label{fig:t_1st_coll}
\efgw

\section{The nature of the gravitational contraction}
\label{sec:nature_of_collapse}

In this section we now discuss a few properties of non-homologous
gravitational contraction that are often overlooked. 

\subsection{The non-homologous, non-simultaneous nature of the collapse}
\label{sec:non_homolog}

A frequent assumption about the process of gravitational contraction is
that it proceeds {\it homologously}; i.e., that it proceeds maintaining
uniform (although not constant in time) density \citep[e.g.,] []
{Chandra39}. This mode of collapse is accomplished if the infall
velocity increases linearly with radius, so that all the shells of a
uniform-density spherical structure reach the center at the same
time. This idealized solution, however, never occurs in practice in
media containing multiple Jeans masses, because any density
perturbations present have shorter free-fall times, and thus collapse
faster, than the background medium. The requirement of having multiple
Jeans masses allows for there to exist a sufficiently large range of
unstable scales (larger than the Jeans length) that the reduction of the
growth rate at scales just over the Jeans length (cf.\ Sec.\
\ref{sec:deconstr}) is negligible. This can be illustrated by recalling
the dispersion relation for the Jeans instability analysis which, for
linear perturbations of the form $\rho_1(x,t) = \epsilon \rho_0 \exp
\left[-i\left(kx -\omega t\right) \right]$, reads \citep[e.g.,] []
{Shu92}
\beq
\omega^2 = \cs^2 k^2 - 4 \pi G \rho_0,
\label{eq:Jeans_disp_rel}
\eeq
where $\omega$ is the growth rate, $k$ is the wavenumber, and $\rho_0$
is the mean density of the medium. As is well known, 
\[
\kj \equiv \left(\frac{4 \pi G \rho_0} {\cs^2}\right)^{1/2}
\]
is the wavenumber corresponding to the Jeans length, for which the
growth rate is zero; i.e., the restoring force from the pressure
gradient balances the gravitational pull in the marginal equilibrium
case. At the other extreme, the growth rate is maximum for $k
\rightarrow 0$; i.e., at the largest physical scales, at which the
growth rate approaches the free-fall value, $\omega \rightarrow
\sqrt{4 \pi G \rho_0}$, so that, in the linear regime, all large-scale
perturbations grow at nearly the same rate.

This linear argument, however, neglects the nonlinear effect of the
reduction of the free-fall time in the perturbation due to the locally
enhanced density. Therefore, when the dynamic range of the large-scale
perturbations (for which the pressure gradient is negligible) is
sufficiently large, the reduction of the free-fall time at the local
density perturbations allows them to outgrow the background. Indeed,
it has been recently shown numerically by \citet{Guszejnov+18} that,
for numerical boxes containing more than $\sim 3$ Jeans masses, and in
the presence of nonlinear turbulent density fluctuations,
fragmentation does occur, while, when the numerical box containing
less than $\sim 3$ Jeans masses, a single focused, monolithic collapse
mode ensues. Moreover, they showed that the number of fragments is
independent of the Mach number of the turbulence, which suggests that
the fragmentation is of purely gravitational origin.

The above discussion suggests then that the collapse of multi-Jeans-mass
clouds in the presence of nonlinear density fluctuations {\it is highly
non-homologous}, and thus the density contrast in the cloud tends to
increase over time, because the denser parts contract faster than the
lower-density ones, and quickly overtake them. 

An alternative way of seeing this is by noting that the standard
similarity treatment of gravitational contraction \citep[e.g.,] []
{Larson69, Penston69, Shu77, Hunter77, WS85} proceeds by defining a
similarity variable $\xi \equiv r/\cs t$, where $r$ is the radial
coordinate, $t$ is the time, and $\cs$ is the sound speed. The essence
of such an analysis is that one radial position in a collapsing sphere
at a given time is equivalent to a different radial position at some
other time. This can be interpreted as meaning that, at a given time
$t$, {\it the gas at a higher density nearer to the center is at a more
advanced evolutionary stage (in its approach to the collapse center)
than the gas at lower density further out}.

In standard spatial and temporal
coordinates, this implies that the gas is {\it flowing} from low- to
high-density regions, so that, as the gas density increases, a gas
parcel {\it moves} from a more distant to a more nearby location around
the collapsing center. Moreover, this process takes a finite amount of
time. 

The non-homologous nature of the collapse has various implications, most
important of which is that, contrary to the implicit assumption in the
idealized reasoning of \citet{ZP74}, {\it not all gas in a cloud reaches
the singularities (i.e., the stars) at the same time.}  Instead, the
densest parts of the cloud terminate their collapse earlier. This is at
the foundation of models for the SFR in clouds, that assume the
instantaneous value of the SFR in a turbulent cloud, characterized by a
certain density probability distribution (PDF), is given by the mass in
fluctuations with densities above a certain threshold, divided by a time
of the order of the free-fall time for the threshold \citep[e.g.,] []
{KM05, PN11, HC11, FK12, ZA+12, Volschow+17, Burkhart18}, so that the
stars formed by the earliest collapses may prevent the collapse of the
rest of the material. We return to the application of this reasoning to
the GHC scenario in Sec.\ \ref{sec:evol_SF}.

\subsection{The difference between the pre- and protostellar collapse
regimes} \label{sec:diff_pre_proto}

Another feature of the gravitational collapse process is that the pre-
and protostellar stages have important qualitative differences in the
shape of the radial infall velocity profile. As it is well known from
isothermal, spherical collapse calculations \citep[e.g.,] [] {Larson69,
Penston69, Hunter77, WS85, Naranjo+15}, the collapse process consists of
two stages, separated by the formation of a singularity (i.e., a
protostar). These correspond to the ``prestellar'' and the
``protostellar'' stages of cores.

During the prestellar stage, the gravitational contraction of
isothermal, spherically symmetric objects with a central density
enhancement asymptotically approaches a radial configuration consisting
of two distinct regions (see, e.g., Fig.\ C1 of \citet{Larson69} or
panels (a)-(c) in Fig.\ 5 of \citet{WS85}): 

\medskip
\noindent
{\bf Prestellar stage:}
\[
\mbox{Inner core:} \left\{\begin{array}{ll}
\rho(r) =  {\mbox cst.} \\
v(r) \propto -r.
\end{array} \right.
\]
\beq
\mbox{Envelope:} \left\{\begin{array}{ll}
\rho(r) \propto r^{-2}, \\
v(r) = \mbox{cst.} < 0.\\
\end{array} \right.
\label{eq:prestellar_profiles}
\eeq

Note that the infall speeds are negative because they point in the
negative radial directions. Note also that the transition between the
``inner core'' and the ``envelope'' occurs roughly at a radius equal to
the Jeans length corresponding to the central density and temperature
\citep{Naranjo+15, Keto+15}.

The transition between the two regions occurs around a radius
equal to the Jeans length at the central density \citep[e.g.,] []
{KC10}, and approaches the core's center. The uniform central density
inside this radius increases without limit with time, becoming infinite
at the time of protostar (singularity) formation, at which time the
transition radius becomes zero. That is, at the time of singularity
formation, the density profile has an $r^{-2}$ shape at all radii,
resembling a singular isothermal sphere, but with nonzero, uniform, and
supersonic infall speed at all radii \citep{Naranjo+15}.

Conversely, during the protostellar stage, the density profile increases
monotonically inwards at all radii, diverging toward the core's center,
where a finite-mass object (the protostar) is located. Two
radial regions still exist in the core, but they are now mediated by a
shock front, located at a radius $\rs = \cs t$, where $t$ is the time
since the formation of the singularity. The density and infall velocity
profiles in the two radial regions in the protostellar stage have
asymptotic regimes given by

\medskip
\noindent
{\bf Protostellar stage:}
\[
\mbox{Inner core:} \left\{\begin{array}{ll}
\rho(r) \propto r^{-3/2}, \\
v(r) \propto -r^{-1/2}.
\end{array} \right. 
\]
\beq
\mbox{Envelope:} \left\{\begin{array}{ll}
\rho(r) \propto r^{-2}, \\
v(r) = \mbox{cst.} < 0.\\
\end{array} \right.
\label{eq:protostellar_profiles}
\eeq

It is important to notice that this solution differs strongly from the
canonical ``inside-out collapse'' (IOC) \citep{Shu77}. The IOC solution
results from assuming that the initial condition for the collapse is a
hydrostatic singular isothermal sphere (SIS). This assumption requires
the pre-stellar contraction to be quasi-static, with the clump's weight
being balanced by some other force(s), such as magnetic forces or
turbulent pressure, and a tiny bit of contraction allowed by, for
example, ambipolar diffusion (AD) in the case of magnetic
support. Although \citet{Shu77} gave a number of arguments in favor of
this prestellar quasistatic contraction, \citet{Hunter77} pointed out
that numerical simulations in general tend to be more closely described
by Larson-Penston solution than by Shu's solution.  This can be
understood because, in practice, an SIS configuration is unattainable,
since it is the {\it most unstable} possible hydrostatic equilibrium in
spherical symmetry, as follows from the fact that it is the limit of
unstable Bonnor-Ebert (BE) spheres when the ratio of the
central-to-peripheral density tends to infinity. As discussed by
\citet{Whitworth+96} and \citet{VS+05}, such an unstable equilibrium is
not expected to arise spontaneously within a significantly turbulent
molecular cloud. Moreover, magnetic support is also not expected to
provide sufficient support, as GMCs are now known to be magnetically
supercritical in general \citep{Crutcher12}, and so their gravitational
contraction must proceed dynamically rather than quasistatically.
Finally, turbulent support that dissipates gradually, as often assumed
\citep[e.g.,] [] {Goodman+98, BT07, Pineda+10}, is not feasible 
because at every scale, the dominant modes of the turbulence are of
the same scale, and thus their effect is not to maintain the structure
in near-equilibrium, but rather to shear, compress or expand the
structure, as discussed in Sec.\ \ref{sec:probs_turb}. 

A common argument against the possibility of cores being in the process
of dynamical contraction already during their prestellar stage is that
the final parts of this stage produce supersonic infall speeds at the
core's envelope, while typical estimators of the infall speed, such as
blue-excess line profiles of moderately optically thick transitions
often imply subsonic infall velocities \citep[e.g.,] [] {Lee+01,
Campbell+16}. However, as shown by \citet{Loughnane+18}, the fact that
the largest infall speeds in collapsing cores occur at radii where the
density is already decreasing (see eq.\ \ref{eq:prestellar_profiles})
may cause a systematic underestimation of the infall speed by these
indicators.

\section{The complete scenario} \label{sec:model}

In this section we now compile all the evolutionary stages and
byproducts of the GHC scenario, as they have been reported in several
papers over the last decade.

\subsection{Cloud formation} \label{sec:formation}

Observationally, ``clouds'' are often defined as regions detected in
some tracer, such as $^{12}$CO. However, in numerical and analytical
studies, clouds are often defined as cold density enhancements over
some warm, diffuse background, which may constitute a different
thermodynamic phase of the gas. Here we will adopt this definition, so
that clouds can contain both cold atomic and molecular gas.

Under solar galactocentric radius conditions, the azimuthally- and
vertically-averaged molecular mass fraction is only 10-20\%, and most of
the Galactic disk volume is occupied by warm atomic gas, except in the
spiral arms, where the gas is mostly molecular \citep{Koda+16}. At the
turn of the century, several studies showed that dynamic compressions in
the warm atomic phase can nonlinearly trigger a phase transition to the
cold atomic phase \citep[e.g.,] [] {HP99, KI02}, while generating
turbulence in the cold, dense layer that forms as a result of the
compression by the combined action of the nonlinear thin-shell
\citep{Vishniac94}, thermal \citep{Field65} and Kelvin-Helmholtz
instabilities \citep[see] [for a summary of the instabilities active in
cloud formation] {Heitsch+06}.  This turbulence consists mostly of the
velocity dispersion of the cold cloudlets formed by the thermal
instablity (TI), which is a fraction of the sound speed in the warm
phase. Since the latter is $\sim 10\times$ the sound speed in the cold
phase, the cloudlets move with a velocity dispersion that is moderately
supersonic (Mach numbers of a few) with respect to the sound speed in
the cold phase \citep[e.g.,] [] {KI02, VS+06, HA07, Banerjee+09}. In
particular, \citet{Banerjee+09} showed that, if the conditions of
thermal- and ram-pressure balance are imposed separately between the
warm and cold phases, then the turbulent Mach numbers in the warm and
the cold phases are similar.

Finally, an important result is that, at the earliest stages, the
condensed layer is very thin, with column densities $\sim 10^{19} \psc$,
and therefore must consist almost exclusively of cold atomic gas, thus
being highly consistent with the conclusion of \citet{HT03} that CNM
clouds appear to be large, thin sheets, with aspect ratios of up to
$\sim 100$ \citep{VS+06}. Thus, GMCs may start their lives as thin
sheets of cold atomic gas.

It is important to remark that this formation mechanism implies that,
even though the clouds do constitute a cold, dense phase in the ISM
and are surrounded by a warm, diffuse substrate, there is a continuous
process of {\it accretion} of gas from the warm phase into the
clouds. The gas suffers a phase transition from the warm to the cold
phase upon entering the clouds, and thus the cloud boundaries are {\it
phase transition fronts} \citep{Banerjee+09}, but no restriction to
the flow of gas across these fronts exists \citep{BP+99a, HP99, KI02,
AH05, Heitsch+05, VS+06, Banerjee+09}. This kind of boundary is in
sharp contrast with the classical picture that the cloud boundaries
are contact discontinuities, where the warm phase ``confines'' the
cold one, but there is no motion across the boundaries. This also
implies that, in general, {\it cloud masses vary over time} rather
than being fixed. During the early epochs of the clouds, their masses
in general grow over time, until they begin to be eroded by their
newly formed stars (Sec.\ \ref{sec:evol_SF}). Observational evidence
for this mass evolution of the clouds has been presented by
\citet{Kawamura+09}, who showed that GMCs in the Large Magellanic
Cloud with more evolved stellar populations tend to have larger
masses.

\subsection{Onset of global gravitational contraction}
\label{sec:grav_contr}

An often neglected fact is that, as pointed out by \citet{GV14}, when
the atomic gas undergoes a transition from the warm, diffuse phase to
the dense, cold one, the temperature decreases by a factor $\sim 100$
and the density increases by the same factor, so that the Jeans mass,
$\MJ \propto \rho^{-1/2} T^{3/2}$, decreases by a factor $\sim 10^4$,
from $\MJ \approx 7 \times 10^7\, \Msun$ at $T = 7000$ K and $n = 0.3
\pcc$, to $\MJ \approx 7
\times 10^3\, \Msun$ at $T = 70$ K and $n = 30 \pcc$. 

Another phase transition occurs when the gas becomes molecular
\citep[e.g.,] [] {KI00}, which,
at the same thermal pressure, typically has $T = 10$ K and $n = 200\,
\pcc$, and furthermore the number density $n$ drops by a factor of 2, so
that the Jeans mass is another two orders of magnitude lower, at $\MJ
\approx 30\, \Msun$, in this gas. That is, {\it in molecular gas the
Jeans mass is six orders of magnitude lower than in the WNM}. This
implies that the phase transition from WNM to molecular gas provides a
mechanism for creating strongly Jeans unstable gas, if the transition is
coordinated by large-scale compressions that coherently produce large
masses of cold, dense gas. As mentioned in Sec.\
\ref{sec:formation}, the initial step of condensation produces thin
sheets of CNM which in the simulations are observed to fragment into a
network of clumps and filaments \citep[see, e.g., Fig.\ 10 of] []
{VS+06}. {\it It is this ensemble of clumps and filaments that is observed
to be globally gravitationally unstable} in numerical simulations of
cloud formation and collapse \citep{VS+07, VS+10, HH08, Colin+13,
Carroll+14}. In addition, those simulations also show that, in
general, the turbulence (mostly clump-to-clump velocity dispersion)
generated in the clump ensemble (``the cloud'') produced by the large-scale
compressions is {\it not} sufficient to support it against
its own self-gravity once it becomes Jeans unstable, and therefore the
cloud begins to undergo gravitational contraction.

\subsection{Molecule and filament formation} \label{sec:molec_fil_form}

It has been known for a few decades \citep[] [] {FC83, HBB01, Bergin+04}
that the column density in the cold phase at which the gas becomes
dominated by gravity is very similar (again, for typical solar
neighborhood conditions) to that required for H$_2$-molecule
self-shielding. Thus, as the gas begins to contract gravitationally, as
described in the previous section, it also begins to become molecular,
at column densities $N \sim 10^{21}\, \psc$. It is important to recall
that the presence of molecular gas does not appear necessary to allow
the gas to cool and thus keep contracting; at densities lower than $10^4
\pcc$, C$^+$ fine-structure emission appears to provide the necessary
cooling \citep{GC12}.

Thus, when the gas begins to contract, it is likely to consist mostly of
cold atomic gas, and is expected to gradually become molecular during
the contraction process.  In turn, the transition to molecular
contributes to the second reduction stage of the Jeans mass discussed in
Sec.\ \ref{sec:grav_contr}, and thus strengthens the gravitational
contraction by further reducing the temperature \citep[see the phase
diagrams in Fig.\ 8 of] [] {KI00}. The late appearance of molecular gas
during the collapse has been observed in numerical simulations of cloud
contraction including a prescription for the formation of H$_2$ and
$^{12}$ CO molecules \citep[e.g.,] [] {HH08, Heiner+15}. Thus, a
``molecular cloud'' does not form suddenly, but rather gradually, so
that, as the cold gas cloud contracts, its molecular content increases
over time, as described by \citet{VS+18}. Nevertheless, in the
simulations, the gravitational potential driving the collapse is that of
the combined molecular and cold atomic gas. This fact must be taken into
account when the gravitational binding of clouds is considered.

In addition, as pointed out by \citet{GV14}, the fact that the Jeans
mass decreases continuously during the gravitational contraction as
long as the gas remains nearly isothermal, implies that the thermal
pressure becomes progressively less important, and the collapse proceeds
in an increasingly pressureless manner. In turn, this implies that the
collapse tends to enhance any anisotropies initially present in the
cloud, since the collapse proceeds faster along the shortest dimensions
of the cloud, so that triaxial ellipsoids evolve into sheets, and sheets
into filaments \citep{Lin+65}. Moreover, since the cloud is likely to
have started out with a sheet-like geometry due to its formation at the
interface of converging flows, the pressureless collapse is expected to
produce filamentary structures, as verified in numerical simulations of
collapsing sheet-like clouds \citep[e.g.] [] {BH04, HB07, VS+07,
Heitsch+08b, Heitsch+09, GV14, Carroll+14}.

In particular, \citet[] [see also Smith et al.\ 2011] {GV14} observed
the formation of filaments in numerical simulations of contracting
clouds, and pointed out that {\it the filaments constitute the
collapse flow itself from the large to the small scales}, in which the
velocity field is such that material from the sheet-like cloud flows
onto the filaments roughly perpendicularly to them. {Then the} flow
changes direction at the filament, becoming longitudinal and directed
towards cores in the filaments, or at filament intersections. That is,
the filaments are akin to rivers, funneling the gas from the cloud to
the dense cores where stars form, as illustrated in the {\it left panel}
of Fig.\
\ref{fig:GV_fil}, where it is seen that the velocity field {\it around} the
filament is directed mainly toward the filament, while {\it inside}
the filament the flow is mostly longitudinal, pointing toward the
dense core.

It is important to note that the change in direction of the flow along
the radial direction occurs smoothly, without the formation of
noticeable shocks. This can be observed in the {\it right panel} of
Fig.\ \ref{fig:GV_fil}, which shows that the inwards radial velocity
decreases smoothly towards the filament's axis, while the longitudinal
component grows also smoothly. No jump in the radial velocity is
observed at the axis. This is similar to the smooth nature of the
velocity field in a collapsing core during its prestellar stage, since
a shock does not form until the time of the formation of the singularity
\citep[i.e., the protostar; e.g.,] [] {WS85}. In the filament, no
singularity forms precisely because the flow changes direction as it
enters the filament and is redirected towards the core. Therefore, the
filament is ``drained'' by the longitudinal flow, and in fact, the
simulations indicate that it does not exceed a certain maximum central
density, which probably depends on the total amount of mass in the
contracting potential well. Thus, in a sense, the filament becomes
``frozen'' in a state analogous to the prestellar stages of core
collapse, albeit in cylindrical geometry.

\bfgw
\includegraphics[scale=0.48]{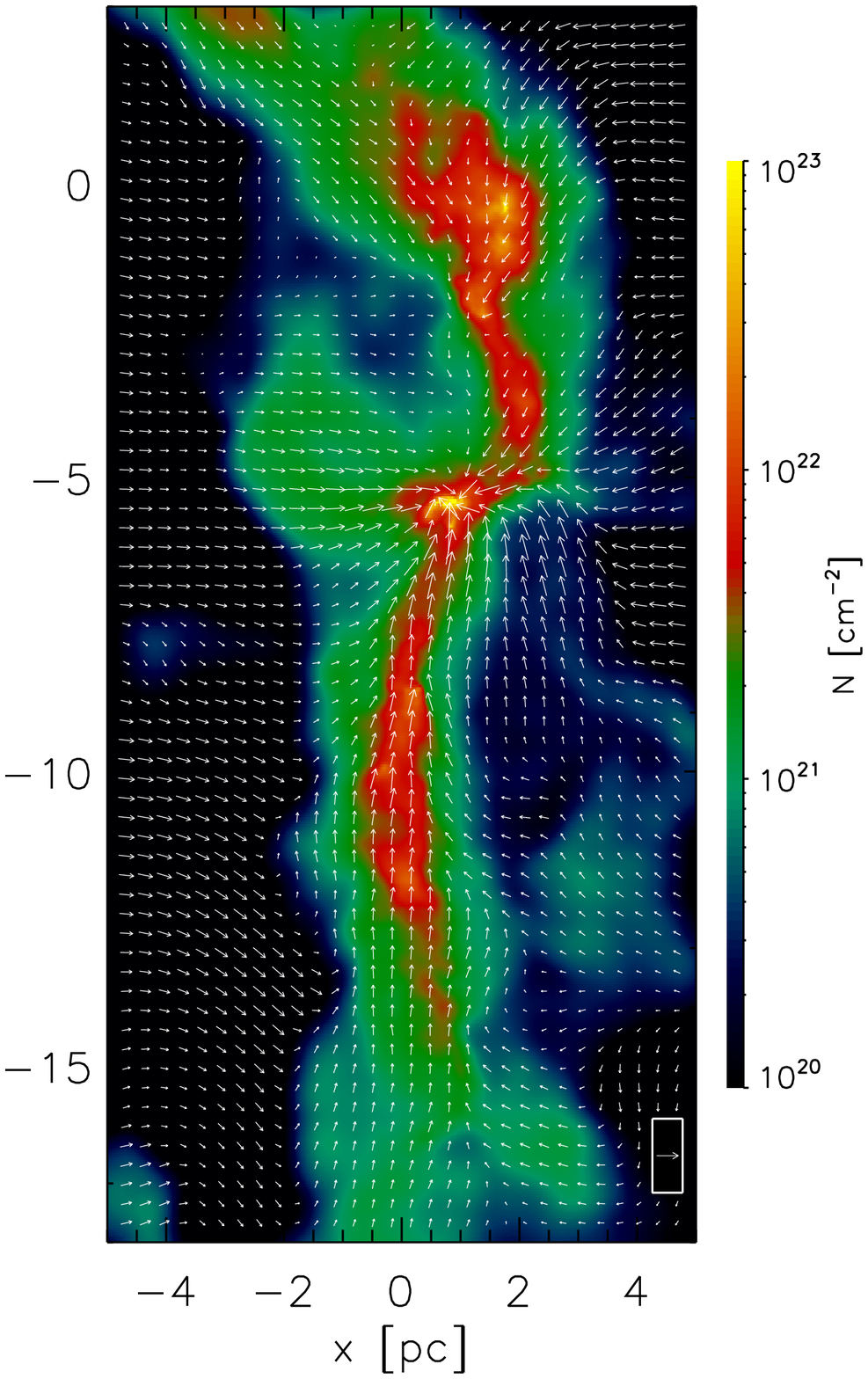}
\includegraphics[scale=0.48]{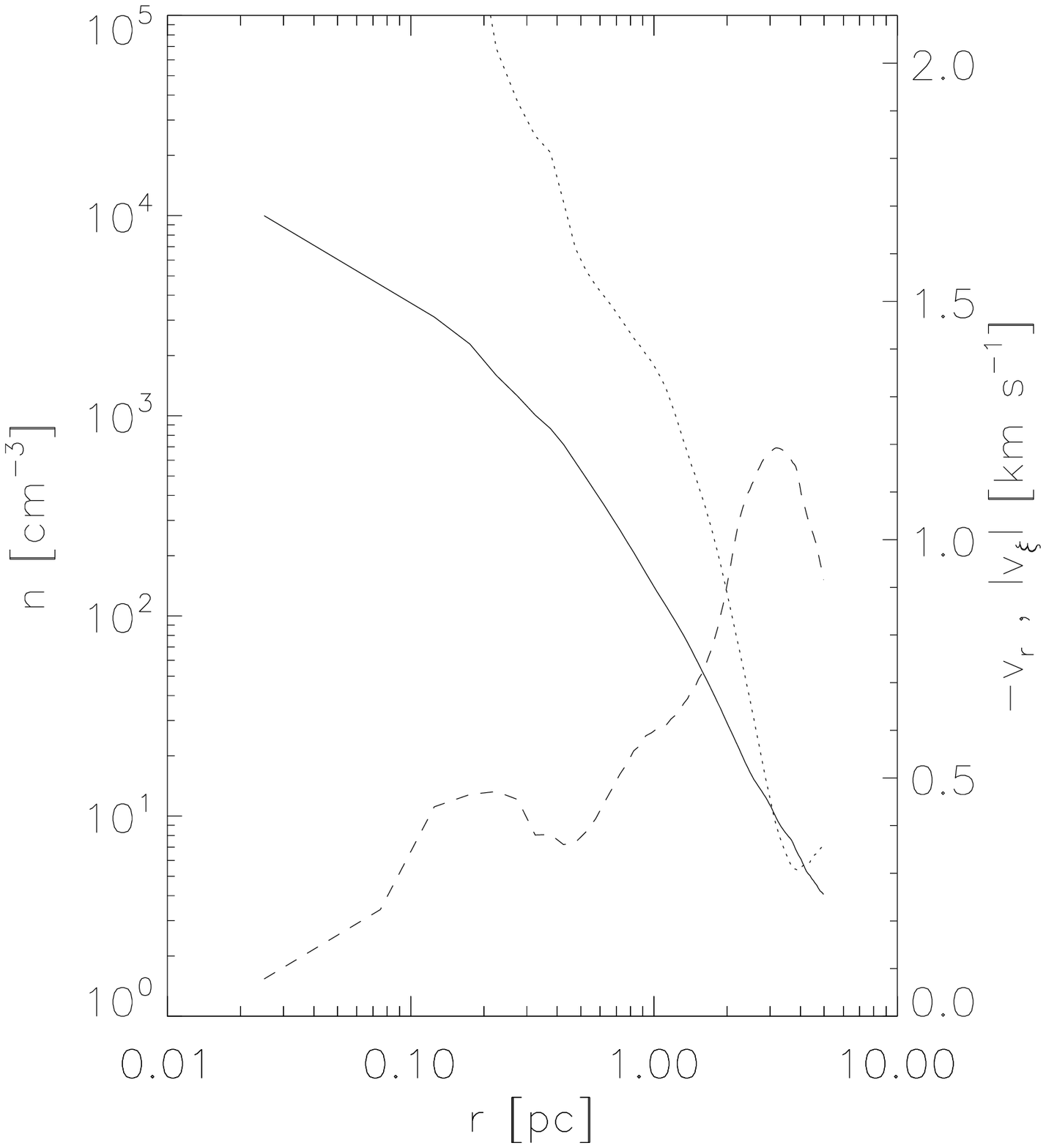}
\caption{{\it Left:} Column density and projected velocity field of a filament-core
system, which formed spontaneously in a simulation of a contracting,
moderately turbulent cloud. The velocity field is seen to point from
the cloud onto the filament, and then to smoothly change direction
within the filament, so that it flows longitudinally along the
filament toward the core, so that the accretion flow proceeds from the
cloud onto the filament, and from the filament onto the core
\citep[from] [] {GV14}. {\it Right:} Radial profiles of the density
({\it solid line}) and the longitudinal ({\it dotted line}) and radial
({\it dashed line}) velocities in the filament. The radial velocity is
seen to smoothly decrease toward the spine of the filament, impying
that there is no strong shock at the filament axis.
}
\label{fig:GV_fil}
\efgw

\subsubsection{The magnetic field in the filaments} \label{sec:mag_fil}

An implication of the mechanism of filament formation by anisotropic
gravitational contraction has been investigated by \citet{Gomez+18}, who
found, in magnetized simulations, that the accretion flow from the cloud
onto the filaments drags the magnetic field lines along the flow,
causing it to be oriented mostly perpendicularly to the filaments
around them, as observed \citep[e.g.,] [] {Palmeirim+13, Cox+16}.
However, the longtudinal flow along the filaments onto the main hubs
bends the field inside the filament, until the ram pressure of the
longitudinal flow is balanced by magnetic diffusion, causing a
characteristic ``U'' shape. The structure seen in the simulations is
qualitatively consistent with that observed in polarization maps
\citep[e.g., ] [] {PlanckXXXV}.

\subsubsection{A quasi-stationary state in the filaments?}
\label{sec:stationary_fils}

In fact, an argument can be made that gravitationally-formed filaments
should approach a quasi-stationary state in which the mass flux from the
cloud onto the filament is balanced by the flux from the filament to the
core, which should vary only on the characteristic variability timescale
for the accretion flow onto the filament. This can be seen as
follows. Let us assume that, at some time, the filament has a certain
charateristic radius $\Rf$, longitudinal velocity $\vl$, and mean
density $\rhomeanf$. The longitudinal mass flux is then $F_\| = \pi
\Rf^2 \vl \rhomeanf$. Let us now assume that the perpendicular flux, $F_\bot$,
exceeds $F_\|$. This will cause the filament to accumulate mass, and
thus cause an increase in its mean density and/or radius.  Therefore,
$F_\|$ must increase as well, provided the longitudinal velocity does
not decrease. Conversely, let us now assume that $F_\bot < F_\|$. In
this case, the filament loses mass, causing its density and/or radius to
decrease, and so $F_\|$ decreases. In both cases, then, the change in
$F_\|$ makes it approach $F_\bot$.

This argument suggests that the condition $F_\| = F_\bot$ may be an
attractor for the filament, which may then approach a quasi-stationary
flow regime. Note that we refer to it as a {\it quasi}-stationary regime,
because we have implicitly assumed that $F_\|$ has time to adjust to the
value of $F_\bot$; i.e., that $F_\bot$ varies slowly in comparison to
the variation timescale of $F_\|$. This is a reasonable assumption,
since $F_\bot$ is the result of the large-scale accretion flow, while
$F_\|$ represents a smaller-scale accretion flow. But nevertheless, the
quasi-stationary regime must vary on the variability timescale of
$F_\bot$. Numerical experiments exploring the approach to this
quasi-stationary regime will be presented elsewhere (Naranjo-Romero et
al., in prep.).

If the central density of the filament saturates because of the
establishment of this quasi-stationary regime, then no strong shocks
arise in the central parts of the filament due to the accretion flow
(except at the cores, which represent local collapse sites in
supercritical filaments). The absence of a strong central shock should
thus be a unique signature of filaments formed by gravitational
contraction, in contrast to filaments formed by strongly supersonic
turbulence. {\it This may be used as a diagnostic to distinguish the
origin of MC filaments.} The only shocks present in the filaments may be
weak shocks due to the residual turbulence in the filamentary flow,
which is likely to be only mildly supersonic at most in MC
filaments.

\subsection{Onset and acceleration of the star formation activity}
\label{sec:evol_SF}

\subsubsection{Can turbulence alone induce local collapses?}
\label{sec:can_turb?}

Numerical simulations of the contraction of MCs after their formation as
cold atomic clouds including sink particles \citep{VS+07, VS+10,
Colin+13, Carroll+14, FK14} systematically show that the sink particles
appear late in the evolution of the cloud,  typically $\sim 5$--10
Myr  \citep[a few free-fall times, due to geometrical effects;] []
{Toala+12, Pon+12}
after the onset of global contraction in the cloud, which in turn occurs
$\sim 10$ Myr after the first appearance of cold atomic gas triggered by
compressions in the warm atomic gas.\footnote{ Note that newly formed
cold atomic clouds in simulations with physical conditions
characteristic of the Solar neighborhood, typically take up to 10 Myr to
become Jeans unstable. Note also that the shorter timescales of 1--3 Myr
for collapse calculated by \citet{HBB01} correspond to the duration of
the collapse process for clouds that are already molecular and Jeans
unstable. Instead, the timescales referred to here correspond first to
the timescale for cold atomic gas to {\it become} Jeans unstable, and
then to the duration of the collapse starting out from cold atomic gas
conditions, which is typically 10$\times$ warmer and less dense than the
molecular gas considered by \citet{HBB01}.}  However, the
appearance of these first stars occurs $\sim 5$ Myr before the
large-scale collapse has advanced sufficiently to form massive stars.

This clearly implies that the primordial turbulence injected to the
clouds by various instabilities during their assembly (Sec.\
\ref{sec:model}) is too weak to directly generate density
fluctuations of large enough amplitude for the local Jeans mass to
become smaller than the fluctuation mass, triggering local
collapse. {This is true even in simulations starting with turbulent
fluctuations with kinetic energies comparable or even larger than the
gravitational energy \citep[e.g.,] [] {Bate+03, Bate09, Dale+12, Howard+18,
Grudic+18}. These simulations also do not form stars until the
turbulence has been dissipated, and gravitational contraction has set
in. That is, the initial turbulence does not form stars by itself. In
fact, \citet{CB05} have suggested that isothermal turbulent
compressions do not effectively reduce the local Jeans mass, because
they are preferentially one-dimensional.}

Instead, as described in Secs.\ \ref{sec:seq_coll} and
\ref{sec:1st_coll}, the turbulent density fluctuations begin to become
unstable at later times, when the {\it average} Jeans mass in the cloud
has decreased sufficiently due to the large-scale collapse, so that the
masses of the local fluctuations can match it, at which point local
collapses start.  However, as also described in those sections
\citep[see also] [] {VS+09}, the turbulent density fluctuations are able
to terminate their collapse earlier than the whole cloud because they
have nonlinear amplitudes, and therefore they have significantly shorter
free-fall times than that of the whole cloud.

\subsubsection{Acceleration of the star formation activity}
\label{sec:SFR_increase} 

As the average Jeans mass in the parent cloud decreases, the fraction of
mass at high enough densities increases, implying an acceleration of the
SF process; i.e., an increase in the SFR. This acceleration of the SF
continues until a large enough mass in stars exists that the IMF is
sampled up to stellar masses large enough that their feedback begins to
evaporate and/or disrupt the dense gas around them, and the SFR begins
to decrease again.

The evolution of the SFR has been modeled by \citet[] [hereafter
ZA+12] {ZA+12} and \citet[] [hereafter ZV14; see also
\citet{Volschow+17} and \citet{Burkhart18}] {ZV14}. The model,
hereafter referred to as the ZV14 model, considers, similarly to other
models of the SFR \citep{KM05, HC11, PN11, FK12}, that the turbulence
generates a probability density distribution (PDF) of the density
field, which in particular for nearly isothermal flows (such as
molecular clouds) adopts a lognormal form \citep{VS94, PV98}, whose
characteristic width depends on the turbulent Mach number
\citep{PV98, Federrath+08}. However, those other models assume that the
turbulence opposes the gravitational contraction and is strong enough to
prevent the global collapse of the clouds, thus being stationary in
nature, and giving predictions, for example, for the star formation
efficiency {\it per free-fall time}, $\sfeff$, since the final
efficiency depends on the lifetime of the clouds. Those stationary
models then compute the instantaneous SFR in the cloud as the mass at
high densities divided by a suitable free-fall time representative of
those high densities. The various models differ in their choice of the
``high densities'' and their characteristic timescales.

Instead, the ZV14 model assumes that the turbulence is relatively weak
(sonic Mach number of the order of a few, rather than of order 10, as is
often assumed) and insufficient to support the cloud, so that the cloud
can proceed to global and hierarchical gravitational contraction. Thus,
this model can predict the temporal evolution of the SFR and the SFE,
from which suitable temporal and ensemble averages can be computed that
correctly match observations (ZV14). A brief summary of the ingredients
of the model has been given by \citet{VS+18}.

The ZV14 model then assumes that, due to the global cloud contraction,
the density PDF shifts to higher density, as the cloud's mean density
increases.\footnote{The model assumes that the turbulent motions proper
remain at a roughly constant moderate level throughout the evolution,
and that the strongly supersonic motions routinely observed in MCs are
dominated by infalling motions that do not oppose the collapse, but
rather result from it.} This in turn implies that the mass at high
densities, and consequently the SFR, initially increase over time. This
is because, as the cloud contracts, a progressively larger fraction of
its mass is at sufficiently high densities to undergo ``instantaneous''
collapse.

The model also considers a standard IMF to compute the mass in massive
stars produced each timestep, given the total stellar mass formed up to
that time. Next, from the total number of massive stars present at every
time, the model calculates the photoionizing radiation produced, which
begins to disrupt the cloud. So, eventually the cloud's mass begins to
decrease again.

It is important to note that the ZV14 model assumes that the initial
conditions for the clouds are those of the cold atomic gas \citep[e.g.,]
[] {HT03} in the conditions under which it arises from nonlinear
triggering of the thermal instability by converging flows 
in the warm medium that occupies most of the Galactic disk at Solar
Galactocentric radii. Thus, contrary to the stationary models, the
initial density and turbulent Mach numbers are fixed, rather than free
parameters of the model. The only remaining parameter is the total mass
of gas that becomes Jeans unstable and proceeds to collapse.

\begin{figure*}
\includegraphics[width=0.45\linewidth]{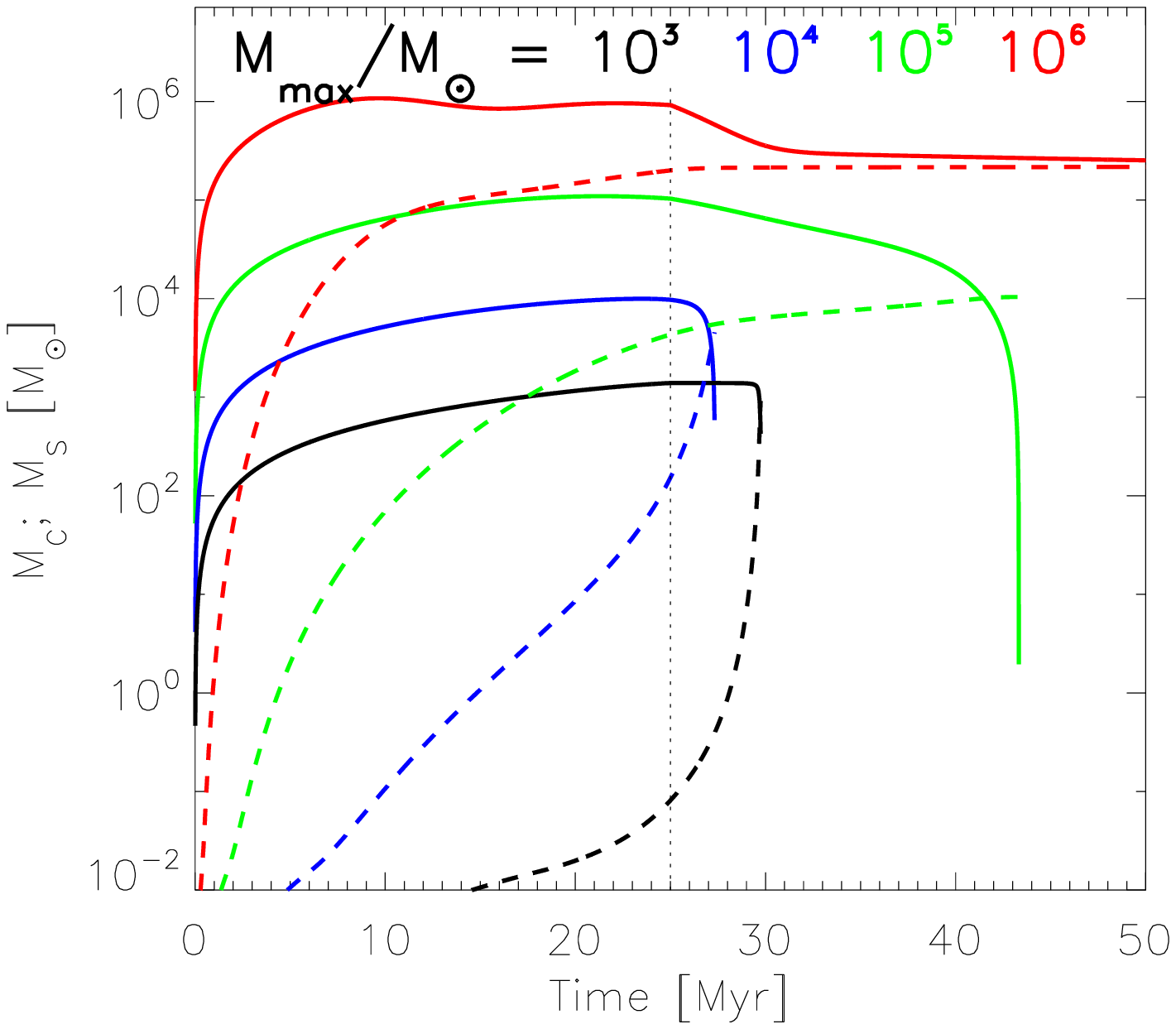}
\includegraphics[width=0.45\linewidth]{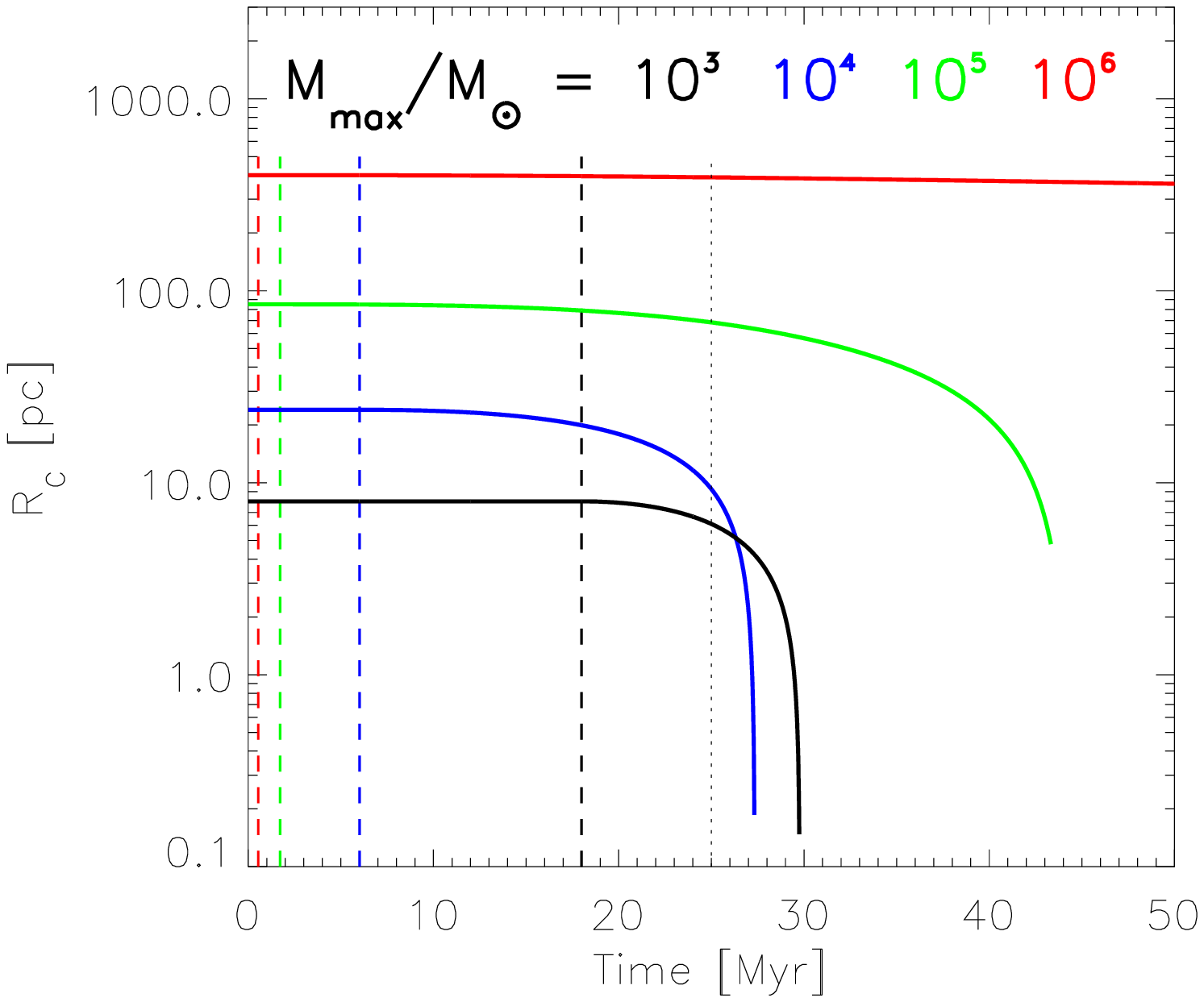}
\includegraphics[width=0.45\linewidth]{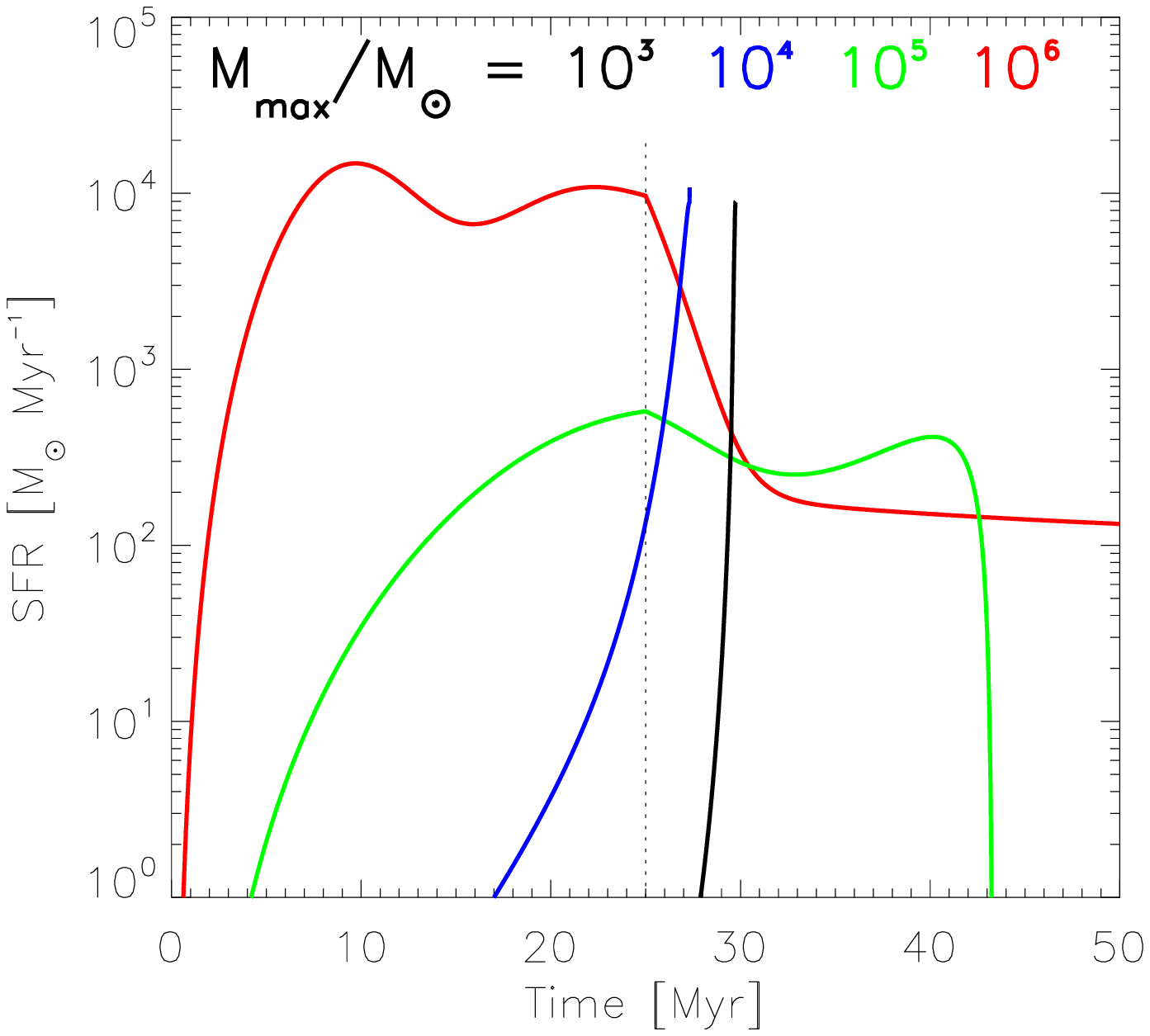}
\includegraphics[width=0.45\linewidth]{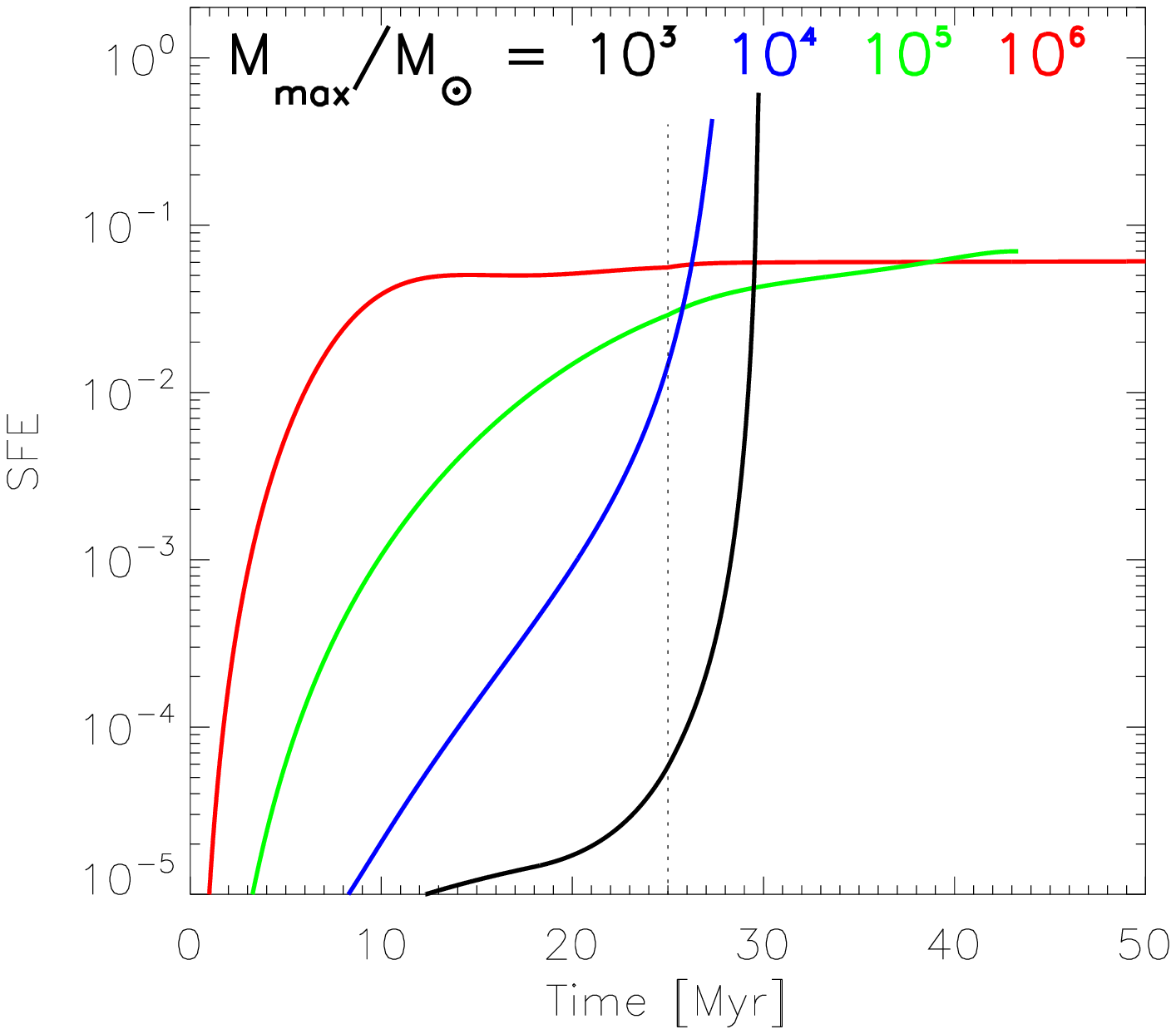}
\caption{Time evolution of the cloud mass and mass in stars (top left panel,
solid and dashed lines, respectively), radius (top right panel), SFR
(bottom left panel), and SFE (bottom right panel) for clouds whose
masses reach values $M_{\rm max}=10^3$, $10^4$, $10^5$, and $10^6 \,
\Msun$ (black, blue, green and red lines, respectively) according to the
ZV14 model. The vertical dotted black line is the time at which the
accretion stops ($t=25 \, \Myr$). (Plots reproduced from ZV14).  } 
\label{fig:model}
\end{figure*}

Figure \ref{fig:model} shows the evolution of clouds of masses $10^3$,
$10^4$, $10^5$, and $10^6 \Msun$ according to the ZV14 model. It is
important to note that, by construction, the ZV14 model matches the
evolution of the SFR in the numerical simulations of \citet{VS+10},
implying that this evolution is representative of that observed in
numerical simulations.

This kind of evolution has several important implications. First, the
model predicts that young clouds will have low SFRs. This eliminates the
supposed need for a supporting agent in order to keep low-mass
star-forming clouds from having too large SFRs. The clouds {\it are}
proceeding at free-fall, but, as discussed in Sec.\
\ref{sec:seq_coll}, and indicated by eq.\ (\ref{eq:rho_of_t}), this
process takes time, and develops very slowly at its early stages
\citep[see also] [] {BH13, Girichidis+14}. As a consequence, the SFR is
very low at early stages.

\begin{figure}
\includegraphics[width=0.48\textwidth]{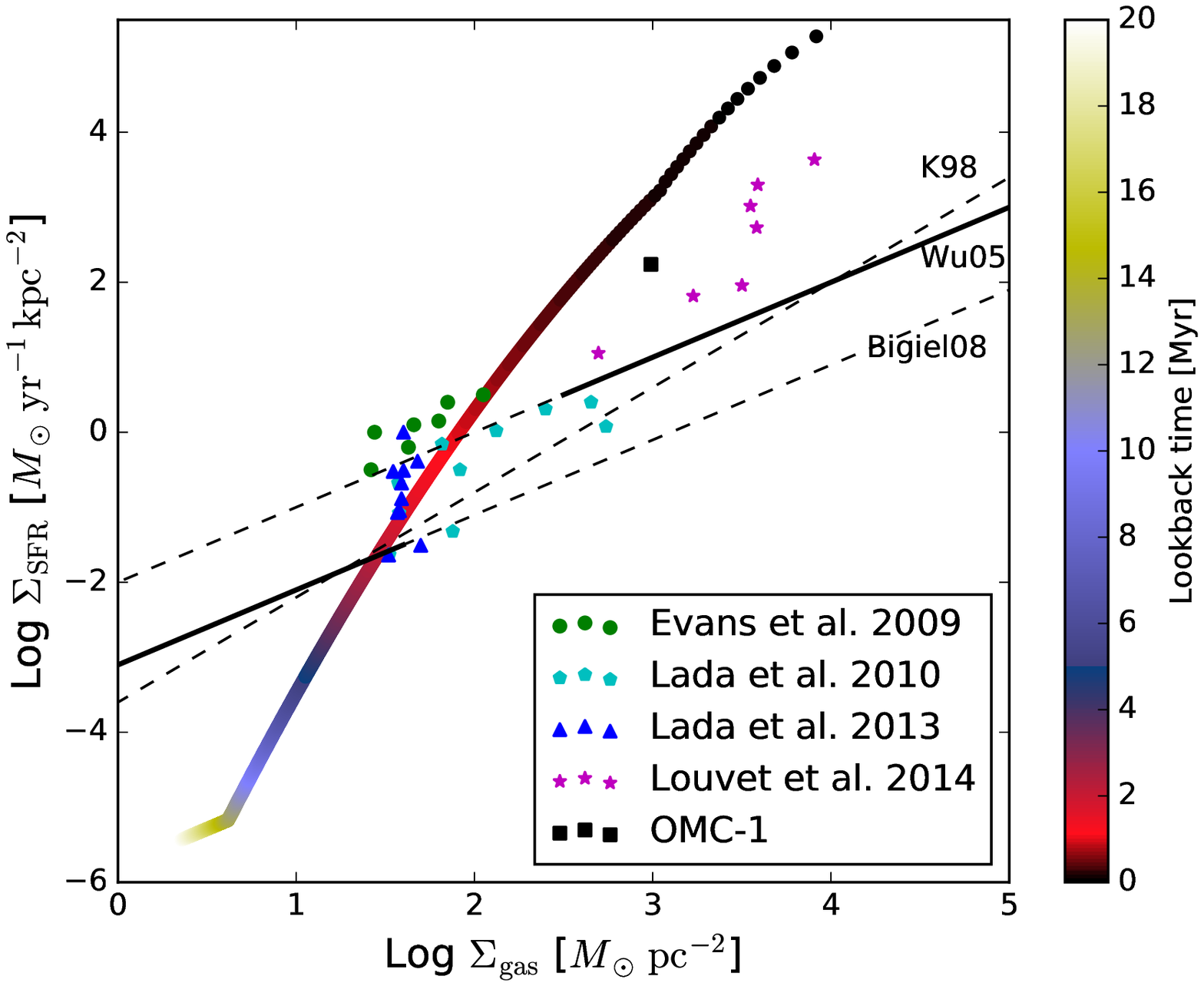}
\caption{{\it Colored line:}  Evolutionary diagram for a
2000-$\Msun$ cloud in the Kennicutt-Schmidt diagram of SFR surface
density $\Sigma_{\rm SFR}$ vs. gas surface density $\Sigma_{\rm gas}$,
together with data for low-mass star-forming clouds \citep{Evans+09,
Lada+10, Lada+13} as well as mini-starbursts \citep{Louvet+14}. 
The color indicates the (``lookback'') time since the cloud passed by a
given point in the evolutionary track, viewed from the moment when the
cloud is destroyed by feedback. The straight {\it
solid-dashed lines} indicate the trends determined by
\citet{Kennicutt98}, \citet{Wu+05} and \citet{Bigiel+08}, as
indicated. The model is seen to spend the last few 
{megayears} hovering around the region of the low mass clouds, and the last
fraction of a {megayear} near the region of the dense, high-mass star-forming
clumps, after it has contracted from cloud to clump scales.}
\label{fig:KS_ZA} 
\end{figure}

Indeed, ZA+12 calculated the evolutionary track of a 2000-$\Msun$ cloud
in the Kennicutt-Schmidt (KS) diagram of $\SSFR$ {\it vs.} $\Sigma_{\rm
gas}$ \citep{Kennicutt98}. We show this track in Fig.\ \ref{fig:KS_ZA},
indicating the times before the final burst that destroys the cloud,
together with observational data for various types of clouds, from
low-mass star forming clouds \citep{Evans+09, Lada+10, Lada+13} to
high-mass regions like the OMC1 cloud and other mini-bursts \citep{Louvet+14}. 

Figure \ref{fig:KS_ZA} shows that, during its early stages, the cloud
passes through the locus of low-mass star-forming clouds, and then,
during its later stages, passes near the locus of high-mass
star-forming clouds, such as the Orion OMC1 cloud, and finally, near
the locus of mini-bursts, clearly indicating the increase in its star
formation activity. It is important to remark that, as the SFR
increases, the cloud shrinks, as indicated by the top-right panel of
Fig.\ \ref{fig:model}, so that the ``cloud'' is already a dense,
compact, massive star-forming clump when it reaches SFRs characteristic
of massive star-forming regions. Also, the clump at this time is
expected to be embedded in a more massive cloud, as accretion is a
fundamental part of the evolutionary process, and so clouds are expected
to continue accreting from their surroundings during their evolution.

It is also important point to note that, according to the ZV14 model,
all parameters of a cloud, such as its dense mass  fraction, mean
density, size, and SFR, evolve simultaneously. Therefore, pairs of
parameters, such as the total instantaneous molecular mass and the dense
mass fraction, can be used to completely determine the evolutionary
state of a cloud of a given total dense gas mass. This property of the
model has been used by \citet{VS+18} to estimate the ages of the clouds
in the sample studied by \citet{Lada+10}, using the dense gas mass
fraction as a proxy for evolutionary time. This study concluded, for
example, that clouds of similar masses but very different instantaneous
SFRs and SFEs, such as the California and the Orion B clouds, can be
interpreted as being in very different evolutionary stages. For these
two clouds, the California cloud is described by the model as being only
a few {megayears} old, while the Orion B cloud is described as being between 15
and 20 Myr old.

{Another point to remark is} that ample observational evidence
exists for the acceleration of star formation, either from the presence
of tails of old-age stars in the age histograms in young, embedded
clusters \citep{PS99, PS00, DaRio+10}, or by the more numerous and more
centrally concentrated nature of the younger members of a cluster
\citep[e.g.,] [] {Povich+16, CC18, Grossschedl+19}.

\subsubsection{On the usage of the lognormal form of the PDF}
\label{sec:logn} 

{It is important to emphasize that the ZV14
model assumes that the density PDF has a {\it lognormal} form, in spite
of the fact that it is now well known that the PDF develops a power-law
tail at high densities as various collapses proceed throughout the cloud
\citep[e.g.,] [] {Kainulainen+09, Kritsuk+11, BP+11b, Girichidis+14}. A
similar model using the growth of the power-law tail of the PDF has been
presented by \citet{Burkhart18}. However, it is not obvious that the
power-law form is actually the most appropriate form to use for the
purpose of the model. This is because, as suggested by
\citet{Kritsuk+11}, the power-law tail of the PDF appears to be related
to the development of power-lar radial density profiles in collapsing
regions. In this interpretation, the power-law part of the PDF
represents the {\it already-collapsing material in the cloud}. Using the
lognormal PDF attempts to represent the density fluctuations produced
exlusively by the turbuelence. This is equivalent to assuming that the
velocity field in a turbulent, self-gravitating cloud can be assumed to
consist of a purely turbulent component plus an infall component, so
that the infall velocities are associated with the power-law PDF, and
the purely turbulent velocities are associated with the lognormal form,
and that it is the latter that constitute the seeds for the subsequent
collapse.

}

\subsection{Combining fragmentation and the acceleration of star
formation. The assembly of clusters} \label{sec:clusters}

Numerical simulations of the global contraction of clouds show that
the first fragments to complete their collapse have lower masses than
those that complete their collapse later. This is because the
fluctuations that terminate their collapse much earlier than the bulk
of the cloud are the ones that have much shorter free-fall times
(i.e., much larger densities) than the mean values in the
cloud. According to the lognormal form of the turbulent density PDF,
the total mass in these high-density fluctuations is a small fraction
of the total cloud mass, and therefore, these first collapses involve
low-mass objects. This is illustrated in Fig.\
\ref{fig:m_abv_nsf_vs_t}, which shows the evolution of the mass
fraction at densities above $\nsf = 10^3\, n_0$,  where $\nsf$ is
defined in Sec.\ \ref{sec:1st_coll}, using the same approach
as in Sec.\ \ref{sec:1st_coll}. This mass fraction is seen to increase
in time, so that initially it is very small, implying that the first
star-forming sites within the cloud that appear must be of low mass. In
turn, this suggests that the stars that form should also be of low mass. As
time progresses, the mass in these sites increases, and thus they may be
more massive themselves, as well as the stars they contain, assuming
that each collapsing site forms a distribution of stellar masses bounded
by the mass of the site itself \citep[e.g.,] [] {Oey11}.

\bfg
\includegraphics[width=0.48\textwidth]{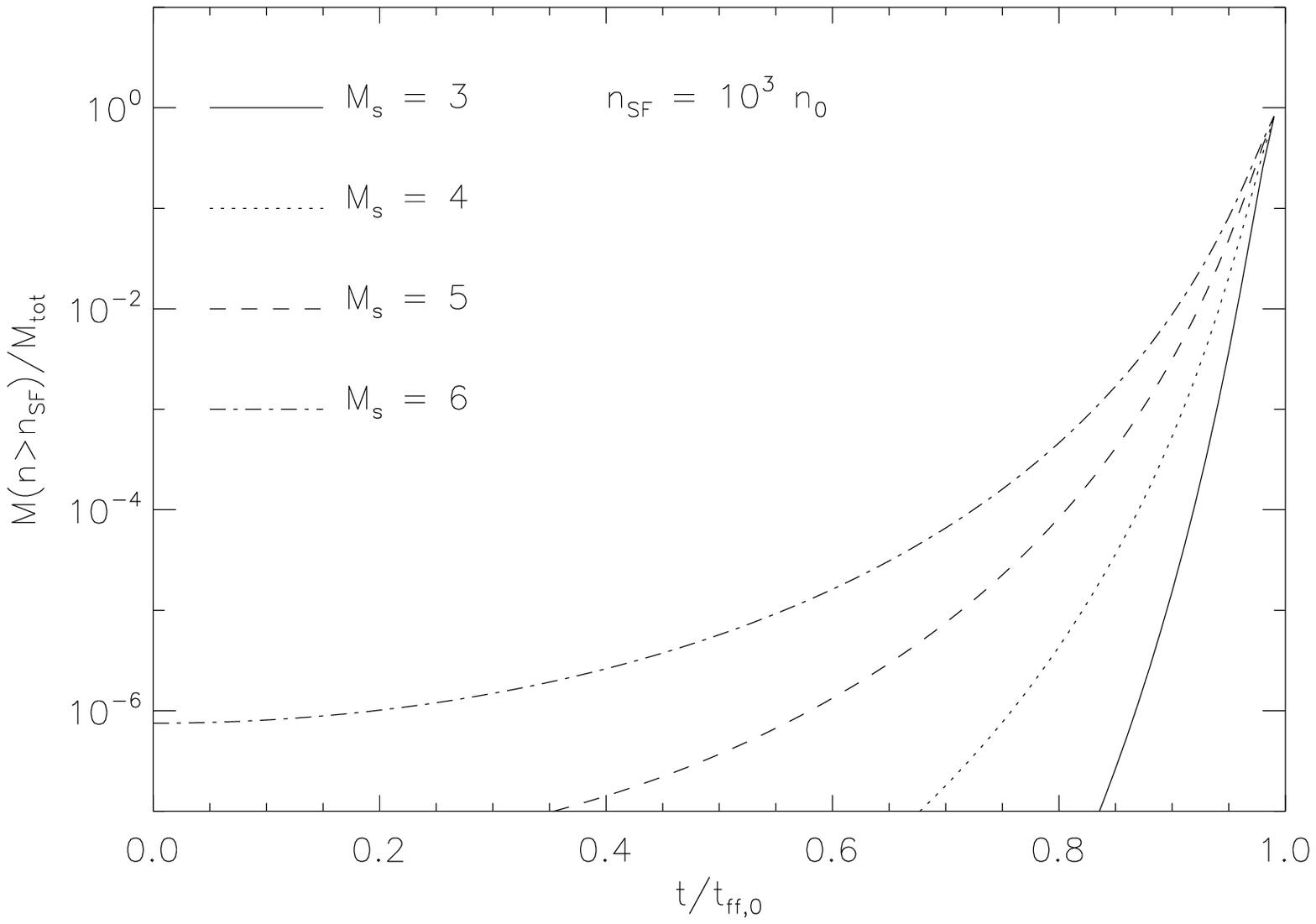}
\caption{Evolution of the mass fraction at densities above $\nsf$, for various
values of the turbulent Mach number, $\Ms$, assuming $\nsf = 10^3\, n_0$.} 
\label{fig:m_abv_nsf_vs_t}
\efg

The temporal increase of the mass involved in star formation within a
cloud also implies a temporal increase of the SFR, as observed in
numerical simulations \citep[e.g.,] [] {HH08, VS+10, Hartmann+12, Colin+13, Koertgen+16},
and predicted by the ZV14 model. Moreover, numerical simulations of
whole GMCs (at the hundred-pc scale) in which the sink particles
represent individual stars rather than stellar groups, also exhibit a
temporal increase of the mean mass of the stellar particles. This is
shown in Fig.\ \ref{fig:cumul_mass_distr}, reproduced from
\citet{VS+17}, where it is seen that the stellar-particle mass
distribution evolves by developing a larger fraction of high-mass
particles as time progresses, in tandem with the increase of the
SFR. {Finally, the formation of subsequently more massive stars
while the cluster itself becomes more massive implies a correlation
between the mass of the most massive star and that of the cluster
itself, as is indeed observed \citep{WK06}.}

\bfg
\includegraphics[width=0.48\textwidth]{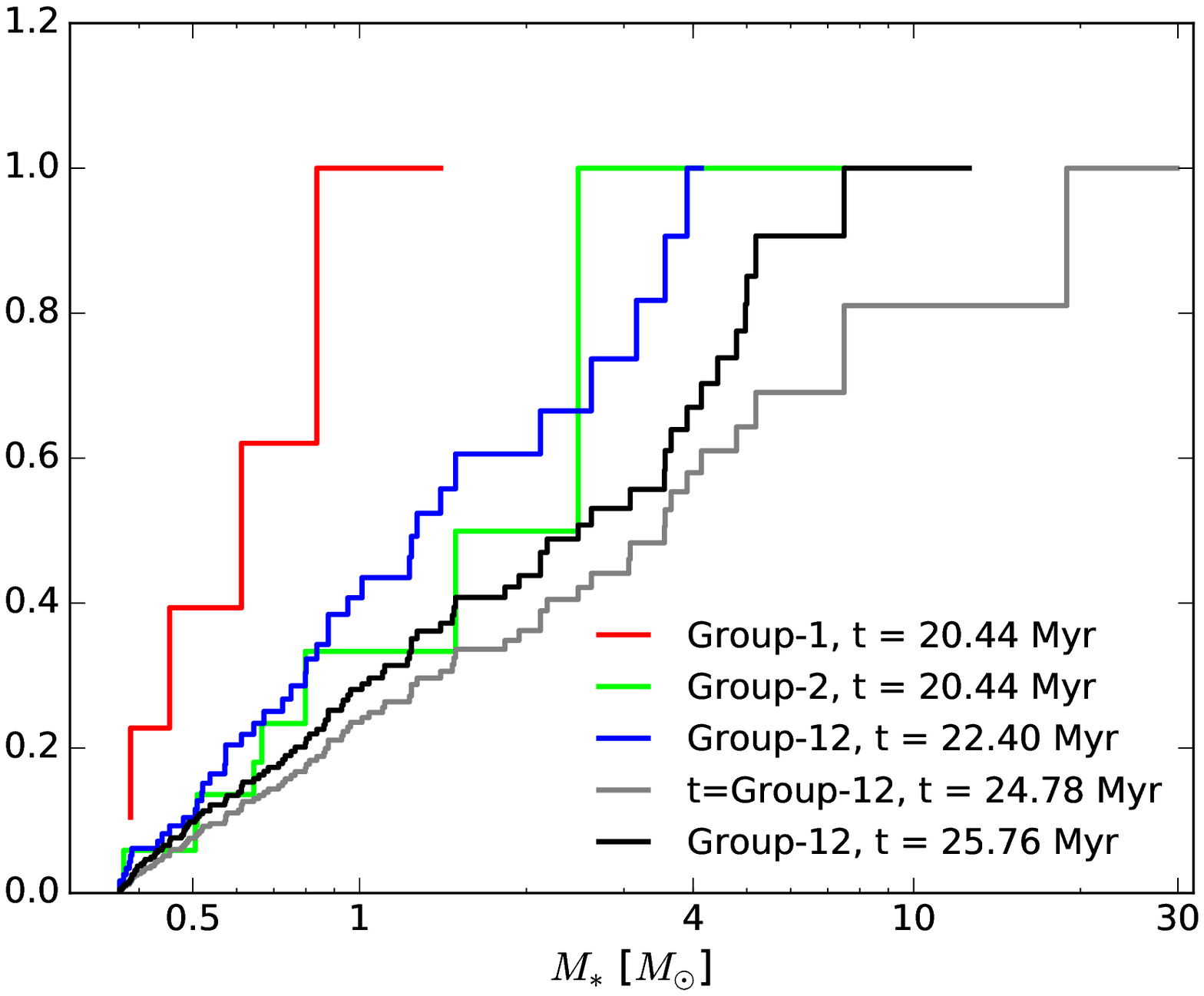}
\caption{Evolution of the {normalized} cumulative mass distribution of the stellar
particles in one cluster of the simulation of cloud and cluster
formation by \citet{VS+17}. The fraction of massive stars is seen to
increase over time {at first, and then to decrease again at the
last step shown. Note that the first three curves refer to the same
time, the first two representing two stellar groups that later merge,
and the third one showing their combined stellar mass distribution.}}
\label{fig:cumul_mass_distr}
\efg

Once the SFR is high enough, the feedback from the massive stars that
appear in the cloud begins to erode the cloud, and causes the SFR to
decrease again, {together with the high-mass fraction}. This evolution of the SFR implies that the
stellar age histograms of star-forming regions  must peak at the age
corresponding to the maximum of the SFR and to decrease toward both
larger and smaller ages, very much in agreement with observed age
histograms of embedded clusters \citep[e.g.,] [] {PS99, PS00, DaRio+10,
VS+17}. 

The spatial distribution of the stars in young clusters is also
predicted to be affected by the GHC mechanism. Recall that filaments
constitute the large-scale accretion flow (cf.\ Sec.\
\ref{sec:molec_fil_form}), and that small-scale collapses are the first
ones to terminate, but that they also accrete. Thus, the small-scale
collapsing sites can continue to form stars as they fall into the
large-scale potential well, feeding the latter with both stars {\it and}
gas. 

The stars that fall into the central hub inherit the velocity of the
infalling gas. However, when both the stars and gas reach the hub, the
gas is shocked \citep[see Fig.\ 4 of] [] {GV14}, and loses kinetic
energy. Therefore, stars formed later in the hub from the shocked
gas, inherit this velocity dispersion, and thus have lower typical
velocities than the stars that fell from the outside. Thus, the newer
stars tend to have smaller orbits than the ones that fell in from the
outside, generating an age gradient in the cluster, as shown in Figs.\ 8
and 9 of \citet{VS+17}. Recent young cluster studies suggest that such
age gradients are indeed observed \citep[e,g,] [] {Getman+14a,
Getman+14b}.

Finally, not all the material in the filaments feeding the hub has time
to reach it before it is destroyed by the feedback from the massive
stars formed there. For example, the time for massive stars to appear
after the formation of the first stars observed in the simulation
studied by \citet{VS+17} is $\sim 5$ Myr. After that, an \hii\ region
begins to expand, with a characteristic speed $\sim 10$-$20\, \kms$.  On
the other hand, material in the filaments, at least for the range of
masses of the structures formed in the simulation, has typical infall
speeds $\sim$ a few$\kms$. Therefore, material in the filaments located
at distances greater than $\sim 5$-10 pc from the main hub is likely to
be exposed to the ionizing radiation in the \hii\ region before it can
fall into the hub. This radiation evaporates first the filament,
interrupting the supply of gas to the cores within it (or
``starving'' them), leaving for a while a trail of dense cores embedded
in the warm ionized gas, similar to Bok globules. Later, the cores are
consumed by their internal star formation and/or by evaporation by the
ambient radiation. This is illustrated in Fig.\ \ref{fig:fil_erosion},
which shows a filamentary cluster in the simulation LAF1 of
\citet{Colin+13} at 1.6 and 8.6 Myr after the formation of the first
star. The \hii\ region begins to expand  some $\sim 5$ Myr after the
formation of the first star. The filament in the lower part of the
images is seen to be evaporated, leaving behind a chain of isolated
dense cores (``globules''). The cores are finally  consumed too,
roughly 1 Myr later. {A similar behavior is observed in the
simulations including full radiative transfer of \citet{ZA+19}.}

\bfgw
\includegraphics[width=0.48\textwidth]{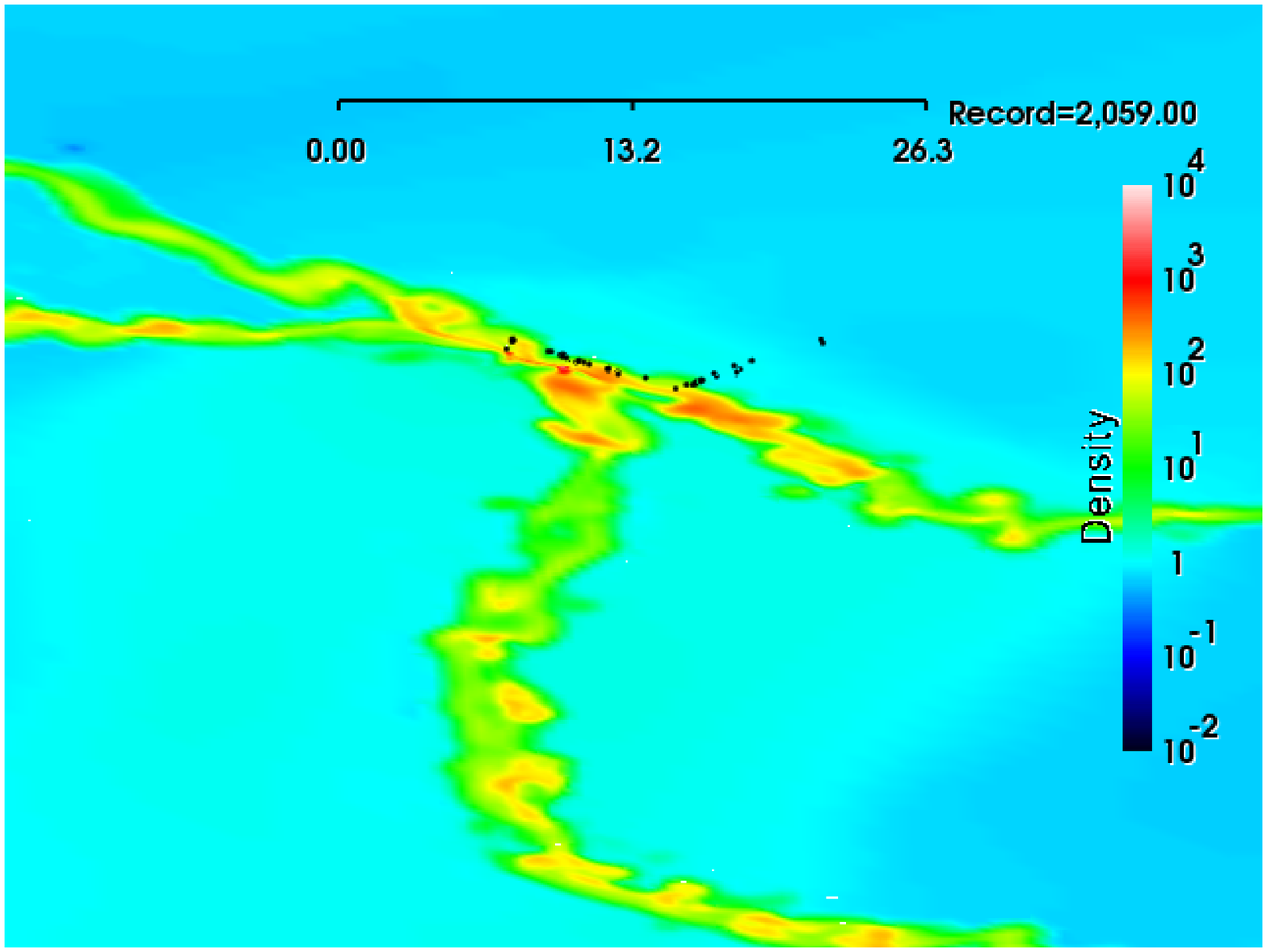}
\includegraphics[width=0.48\textwidth]{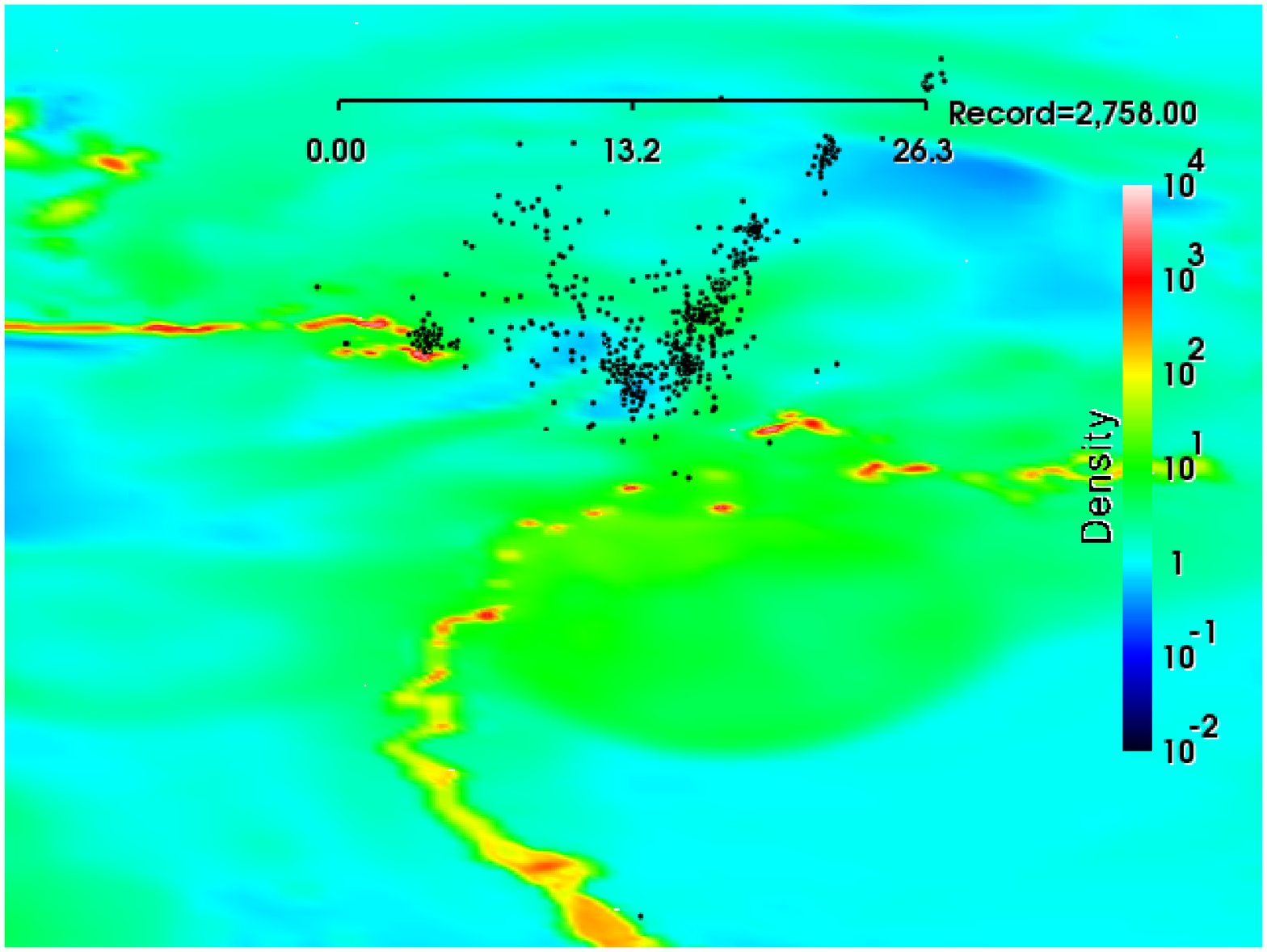}
\caption{Inclined cross-sectional images of the density field in the
neighborhood of a filamentary cloud and its central hub, leading to the
formation of a cluster, in the simulations labeled LAF1 of
\citet{Colin+13}, including a simplified treatment of radiative transfer
for the ionizing radiation from massive stars. The black dots represent
stellar particles, whose masses correspond to individual stars, with a
realistic Salpeter-like IMF, and are shown in the volume in front of the
cross-sectional images of the density field. The {\it left} panel shows
the cloud and cluster at time $t = 20.6$ Myr (1.6 Myr after the onset of
SF), and the {\it right} panel shows the system at $t = 27.58$ Myr, 8.6
Myr after the onset of SF. The filaments are seen to be eroded by the
ionizing radiation from the cluster, leaving their densest parts
(cores) as chains of apparently isolated cores.}
\label{fig:fil_erosion}
\efgw

If the distant cores along the filaments
have already formed some stars, then, when they are 
consumed/evaporated, the stellar groups formed there exhibit a
hierarchical and fractal structure similar to that established
originally in the gas, as shown in Fig.\ 12 of
\citet{VS+17}.

\section{Discussion} \label{sec:disc}

In the previous sections we have outlined the mechanism of {GHC} and
fragmentation, extending the scenario originally proposed by
\citet{Hoyle53} to the case of clouds containing nonlinear density
fluctuations produced by (moderately) supersonic turbulence, as
motivated by the observed evolution of numerical simulations of the
formation and self-consistent turbulence generation and subsequent
{GHC}. In what follows, we discuss some important insights and
implications of the scenario, as well as some caveats of our approach.

\subsection{The IMF: Minimum fragment mass and the formation of brown dwarfs}
\label{sec:min_mass_bds}

The minimum mass that can be reached through the GHC mechanism is an
important question that needs to be addressed in order to understand
whether brown dwarfs can be produced by this mechanism. Already in his
original paper, \citet{Hoyle53} gave an estimate for this minimum
mass, assuming that it would be determined by the moment at which the
collapsing gas becomes dense enough to become optically thick, and
thus suffer a transition from a nearly isothermal to a nearly
adiabatic regime. From this reasoning, he concluded that the minimum
mass should be of order $\sim 0.3$--1.5 $\Msun$. Further refinements
on the cooling mechanisms yielded significantly lower masses, down to
$\sim 0.01\, \Msun$ \citep[] [and references therein] {Rees76}, while
refinements on the assumed geometry, such that the fragmentation
occurs in a shock-compressed layer and that accretion causes a
fragment to grow while still condensing out, lead to more modern
estimates of the minimum mass $\sim 3 \times 10^{-3}\, \Msun$
\citep{BW05}. 

Numerical simulations including this so-called {\it opacity-limited
fragmentation} also show an efficient formation of brown dwarfs
\citep{Bate+02}, and in fact the problem then becomes to avoid an
excessive formation of these objects by inclusion of radiative
transfer to adequately model the gas heating from the accretion shock
onto the hydrostatic core, in order to prevent {\it excessive}
fragmentation \citep[e.g.,] [] {Krumholz+07, Bate09, Offner+09}.
Although these simulations refer to fragmentation in cores within 
initially turbulence-supported clouds according to
the ``gravoturbulent'' scenario, the last stages of fragmentation must
be very similar as in our scenario, so their conclusions also apply to
the GHC scenario. Therefore, it is safe to conclude that the GHC
fragmentation scenario can readily produce objects down to the brown
dwarf regime. See Sec.\
\ref{sec:comparison} for a comparison between GHC and the
gravoturbulent scenario.

\subsection{The IMF: the upper-end as a consequence of chaotic {\bf
GHC}} \label{sec:IMF_chaotic_GHC}

The GHC scenario also provides a natural framework for the
development of the \citet{Salpeter55} slope of the stellar initial
mass function (IMF). The accretion onto stars, or protostellar
objects, constitutes the last step of the mass (accretion) cascade
that must start at the scale of the contracting cloud. Numerical
simulations using a variety of
schemes and setups have successfully reproduced the high mass end of
the IMF, with a slope $\sim -1$, close to the $-1.3$ canonical value
of Salpeter \citep[e.g.,] [] {Bate09a, Bonnell+11}. The usual approach
in those simulations has been to create sink particles---particles
that continue interacting gravitationally, but not hydrodynamically,
with their environment--- in places where the conditions are believed
to be adequate for star formation,  such as large densities,
gravitational binding against all possible sources of support
(magnetic fields, turbulence, thermal pressure), and local inward
motions.  Once created, the sink particles are allowed to
continue accreting, increasing  their mass with time, according
to the local conditions \citep[e.g., ][]{Bate+95, Jappsen+05,
Federrath+10}.  In principle, when the simulation is properly
resolved, and the relevant physics is included, the distribution of
sink-particle masses should reproduce the IMF, since the accretion
onto the sink particles should mimic the actual physical process
occurring in stars. In the rest of this section we discuss how the GHC
scenario allows the accretion processes that may lead to the Salpeter
slope. We do not discuss the turnover of the IMF, which may be more
related to local stellar physics than to the global gravity of the
medium \citep[e.g.,] [] {Krumholz+07, Bate09}.

\subsubsection{The kinetic or Boltzmann approach.  The last stage of
accretion}

While the thermal and kinetic physics of the numerical simulations
that have obtained a sink mass function with slopes close to the
Salpeter value varies from one work to another, all of
them have one common ingredient: gravity.  In particular,
\citet{Zinnecker82}  proposed an approach to the Salpeter slope
 based on gravitational accretion assuming that the process of
accretion of mass  onto the {protostellar objects} is of a
Bondi-Hoyle-Littleton \citep[BHL;] [] {HL39, Bondi52} type. In
this scheme, a star of mass $M$ traveling with a relative velocity $v$
 through a homogeneous medium of density $\rho_0$ presents an
effective cross section 
determined by the star's gravitational potential well and
its velocity relative to the medium.
The resulting mass accretion rate is

\begin{equation}
 \dot M = \frac{4\pi G^2 M^2 \rho_0}{(c_s^2 + v^2)^{3/2}} \equiv
\alpha M^2, \label{eq:bondi}
\end{equation}
with $c_s$ the sound speed of the medium.  \citet{Zinnecker82} showed
that, for a given population of {protostellar objects} and $\alpha$ constant,
the mass distribution of that population will evolve to an asymptotic
power-law slope of

\begin{equation}
    \Gamma \equiv \frac{d\log N}{d\log M} = -1 \label{eq:imfslope}
\end{equation}
where $\Gamma$ is the exponent in the logarithmic mass distribution of
the stars, i.e.,
\begin{equation}
   \frac{dN}{d\log M} = M^\Gamma,  
\label{eq:IMF}
\end{equation}
 and $dN/dM \propto M^{\Gamma-1} = M^{-2}$.

Although this approach  gives a slope  $\Gamma = -1$ close to the
Salpeter value, it nevertheless has some problems:
\begin{enumerate}
    \item The environment where the star is being formed is far from
    homogeneous.  

    \item It neglects the self-gravity of the medium.

    \item The solution to eq.\ (\ref{eq:bondi}) diverges in
    a finite time.
\end{enumerate}
Indeed, numerical simulations \citep[e.g.,] [] {KlessenBurkert00, Bonnell+06,
Maschberger+14, BP+15} exhibit departures from the pure BHL accretion:

\begin{enumerate}
  \item The mass  accretion histories of the sink particles have
  time-variable slopes, starting very steep, and then decreasing to
  being almost flat in most cases \citep[see, e.g., Fig.\ 13 of] []
  {BP+15}.

  \item The ensemble of stars in simulations does not exhibit a clear
  $\dot{M} \propto M^2$ accretion law, but 
  instead exhibits a scattered distribution
  in the $\dot M$--$M$
  plane, the scatter being larger at lower $M$ \citep[see, e.g., Fig.\
  2 of] [] {BP+15}. The lower  boundary of the distribution has
  a slope $\sim 2$, but at the centermost parts of the
  distribution, it has a flatter slope.

  \item The individual mass accretion rates exhibit large oscillations
  during the accretion process, likely due to the fluctuations in the
  environmental conditions \citep{Maschberger+14, Kuznetsova+18b}.

\end{enumerate}

\begin{figure} 
\includegraphics[width=\columnwidth]{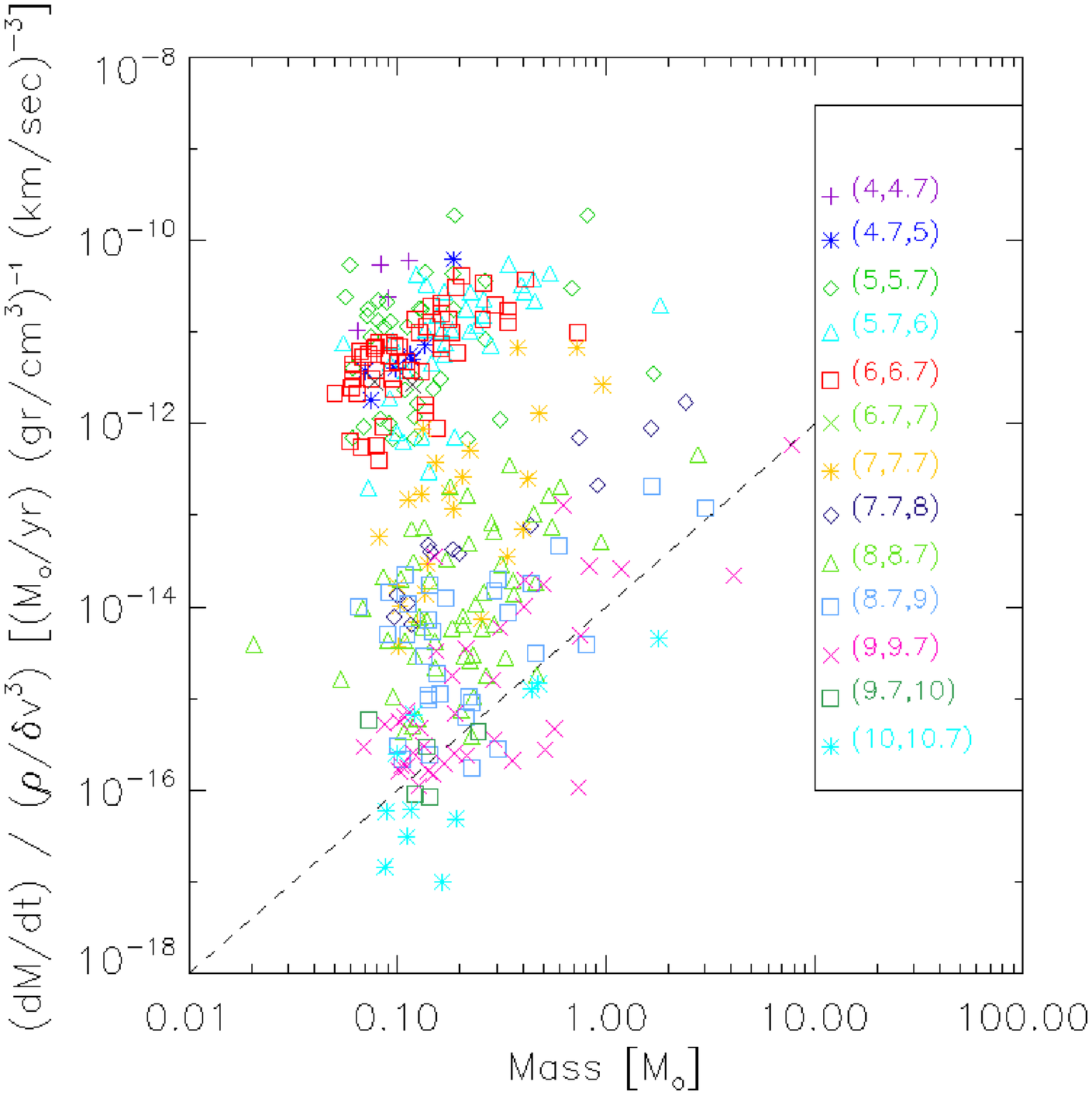} 
\caption{Mass accretion rate of each sink particle divided by the
local $\alpha$, vs Mass.  Each symbol denotes different values of
$\alpha$.  The dashed line shows a slope of $2$
\citep[from][]{BP+15}.} 
\label{fig:BHaccretion}
\end{figure}

Even though these points suggest that BHL is not a complete model for
the mass accretion onto the stars, it comes reasonably close. Indeed, as
shown by \citet{BP+15}, the dependence $\dot{M}\propto M^2$ is a good
approximation if objects  are collected into groups with similar
values of $\alpha$; i.e.,  groups of objects  forming in
regions where the density and velocity dispersion are similar. This is
shown in Fig.~\ref{fig:BHaccretion}, where the mass accretion rate
divided by the local $\alpha$  is plotted against the mass of the
sinks, in simulations of {GHC}, after one free-fall time of
the initial density. Each range in $\alpha$  is denoted with a
different symbol and color. The dashed line has a slope of 2,  which
is seen to be close to the slope defined by objects depicted with the
same color/symbol.

 In view of this, \citet{BP+15} argue that, since the environment
of the sinks is temporally and spatially variable, one may write the
accretion rate in a more general form as
\begin{equation}
    \dot{M} = \alpha(r,t) M^2,
\end{equation}
{and therefore,} that \citet{Zinnecker82}'s approach
is applicable even if $\alpha$ is not constant, as long as it does not
depend on $M$. 
%
This generalization of the formalism
to describe environmental variability can account for the
mass accretion rates observed in
simulations, in comparison to those predicted by the BHL approach.

\subsubsection{The core mass function and the IMF: the
next-to-last accretion stage} \label{sec:IMF_hub_accr}

It is well known that the mass function of the dense cores that form
stars (core mass function, or CMF), bears a striking similarity to the
stellar IMF \citep{Simpson+08, Andre+10}, which has often been
interpreted as evidence that the final stellar masses are determined
already at the dense core stage \citep[e.g.,] [] {Andre+14}.  The
connection between the slope of the core mass function (CMF) and the IMF
becomes natural in the global-collapse scenario, since one of its
essential features is that all scales accrete from their parent
structures. Indeed, as shown by \citet{Kuznetsova+17}, $n-$body
simulations of cluster formation by gravitational contraction without
the effects of gas physics, develop both the quadratic dependence of the
mass accretion rate on mass, $\dot{M}\propto M^2$ and the mass function
of clusters with slope $\Gamma = -1$.  This is in agreement with the
well-known result that the cluster mass function also exhibits a slope
of $-1$ \citep[see, e.g.,] [and references therein] {Oey11}. More
relevant for our interests here, is the fact that a similar result is
found by \citet{Kuznetsova+18b} when the gas physics is included,
suggesting that the effect is purely gravitational. This suggests that
the BHL approach is applicable, as a first approximation, to the
accretion onto the dense cores where the stars form, as well as to the
stars themselves.

Thus, in the GHC scenario, the resemblance between the CMF and the IMF
may be the result of a similar accretion mechanism operating both at the
core scale and at the protostellar object scale. This is in contrast
to  ``core
collapse'' models \citep[e.g.,] [] {PN02, MT03, HC08, Hopkins12}
where the star simply ``inherits'' a fraction of the the mass of its
parent core. While compact pre-stellar cores may be gravity-dominated,
they still do fragment into several smaller units, frequently simply
termed ``fragments'' \citep[e.g.,] [] {Palau+14, Palau+15, Lee+15,
Beuther+18}. But both the cores and the stars will tend to have a
Salpeter-like mass distribution as long as their accretion process is
dominated by  the local gravity.

\subsubsection{The IMF in strongly turbulent environments}

 In ``core collapse'' or ``gravoturbulent'' models of star
formation and the IMF \citep[e.g.,] [] {PN02, MT03, HC08, Hopkins12},
the CMF is assumed to be the result of pure turbulent fragmentation,
and the cores then simply contract to form the stars, while the cloud
continues to be globally supported by turbulence. However, the
results of several numerical simulations
\citep{Clark+08, BM+16} suggest instead
that, when turbulence dominates (i.e., for relatively large values of
the virial parameter, $\av \gtrsim 3$), the number of low-mass stars
decreases, the mass distribution becomes more top-heavy, and thus, the
slope of the mass function of sinks departs from the Salpeter slope. The
reason for this behavior is that in the presence of strong turbulence,
only high mass pre-sink entities have enough gravitational energy to
overcome the kinetic energy and proceed to collapse. We conclude that,
rather than turbulence, what sets the masses of cores and stars is a
BHL-type accretion mechanism. Stars may have a slightly steeper slope
because the higher masses are limited by the mass of the parent clump
where they form \citep{Oey11}.  On the other hand, uncertainties in the
measured IMFs are often large enough that it may be impossible to
distinguish between a slope of $-1$ and a slope of $-1.3$. In fact,
slopes of $-1$ (or $-2$ in $dN/dM$) are often reported \citep[e.g.,] []
{De_Marchi+10}. 



\subsection{Simultaneous evolution of cloud physical properties and the
star formation activity} \label{sec:simult_evol}

An essential feature of the GHC scenario is that it is  {\it
evolutionary} at the cloud level (see further discussion in Sec.\
\ref{sec:comparison}). The evolution occurs at the level of all the
cloud properties: their masses, densities, dense mass fractions, energy
balance, and star formation activity. In particular, in Sec.\
\ref{sec:SFR_increase} we discussed the increase of the SFR in a cloud
due to the increase of its mean density and, as a consequence, its dense
gas fraction. This result offers a straightforward explanation for the
observed correlation between the fraction of clumps associated with
massive stars and their peak column density \citep[e.g.,] []
{Urquhart+18} or between the surface density of dense gas and of the
star formation rate \citep[e.g.,] [] {GS04, Lada+10}, as shown by
\citet[] [see also Secs.\ \ref{sec:SFR_increase} and
\ref{sec:dep_frag_lev_dens}, and Camacho et al. 2019,
in prep.] {VS+18}.

\subsection{Evolution of the energy balance of clumps and cores}
\label{sec:energy_balance}

As discussed by \citet{BP+11a} and in Sec.\ \ref{sec:seq_definitions},
gravitationally contracting objects are expected to {attain}
contraction velocities of the order of the free-fall, or gravitational,
speed, eq.\ (\ref{eq:vff}) {after roughly one free-fall time (cf.\
Sec.\ \ref{sec:seq_definitions})}.
Because this speed differs by only a factor of $\sqrt{2}$ from the
``virial'' speed
\beq
\vvir = \sqrt{\frac{GM}{R}}, 
\label{eq:vvir}
\eeq
the observation that clouds and their substructures tend to appear
``virialized'' can be understood if these structures are
contracting at approximately the gravitational speed. 

When the velocity is originated by the gravitational contraction of the
objects, it is of the order of the gravitational speed, eq.\
(\ref{eq:vff}). In this case, a scaling relation between the Larson ratio
$\LR$, (cf.\ Sec.\ \ref{sec:probs_turb}) and the column density of the
form
\beq
\LR \approx \sqrt{G \Sigma}
\label{eq:KH_rel}
\eeq
is expected, and indeed is observed in general \citep[e.g.,] [] {KM86,
Heyer+09, BP+11a, Field+11, Leroy+15}. We refer to this condition
generically as equipartition between the kinetic and self-gravitating
energies.

\subsubsection{{The approach to equipartition and the inverse
$\av$-mass relation}}
\label{sec:alpha-mass}


Although in general most MCs and their substructures appear to be near
the equipartition relation, eq.\ (\ref{eq:KH_rel}), significant
deviations are routinely observed around this scaling. {In
particular, an inverse $\av$-mass correlation is often observed in
surveys ranging from GMCs \citep[e.g.,] [] {KM86, Leroy+15,
Traficante+18a} to dense molecular cores \citep[e.g.,] [] {Kauffmann+13,
Ohashi+16, Sanhueza+17, Traficante+18b}. Because of the variety of
objects considered in this section, we will generically refer to them
simply as ``the objects''. This inverse correlation can be
understood within the context of the GHC scenario, in combination with a
common selection effect inherently introduced by the survey
observational procedure. 

A fundamental point to understand the $\av$-mass correlation is that the
turbulent (or ``inertial'') motions, characterized by a velocity
dispersion $\sturb$, are independent of the self-gravity-driven motions
in the objects, which in turn are characterized by the gravitational
velocity $\vg$ (see Sec.\ \ref{sec:seq_definitions}, and eq.\
[\ref{eq:vg}] therein). This consideration differs essentially from the
standard assumption that the turbulence somehow tend to be
``virialized'', an assumption that we have questioned in Sec.\
\ref{sec:probs_turb}. Indeed, throughout this paper we assume that the
turbulent motions do not automatically adjust to the local gravitational
potential to ``virialize'', but rather constitute just a background on
top of which the self-gravitational motions develop, and create a
tendency toward equipartition.

This coexistence of the turbulent and gravitational motions is
incorporated into the virial parameter in eq.\
(\ref{eq:alpha_vir_composite}). Combining this equation with eq.\
(\ref{eq:r_of_t}), we can obtain an equation for the prestellar
evolution of the virial parameter of a gravitationally contracting
object of constant\footnote{{\mgta Recall the derivation of eqs.\
(\ref{eq:LRtot}) and (\ref{eq:alpha_vir_composite}) in Sec.\
\ref{sec:seq_definitions} was made assuming a {\it constant} mass of the
object. This treatment neglects the accretion onto the object, and has
the consequence that the predicted trajectories of objects in the \LS\
and $\av$-mass diagrams differs somewhat from those oberved in
simulations \citep[] [see also Camacho et al.\ 2019, in prep.] {BP+18}. 
We discuss this some more in Sec.\ \ref{sec:caveats}.}} mass $M_i$,
where the subindex $i$ denotes the $i$-th object of a survey:
\begin{eqnarray}
\alpha_{{\rm v},i} &=& 2 \left\{\frac{\sturb^2} {v_{{\rm ff},i}^2} + 1 -
\left[1 - \left(\frac{t} {t_{{\rm ff},i}}\right)^2 \right]^{a/3}
\right\} \nonumber \\
&=& 2 \left\{\frac{\sturb^2} {v_{{\rm ff},i}^2} + \left[1 - \left(1 -
\frac{32 G \rho_{0,i} t^2} {3 \pi} \right)^{a/3} \right] \right\},
\label{eq:alphavir_vs_rho}
\end{eqnarray}
where $t$ is the time since the object began contracting, and we have
used the facts that $\tau_i = t/t_{{\rm ff},i}$ and $t_{{\rm ff},i} =
\sqrt{3 \pi/32 G \rho_{0,i}}$, where $\rho_{0,i}$ is the density of the
object when it began contracting.  Note that the first term in the
{\mgta right-hand side (RHS)} of eq.\ (\ref{eq:alphavir_vs_rho}) can be either
smaller or larger than unity, while the remainder of the RHS varies
between 0 and 1 for $0 \le t \le \tff$.

Note also that, due to the standard definition of the virial parameter,
the denominator contains the final free-fall speed $\vff$, given by eq.\
(\ref{eq:vff}), rather than the actual (smaller) gravitational speed
$\vg$, given by eq.\ (\ref{eq:vg}). The first term within the curly
brackets then reflects the contribution of the turbulent (or
``inertial'', as opposed to driven by self gravity) velocity to $\av$,
normalized by the {\mgta final} free-fall speed, while the rest of the
right hand side reflects the contribution of the true
self-gravity-driven speed to $\av$, and shows that a given time $t$
represents a larger fraction of the free-fall time for denser objects,
because their free-fall time is shorter. Therefore, denser objects will
be closer to the final, free-fall speed than lower-density objects of
the same age.

For an individual object, eq.\ (\ref{eq:alphavir_vs_rho}) shows that, as
it contracts gravitationally, it should approach a value $\av \sim
2$. This is because, as the object contracts at constant $M$ (cf.\ Sec.\
\ref{sec:seq_definitions}), its $\vff$
increases, and so the ratio $\sturb^2/\vff^2$ decreases (recall we
assume that the turbulent speed is independent of the gravitational
potential). Also, as $t \rightarrow \tff$, the term within square
brackets in the first equality of eq.\ (\ref{eq:alphavir_vs_rho})
approaches zero, and so $\av \rightarrow 2$. However, it can approach it
from above of from below, depending on the inital value of the ratio
$\sturb^2/\vff^2$, as described in Sec.\ \ref{sec:seq_definitions} and
Fig.\ \ref{fig:Lin_Lff}: If initially $\sturb^2 > \vff^2$, then
initially $\av > 2$, and $\av \rightarrow 2$ from above. Conversely, if
initially $\sturb^2 \ll \vff^2$, then $\av \rightarrow 2$ from below.

In addition, eq.\ (\ref{eq:alphavir_vs_rho}) can also be used to
describe a sample of roughly coeval\footnote{I.e., objects that began
contracting within a time span significantly shorter than their
free-fall times.} objects at a single moment in time, {\mgta such as the
dense cores within a clump, or fragments within a core, whose onset of
collapse may be partially synchronized by the global reduction of their
parent object's Jeans mass, as decribed in Sec.\ \ref{sec:seq_coll}. In
this} case the subindex ``$i$'' denotes the $i$-th object in the
sample. In this case, however, in order to relate the objects' densities
to their masses, the {\mgta selection} effects introduced by the
observational procedure must be considered. It is often the case that
observational samples performed with a given instrument tend to define
the sample members such that their mass scales as some power of their
size, $M \propto R^p$, where typically {\mgta $1 \lesssim p \lesssim 3$
\citep[e.g.,] [] {Kauffmann+10, Kauffmann+13, Simpson+11, Ragan+13,
Wienen+15, Pfalzner+16, Traficante+18b}}.\footnote{{\mgta Slopes $p \gtrsim 2$ can be
understood as the result of the observations requiring a certain minimum
column density in order to be detectable above the noise, which
effectively defines the objects through column density thresholds. In
this case, \citet{BP+12} showed that the mean column density of the
sample members is close to the detectability threshold, because the
typically steep slope of the column density PDFs of molecular clouds and
their substructures implies that most of the material is at the lowest
densities. 
On the other hand, slopes $1 \lesssim p \lesssim 2$ typically arise when
the object samples are created by using multiple thresholds, as in the
case of multi-tracer observations \citep[see Sec.\ 4.1. of] [] {BP+19}}}

Taking into account this selection effect, {\mgta we can consider the
contributions from the inertial and gravitational terms in eq.\
(\ref{eq:alphavir_vs_rho}). Concerning the inertial term,
$\sturb^2/v_{{\rm ff},i}^2$, we note that $v_{{\rm ff},i}$, given by
eq.\ (\ref{eq:vff}) for object $i$, is larger for larger $M_i$ provided
that $p > 1$. But in addition, we must consider the size dependence of
the turbulent velocity dispersion, given by exponent $\eta$ in eq.\
(\ref{eq:turb_scaling}). Plausible values of $\eta$ are zero if the
turbulent background is the same for all members of the sample, $1/3$ if
the turbulence is subsonic and thus follows a Kolmogorov scaling, and
$1/2$ if the turbulence is supersonic and follows a Burgers scaling.
Therefore, the ratio $\sturb^2/v_{{\rm ff},i}^2$ is smaller for larger
$M_i$ when $p > 1$, 1.67 or 2 for $\eta = 0$, 1/3 or 1/2, respectively.
}

Regarding the term within square brackets in eq.\
(\ref{eq:alphavir_vs_rho}), we can find its dependence on $M_i$ as
follows. If we write in general
\beq
M_i \propto R_i^p
\label{eq:mass-size}
\eeq
for the $i$-th member
of the sample, then $M_i \propto \rho_i^{p/(p -3)}$, and so
$M_i$ varies inversely with $\rho_i$ for $0 < p < 3$. Note that,
here, $\rho_i$ is the density if the $i$-th object at time $t$, while
eq.\ (\ref{eq:alphavir_vs_rho}) involves $\rho_{0,i}$, the density of
object $i$ when it began contracting. However, it is clear that, if
$\rho_i(t) > \rho_j(t)$ for two objects $i$ and $j$ of the coeval
sample, then $\rho_{0,i} > \rho_{0,j}$ as well, because the denser
object increases its density faster than the less dense one. Thus, more
massive objects within a sample that satisfies $M_i \propto R_i^{p}$
with $0 < p <3$ have lower values of $\rhoo$. This implies that the
term within the square brackets in the second equality of eq.\
(\ref{eq:alphavir_vs_rho}) is smaller for the more massive objects of a
roughly coeval sample, if the sample satisfies eq.\ (\ref{eq:mass-size})
with $0 < p < 3$.


These results for a sample of objects can be applied to specific types
of objects, as we now discuss.

\bigskip
\noindent
{\it Application to diffuse clouds and clumps}

}

A clear tendency towards exhibiting {values of the virial parameter
$\av$ or the Larson ratio $\LR$ significantly larger than their respective
equipartition values is often reported in surveys of relatively
low-column density clouds and  clumps \citep[e.g.,] [] {KM86, Leroy+15,
Traficante+18a}, implying an excess of kinetic energy over the energy
from the self-gravity of the cloud.

This kinetic energy excess can be interpreted in terms of eq.\
(\ref{eq:alphavir_vs_rho}) mostly as relating to the first term, since for
diffuse objects, which are also generally rather large (of sizes of
parsecs and larger), the gravitational velocity is expected to start out
smaller than the inertial velocity, and so the term $\sturb^2/v_{{\rm
ff},i}^2$ dominates $\av$. Note, however, that this term being
significantly larger than unity does not necessarily imply turbulent
support or disruption. Any external compression that triggers the
formation of a self-gravity-dominated object must start with the
inertial compression velocity being larger than the gravitational
velocity, or otherwise the object would have started being
self-gravitating already. Thus, the formation of self-gravitating
objects by converging flows or ``collect and collapse'' processes must
start with an apparently strongly supervirial object.

This mechanism} has been
investigated by \citet{Camacho+16} in a survey of clouds, clumps and
cores (generically, ``clumps'') in numerical simulations of cloud
formation and evolution by converging motions in the WNM. Those authors
found that, in objects exhibiting such an excess in kinetic over
self-gravitational energy, in roughly half of the cases the excess
kinetic energy was in net compressive motions, while in the rest, the
clumps were {truly} in the process of dispersal. That is, contrary to
suggestions that these objects are in equilibrium, confined by large
external pressures \citep{KM86, Field+11}, \citet[] [see also
Ballesteros-Paredes et al.\ 1999a] {Camacho+16} showed that the objects
are {\it not} in equilibrium. Instead, they are either in the process of
growth, by external compressions whose origin is other than their
self-gravity, or in the process of dispersal. {\it There is no reason to
assume that such structures should in general be in equilibrium}, and
therefore, there is no need for a large confining thermal pressure.

These initially super-virial structures are then expected to evolve
according to the process described by the {\it dash-triple dot lines} in
Fig.\ \ref{fig:Lin_Lff} (if they are defined in a Lagrangian way, with a
fixed mass, as in Sec.\ \ref{sec:seq_definitions}): an external
compression (due to turbulence in the ISM, or a large-scale instability
in the Galactic disk) induces the formation of the cloud, which,
initially, is {\it not} dominated by its self-gravity. Thus, the cloud
appears super-virial in the \LS\ or $\av$-$M$ diagrams. However, as its
density increases due to the compression, and perhaps to thermal
instability, the self gravity of the cloud increases, causing its
gravitational speed $\vg$ (eq.\ [\ref{eq:vg}]) and Larson ratio to also
increase, {and its $\av$ to decrease}. Thus, the energy in the
self-gravitating velocity eventually becomes larger than the inertial
compressive energy. At this point, the cloud becomes dominated by its
self-gravity.  Nevertheless, the cloud may then appear somewhat
subvirial, if the inertial compressive motion dissipates rapidly and
$\vg$ has not yet become close enough to the free-fall value $\vff$
(eq.\ [\ref{eq:vff}]) (the lower {\it dash-triple dot} curve in Fig.\
\ref{fig:Lin_Lff}). Numerical simulations confirm the general shape of
the evolutionary tracks of clumps in the \LS\ diagram \citep[see the solid
curves in the panels labeled ``Mach 16'' and ``Mach 8'' in Fig. 4 of] []
{BP+18}.  Camacho et al.\ (2019, in prep.) show, in addition, the
simultaneous evolution of the clumps' energy balance and of their SFR.

Note that {\it in no case it is necessary to invoke a large external
confining thermal pressure for these apparently super-virial
structures.} If anything, the inflow, either inertial or from
self-gravity, provides a {\it ram} pressure for the inner parts of the
structures, although not really ``confining'' them, since the
structures are not in equilibrium, but rather contracting. In the case
of clumps that do not manage to become self-gravitating and collapse,
then no confining pressure is necessary, either. They will simply
disperse after the transient compression that formed them subsides.

\bigskip
\noindent
{\it Application to subvirial dense cores} 

On the other hand, dense cores are often observed to have
values of $\LR$ and $\av$ {\it smaller} than the gravitational value
\citep[e.g.,] [] {Kauffmann+13, Ohashi+16, Sanhueza+17,
Traficante+18b}. The energy balance of these objects can also be
understood in terms of the onset of their own collapse at a specific
moment in time and with a specified initial radius, as described by eq.\
(\ref{eq:LRtot}), if in this case the initial inertial velocity is
smaller than the object's own free-fall speed. In this case, the
evolution of the object is generically described by the trajectories
depicted by the solid or dashed curves in Fig.\ \ref{fig:Lin_Lff}. It is
important to {point out that, for dense cores that are located
within larger, less dense clumps which are already engaged
in gravitational contraction, the onset of the cores' own} gravitational
contraction is triggered by the global temporal decrease of the average
Jeans mass in the {clump}, rather than by a strong transient, local
reduction of the Jeans mass triggered by the {inertial} compression
\citep{CB05}. But because the contraction begins with a finite radius
and at a specific time, the infall speed is also smaller than the
free-fall speed, and the core appears sub-virial, {even though} the
core is nevertheless proceeding to collapse freely. {Thus,} its
infall speed will asymptotically approach the free-fall speed as the
core evolves. Again, numerical simulations confirm this generic form of
the evolutionary track of cores for which the initial inertial motions
are less than their own free-fall speed
\citep[see panels ``Mach 4'' and ``Mach 0'' in Fig. 4 of] []
{BP+18}. 

{It is important to n}ote that, in this case, {\it it is not
necessary to invoke a strong magnetic field to support the cores}. Their
evolution will self-consistently make them appear sub-virial during the
early stages of their collapse. {Moreover, since the early stages of
the collapse occur rather slowly, the cores will tend to spend more time
in this sub-virial state than in the full-equipartition stage, making
this subvirial state a common feature of these cores.}

{Note that the mechanism described in this section is independent of
the recent suggestion by \citet{Traficante+18c} that the observed
subvirial nature of massive cores may be due to an observational bias
which causes the kinetic energy to be systematically underestimated in
observations of {such} cores, due to the different tracers used
to estimate the mass and the velocity dispersion. Although this effect
may certainly be occurring in several massive core surveys, it is
unlikely that it is the only explanation, since not all surveys employ
the same tracers. Moreover, this effect cannot explain the observation
that dense cores start out subvirial in numerical simulations
\citep[] [see also Camacho et al.\ 2019, in prep.] {BP+18}, since in
those numerical studies the energy determinations have been made
directly from the density and velocity fields in the cores, with no
intermediate synthetic observation step that could introduce the
bias. Thus, we consider that the bias discussed by
\citet{Traficante+18c} can be at play in some observational studies, but
that the subviriality is a real physical property of most dense
cores. 

}

\subsection{Outside-in {\it vs.} inside-out collapse in prestellar cores}
\label{sec:outside_inside} 

Prestellar cores are often observed to 1) have Bonnor-Ebert-like (BE)
column density profiles \cite[e.g.,] [] {Teixeira+05}; 2) have molecular
line profiles with a central self-absorption dip and a blue-peak excess
that seem to imply subsonic infall speeds \citep[e.g.,] [] {Zhou92,
Evans99, Lee+01, Campbell+16}, and 3) to exhibit ``extended inwards
motions'', by which it is meant that they seem to extend beyond the
expected radial location of the ``rarefaction front'' according to Shu's
inside-out collapse model \citep{Lee+01}. The first two of these
properties are often interpreted as evidence of quasi-static contraction
in the cores, as proposed by \citet[] [herafter S77] {Shu77}, although it has
been difficult to reconcile them with the third property \citep[e.g., ]
[] {Lee+01, BT07}. Moreover, low-mass cores often appear to be
gravitationally unbound, thus requiring an external confning pressure to
keep them from dispersing \citep[e.g., ] [] {Lada+08}. In this section
we discuss how these properties can be understood in the framework of
the GHC scenario.

\subsubsection{Dynamic {\it vs.} quasistatic prestellar contraction}
\label{sec:not_quasi-static}

The non-homologous and outside-in nature of the prestellar core
contraction discussed in Sec.\ \ref{sec:nature_of_collapse} allows a
reinterpretation of the observational data. First and foremost, the fact
that dynamical contraction starts much earlier than the formation of the
first singularity (i.e., since the {\it prestellar} stage) is opposite to
the assumption by S77 that the prestellar stage occurs quasi-statically,
and that dynamical collapse begins at the time of the formation of the
protostar (the singularity), leading to the well-known scenario of an
inside-out collapse.

S77 proposed that the prestellar stages should contract quasi-statically
rather than dynamically on the basis of two main arguments: First, that
the conditions necessary for establishing a Larson-Penston \citep[]
[hereafter, LP flow] {Larson69, Penston69} dynamical flow are {\it ad
hoc} and difficult to establish in reality. Second, that the
establishment of an $r^{-2}$ density profile represents the tendency of
an isothermal, self-gravitating gas to approach detailed mechanical
balance, and that this can be accomplished as long as different parts of
the cloud can communicate acoustically with each other, which requires a
subsonic flow. Measurement of apparently subsonic inflow speeds from
infall line profiles \citep[e.g., ] [and references therein] {Evans99}
have provided support to this view.

However, subsequent work has demonstrated the inapplicability of these
arguments. Regarding the first of these, numerical simulations and
analytical studies alike have long suggested that a wide variety of
initial conditions generate flow that asymptotically approaches LP flow
\citep[e.g.,] [] {Hunter77, WS85, FC93, Naranjo+15}. So, rather than
being and artificial and {\it ad-hoc} condition, the LP flow appears to be
an attractor for the collapse flow.

Concerning S77's second argument, various pieces of evidence suggest
that, rather than representing detailed mechanical balance, an $r^{-2}$
profile may simply be the manifestation of nearly pressureless collapse.
Indeed, this profile arises in the outer regions of numerical
simulations of spherical collapse. Far from the center, the internal
mass is much larger than the mean Jeans mass
\citep{Naranjo+15}.
Moreover, it has recently been shown that this density profile can arise
simply as a consequence of pressure-free collapse, under the condition
that the infall speed is generated by self-gravity and that the mass
inflow rate is independent of radius \citep{Li18}, providing an
alternative to the suggestion by S77 that it must arise from detailed
mechanical balance, which in turn would result from acoustic (subsonic)
coupling throughout the core. Thus, the $r^{-2}$ density profile in the
outer parts of prestellar cores can be arrived at through dynamical
collapse during the prestellar stage, rather than requiring a
quasistatic process, as suggested by S77.

\subsubsection{Dynamical contraction with Bonnor-Ebert-like density
profiles in prestellar cores}
\label{sec:BE-like_prof}

The dynamical contraction during the prestellar stage is also known to
produce density structures that resemble a Bonnor-Ebert (BE) profile,
with a flat central part and an $r^{-2}$ scaling in the outer parts (the
{\it envelope}), as denoted in eq.\
(\ref{eq:prestellar_profiles}). However, contrary to true BE spheres,
which are hydrostatic equilibrium configurations, contracting prestellar
cores have a non-zero infall speed at all radii, except at the core
center, as indicated by eq.\ (\ref{eq:prestellar_profiles}).  Moreover,
the decrease of the infall speed towards the core's center (i.e., the
outside-in nature of the profile) implies
that the densest parts of the core do not have very large infall speeds
{\it during the prestellar stage},
while the largest speeds occur at radii where the density is already
decreasing, and therefore those speeds are downweighted in line
profiles, giving the appearance that the cores have smaller infall
speeds than they actually do \citep{Loughnane+18}. This may help
reconcile the supersonic nature of the actual infall speed with the
apparently subsonic values often derived from blue-skewed molecular line
profiles \citep{Zhou92, Evans99, Lee+01, Campbell+16}, as well as
explain the observed BE-like column density profiles in spite of the
configurations {\it not} being in equilibrium.

In addition, if the cores have been undergoing dynamical collapse ever
since the time when the local density fluctuation became unstable, the
radial extent of the contraction motions must be much larger than the
position of the rarefaction front in S77's model, which only starts to
propagate at the time of the formation of the protostar. That is, the
local collapse motions have been propagating outwards for the whole
prestellar stage of the contraction, while in S77's model they only
begin to propagate at the time of the formation of the protostar. This
may explain the observation of ``extended inwards motions'', at radii
larger than the expected position of the rarefaction front in the
inside-out model \citep{Lee+01}.

Finally, in this scenario, the boundary of cores that started as
finite-extent turbulent density fluctuations may be defined in an
observationally-motivated way, as the radius at which the density
fluctuation merges into the background.\footnote{ Observationally,
the background is often defined as the radius at which the
signal-to-noise has become too low, or where the gradient of column
density suffers an abrupt change \citep[see, e.g.,] [] {Andre+14}.}
However, because the fluctuation grows by developing an $r^{-2}$ density
profile and increasing its central density, the boundary of the core
moves outwards as the core evolves. Thus, cores defined in this way grow
both in size and mass.
\citet{Naranjo+15} showed that, when the boundary of a collapsing core
is defined in this way, the core's evolution tracks the locus of
observed cores in The Pipe and the Orion clouds in a diagram of $M_{\rm
c}/M_{\rm BE}$ {\it vs.} $M_{\rm c}$, where $M_{\rm c}$ is the mass of
the core, and $M_{\rm BE}$ is the BE mass for the average density and
temperature of the core. This diagram was first investigated by
\citet{Lada+08} for the cores in The Pipe. Those authors found that both
stable ($M_{\rm c}/M_{\rm BE} < 1$) and unstable ($M_{\rm c}/M_{\rm BE}
> 1$) cores occupied one common locus in this diagram. However, they
concluded that the stable cores are gravitationally unbound and have to
be confined by external pressure, while the unstable cores would be out
of equilibrium, and collapsing. This interpretation, however, leaves the
questions open as to why would both quiescent and dynamic cores occupy
the same curve in this diagram.  In fact, \citet{Lada+08} mention
that the pressure on the cores is most likely due to the weight of the
surrounding MC. This is consistent with the cores just being the
densest, inner part of a large-scale collapsing object, where the infall
speed is low, as dictated by eqs.\ (\ref{eq:prestellar_profiles}), and
the pressure being provided by the ram pressure of the external
infalling material, as suggested by \citet{Heitsch+09}.

Within this context, the suggestion by \citet{Naranjo+15} is that all
cores in \citet{Lada+08}'s sample are collapsing, and that the
``stable'' ones only appear so because they have been observationally
truncated at radii that are too short compared with the extension of the
infall motions.  Therefore, it would be highly desirable for
observations of dense cores, to obtain data, when possible, up to
distances as large as the extension of the inwards motions \citep[e.g.,]
[] {Lee+01}. {Also, \citet{Naranjo+15} suggested that the observed
BE-like shape of observed prestellar cores \citep[e.g., ] [] {Teixeira+05}
is a manifestation of their being in a prestellar stage, and not
indicative of hydrostatic equilibrium.}

Finally, the protostellar stage (i.e., after the appearance of a
singularity, or protostar) has been investigated analytically by
\citet{MC15} and numerically by
\citet{Murray+17}, who also conclude that the cores never go through a
hydrostatic stage.

\subsubsection{The decrease of the linewidth towards the central parts
of cores} \label{sec:decr_linew}

In Sec.\ \ref{sec:diff_pre_proto} we pointed out that the radial
velocity profile in in the inner part of prestellar cores is of
the form $v(r) \propto -r$ (cf.\ eq.\ \ref{eq:prestellar_profiles})
\citep[e.g.,] [] {WS85}, implying that the infall speed decreases
towards the center. This may offer a different interpretation for the
observation that the nonthermal component decreases towards the core
centers \citep[e.g.,] [] {Goodman+98, Pineda+10, Chen+18, Sokolov+18}.
This decrease is sometimes inferred from radial scans of the line
profile moving away from the core's center, in which a progressive
blurring of groups of hyperfine lines is observed towards the outer
regions of the core \citep[e.g.,] [] {Pineda+10}. This blurring is
assumed to be due to an increase in the turbulent component of the
velocity dispersion in the outer parts. In other studies, the decrease
of the hypothetical turbulent component is inferred from the observation
that near the core's center, the linewidth becomes approximately
constant at a value marginally larger than the sound speed, a result
which is interpreted as the turbulent component becoming smaller than
the sound speed
\citep[e.g.,] [] {Goodman+98, Chen+18}.

However, an alternative interpretation of the decrease of the linewidth
towards the central parts of the cores is that the nonthermal motions do
not correspond mainly to turbulence, but rather to the infall motions. For
prestellar cores, then, the amplitude of the infall motions decreases
towards the center, eventually becoming subsonic. Thus, the decrease in
the linewidth towards the cores' centers may correspond to the
decreasing infall speed, rather than to a drop in the turbulent
component. We plan to further investigate this possibility in a future
contribution.

\subsection{Development of the hierarchical collapse and observed
fragmentation properties of cores} \label{sec:implic_t_onset}

\subsubsection{Independence of time for onset of collapse from the
turbulent Mach number} \label{sec:indep_t_Ms}

In Sec.\ \ref{sec:time_to_nth} we found that the time to reach the
$n$th level of fragmentation is independent of the turbulent Mach
number. Although surprising, this result reaffirms the notion that the
role of the turbulence is only to produce the seeds for collapse, but
that the mass of the collapsing objects is determined by the temporal
evolution of the mean Jeans mass in the cloud as it contracts
gravitationally. This result is consistent with the recent numerical
finding by \citet{Guszejnov+18} that the degree of fragmentation of a
turbulent cloud depends only on the number of Jeans masses it
contains, and not on the Mach number of the turbulence.  Finally, it
is also consistent with recent observational results that the
fragmentation level\footnote{Observationally, the fragmentation level
has been defined as the number of fragments --compact sources, of size
$\lesssim 5000$ AU and masses of at least $0.5\, \Msun$-- within a
fixed-size field of view centered on the peak of the clump. This
definition is not exactly the same as ours, but refers to the same
basic process.} observed at scales of 0.1 pc does {\it not} appear to
correlate well with the turbulent Mach number
\citep[e.g.,] [] {Palau+15, Lee+15, Beuther+18}, and instead correlates
with the density of the parent structure, as discussed in the next
subsection.

\subsubsection{Dependence of the fragmentation level on the density of
the parent structure.}  \label{sec:dep_frag_lev_dens}

A natural consequence of the GHC scenario is that the fragmentation
level within a given structure should be directly proportional to the
Jeans number $\NJ$ (mass of the structure divided by the Jeans mass) of
that structure. In addition, given that $\NJ \propto \rho^{1/2}$, it is
then natural that in the GHC scenario a correlation is expected
between the fragmentation level of a structure of a given size and the
density of such a structure (at the corresponding size scale). This is
fully consistent with a number of observational works reporting a
tight relation between the fragmentation level or young stellar
objects counts and the density of their parent structures
\citep{Gutermuth+11, Palau+14, Palau+15, Palau+18, Lee+15, Mairs+16,
Mairs+17, Sharma+16, Hacar+17, Hacar+18, AR18, Murillo+18, Pokhrel+18,
Li+19, Orkisz+19, Sokol+19, Zhang+19}.
However, the aforementioned
correlation between the fragmentation level (assessed by counting
compact fragments) and the density of its parent structure (at the
corresponding scale where the fragmentation level is assessed) is only
expected if the observed structure is already in an advanced stage
after the global collapse started (Fig.\ \ref{fig:tau_n}, right
panel). Otherwise, the last fragmentation stage might have not been
reached yet, preventing the detection of compact fragments (see Sec.\
\ref{sec:time_nth_lack_frag}).

\subsubsection{Time-delay to reach the $n$th level of fragmentation and
observed lack of fragmentation in massive cold clumps} 
\label{sec:time_nth_lack_frag}

Also in Sec.\ \ref{sec:time_to_nth}, we found that higher fragmentation
levels are reached at later times (cf.\ Fig.\ \ref{fig:tau_n}, {\it
right} panel; see also Hoyle 1953). This may explain the apparent lack
of fragmentation in some massive, {\it cold} (infrared-quiet) clumps.
While the masses and scaling of the number of fragments with density has
been found to be consistent with thermal Jeans fragmentation in a number
of clumps \citep[e.g.,] [] {Palau+15, Pokhrel+18}, these clumps are
often very active sites of star formation (infrared-bright). However, in
other works, focusing on colder infrared-quiet clumps at even earlier
evolutionary stages, it is found that the actual masses of the fragments
are often significantly larger than the Jeans mass \citep[e.g.,] []
{Bontemps+10, Wang+14, Zhang+15, Csengeri+17}.

This observation may be interpreted as there existing a ``delay'' in the
appearance of the fragments corresponding to the value of the mean Jeans
mass in the clump  at the time of measurement. The GHC scenario
naturally requires this delay to exist, because the fragments must reach
a significant density contrast with respect to their parent structure in
order to be detected. Indeed, the typical densities of the fragments
observed with interferometers are 2-3 orders of magnitude larger than
that of the parent structure observed with a single-dish instrument
\citep[e.g.,] [] {Palau+13}.

Therefore, the fragments are not expected to be detected immediately at
the onset of their own collapse (even assuming, as we did in Sec.\
\ref{sec:time_to_nth}, that they start with the ``typical'' density of
the rms fluctuation, $\sim 10\times$ the mean density in the parent
structure), but rather after their density has grown by at least one or
two orders of magnitude, so that the Jeans mass in the fragment will
have decreased by a factor of $\sim 3$--10. As a consequence, the measured
Jeans mass will be smaller than that corresponding to the time when the
fragments began to contract locally, and the fragments themselves will
have only recently begun to sub-fragment themselves.

We conclude that the evolutionary nature of the GHC scenario implies
that, by the time when the fragments are sufficiently denser than their
parent structure as to be singled-out observationally, their Jeans
mass will be smaller than that corresponding to the time when they first
began to grow, and that they will only be at an early stage of the next 
level of fragmentation. Another implication is that the new generation of
sub-fragments is not expected to exist at random locations in the parent
clump, but rather within the fragments already present. Therefore, these
lower-mass sub-fragments are predicted in our scenario to have low
density contrasts and to be located within the next-lower-hierarchy
fragments. This suggests that searches for the next fragmentation stage
should be performed within the fragments already present, at high
signal-to-noise and high mass sensitivity.

\subsection{Comparison with the ``gravoturbulent'' scenario}
\label{sec:comparison}

Table \ref{tab:scen_comparo} presents a summary of how the
gravoturbulent and the GHC scenarios deal (or do not) with different
properties and features of MCs and their SF activity. For reference,
the magnetic support scenario \citep[e.g., ] [] {Shu+87,
Mousch91} is also included, as it actually did provide
explanations for some observed features that lack a counterpart in the
gravoturbulent scenario.

\begin{table*}
        \caption{Comparison of the magnetic support, GT, and GHC scenarios}
        \begin{tabular}{P{3.9cm}|P{3.9cm}|P{3.9cm}|P{3.9cm}}
        \hline
{\bf MC feature}		& {\bf Magnetic support$^a$}	& {\bf
Gravoturbulent (GT)$^b$} & {\bf GHC} \\
        \hline
        \hline
Time dependence	& Quasi-stationary	& Quasi-stationary	& Evolutionary at
cloud scale \\
        \hline
Cloud formation	& ...			& ...		&
Large-scale compressions due to turbulence or large-scale gravity
(stellar spiral arms, magneto-Jeans, etc.)\\
        \hline
Cloud-scale supporting pressure   & Magnetic	& Turbulent	& None \\
        \hline
Scale of gravitational contraction& Whole cloud (for magnetically
supercritical clouds). Dense core (for subcritical clouds)& Dense core
& Whole cloud (possibly including cold atomic envelope) \\
        \hline
Filament formation and structure	& ...	& Shocks. No implication
of longitudinal flow & Anisotropic
gravitational contraction with longitudinal flow (river-like structures
running down the gravitational potential) \\
        \hline
Core formation and collapse mechanism	& Slow gravitational contraction
mediated by ambipolar diffusion	& Shock compression and local
reduction of the Jeans mass	& Sequential destabilization of
successively smaller mass scales as global Jeans mass decreases \\
        \hline
Distinction between low- and high-mass star-forming regions 
& Subcritical vs.\ supercritical clouds	& ... & Evolution from low- to
high-mass regions \\
        \hline
SFR	& Stationary. Low (high) in subcritical (supercritical) clouds
& Stationary	& First increasing due to cloud contraction, then
decreasing due to feedback, over timespan $\sim 10$ Myr \\
        \hline
IMF	& ...	& Determined by turbulent core mass function	&
{Bondi-Hoyle-Lyttleton-type accretion can account for $-2$ slope of
power-law part of stellar IMF and core mass function.} Massive stars form during peak SFR\\
        \hline
Reason for dismissal	& Most clouds are supercritical$^c$	& (Proposed)
Inconsistencies between turbulence and observed cloud properties;
effect of turbulence is to form/destroy clouds, not support. Valid on
average over large spatial or temporal scales & TBD \\
        \hline
        \end{tabular}

   \begin{tabular}{l}
$^a$ See, e.g., \citet{Shu+87}.\\
$^b$ See, e.g., \citet{MK04}.\\
$^c$ See, e.g., \citet{Crutcher12}.\\
   \end{tabular}
   \label{tab:scen_comparo}
\end{table*}

\subsubsection{The difference in the flow regimes}
\label{sec:diff_regimes}

The GHC regime differs strongly from the so-called {\it gravo-turbulent}
(GT) scenario \citep[e.g.,] [] {VS+03, MK04, BP+07, HF12, Hopkins12, KG16}
 in that it is intrinsically {\it evolutionary}. In  contrast,
the GT scenario is {\it stationary.} In the latter, the
supersonic nonthermal motions observed in MCs correspond to supersonic
turbulence driven by some external force (e.g., accretion, supernova
explosions, bipolar outflows, etc.). In that scenario, the turbulence
plays a dual role in the clouds and their substructures, providing
global support against the weight of the structures, while locally
producing shocks that generate small-scale density enhancements, in
which the local Jeans  mass (possibly including turbulent pressure)
decreases sufficiently as to become smaller than the fluctuation's mass,
causing its collapse.  Therefore, the clouds as a whole are in a
nearly stationary state, being in approximate virial equilibrium, and
only slowly losing mass to star formation at the rate permitted by the
turbulent formation of collapsing small-scale structures \citep[e.g.,]
[] {VS+03, MK04}.

However, numerical simulations have shown that, in
general, the turbulence generated by the cloud assembly process
\citep[e.g.,] [] {KI02, AH05, Heitsch+05, VS+06, VS+07, VS+10, Wareing+19} is only
moderately supersonic (Mach numbers $\Ms \sim 3$) with respect to the
dense, cold gas, in contrast with the highly supersonic regimes ($\Ms
\gtrsim 10$) indicated by observations of MCs \citep[e.g.,] []
{Larson81, HB04}.  As a consequence, the turbulence
self-consistently generated during the cloud assembly process eventually
becomes insufficient to support the clouds against their self-gravity
\citep[e.g.] [] {VS+07, Heitsch+09}. In addition, \citet{CB05} have
concluded from analysis of the energy budget of turbulent density
flutuations in numerical simulations of decaying turbulence, that the
Jeans mass in the fluctuations does not directly become smaller than the
fluctuation's mass. Instead, the fluctuations grow by the self-gravity
of the larger-scale region.

In view of the above, and in contrast with GT, in GHC the chaotic,
multi-scale infall motions are assumed to dominate at all scales
(except, perhaps, in a fraction of the low-column density clouds and
clumps that appear strongly super-virial; see
Sec. \ref{sec:alpha-mass}), and the truly turbulent motions (those that
are fully disorganized and might {in principle} provide a turbulent
pressure) appear to be
{\it fed} by the gravitational collapse of the structure in which they
are observed, but they do not manage to significantly retard the
collapse, {even though they may approach levels approaching those of
virial balance} (Guerrero-Gamboa et al., in prep.); {this appears to
be a consequnce of the strongly dissipative nature of turbulence, which
is usually not included in virial treatments of the turbulent
contribution. The only role of these turbulent motions} is to
provide nonlinear density fluctuations that, when they become unstable
due to the global reduction of the Jeans mass caused by the global
collapse, begin to collapse on their own. Then the pattern repeats
itself inside these new, smaller and denser objects. The mechanism can
be considered a joint mass and energy cascade driven by self-gravity.

Such a regime had already been envisioned by \citet{Field+08},
although they imagined that the flow was virialized at all
scales. Instead, \citet{BP+11a} proposed that the flow is dominated by
infall motions at all scales, eliminating the need for the flow to
first virialize at each scale and then become unstable again to
proceed to the next stage of collapse. The latter authors pointed out
that the free-fall velocity is only a factor of $\sqrt{2}$ larger than
the virial velocity, and therefore free-falling clouds follow the same
scaling relation between velocity dispersion ($\sv$), size ($R$) and
column density ($\Sigma$) \citep[$\sv/R^{1/2} \sim (G \Sigma)^{1/2}$;]
[] {Heyer+09} as virialized clouds, within typical observational
uncertainties.

Additionally, the fact that any given object only lasts as a
coherent unit (before fragmenting again) while it contains one to a few
Jeans masses implies that very large velocities are in general not
observed within a coherent structure. The large velocities associated
with late stages of the collapse should be observable as
fragment-to-fragment velocities instead.  A similar conclusion has
been reached by \citet{Hacar+16} concerning the CO linewidths in nearby
clouds through an analysis of the correlation between the linewidths and
the line opacity.

\subsubsection{Inconsistency between turbulence and generalized equipartition}
\label{sec:inconsis_turb_equip}


An important point to note is that, in the GT scenario,
turbulence is assumed to be in approximate virial equilibrium with
self-gravity at all scales, and the \citet{Larson81} linewidth-size
scaling relation, $\sv \propto R^{1/2}$ (or, equivalently, $\call
\approx$ cst.) is assumed to be the result of
supersonic turbulence with a Burgers' energy spectrum of the form $E(k)
\propto k^{-2}$ \citep[e.g.,] [] {MO07}, where $E(k)$ is the specific
kinetic energy per wavenumber interval, and $k$ is the wavenumber.
However, this assumption disregards the fact that the Larson
linewidth-size scaling only holds for objects of similar column
density, while, when objects with a wide range of column densities are
considered, a Heyer-like relation of the form $\sv \propto
\sqrt{\Sigma R}$ (or, equivalently, $\call \propto \Sigma^{1/2}$) holds
\citep{Heyer+09, BP+11a, BP+18, Traficante+18a}. This relation arises from
approximate energy equipartition between self-gravity and non-thermal
motions, either because of virialization or free-fall (cf.\ Secs.\
\ref{sec:seq_definitions} and \ref{sec:energy_balance}).
This implies that the column density-dependence of the Larson ratio
$\call$ observed in MCs and their dense substructures {\it is
inconsistent} with Burgers turbulence when objects of a wide range of
column densities are considered, {as was further discussed in Sec.\ \ref{sec:probs_turb}}. 

\subsubsection{The role of accretion: filaments, hub accretion,
development of the massive stellar component, and brown dwarfs}

While in GT filaments are assumed to be the result of shocks, with no
need for developing a longitudinal flow \citep[e.g., ] [] {Padoan+01},
in GHC the filaments constitute the very accretion flow from large to
small scales, akin to rivers carrying the material down the
gravitational potential from high ``altitudes'' to ``lakes'' (the
stars). Because the filaments may extend over scales of up to tens of
parsecs, accretion motions also extend over such large scales, albeit
with an extremely anisotropic structure, rather than through the common
picture of spherical, homologous collapse.

In addition, if the accretion onto the hubs is Bondi-Hoyle-type
with respect to the local environmental conditions
(cf.\ Sec.\ \ref{sec:IMF_hub_accr}), then the accretion rate onto the
hubs should increase over time as the mass of the hubs themselves
increases. This may be at least part of the reason massive stars appear
late in the simulations. 
Higher hub accretion rates are expected to require
stronger stellar feedback in order for the filaments to be dispersed,
and thus result in the formation of more massive stars within more
strongly accreting hubs. This may explain the apparent lack of massive
stars in regions where the star formation activity is still rather
scattered, even if the total mass content is already large, such as the
IRDC M17 SWex \citep[e.g.,] [] {Povich+16}, the apparent excess of
massive stars in regions of extremely high SFRs, such as the Galactic
Center cluster and the Arches cluster \citep[e.g.,] [] {Lu+13,
Hosek+19}, and the observed correlation between the mass of the most
massive star in a cluster and the cluster mass \citep[e.g.,] []
{WK06}. We shall investigate this mechanism in future contributions.

Finally, it is worth noting that the secondary star formation occurring
in the cores along the filaments (rather than the central hubs) must
involve lower local accretion rates, and moreover may have their gas
supply interrupted by the feedback from nearby main hubs already forming
massive stars. Thus, the secondary collapses occurring in the filaments
are consistent with suggestions that brown dwarf formation
may require secondary formation in the periphery of the main
star-forming hubs \citep[e.g., ] [] {Thies+15}.

\subsection{Comparison with other scenarios} \label{sec:other_scen}

The GHC scenario is much closer to the {\it competitive accretion} one
\citep[e.g.,] [] {Bonnell+01, BB06}, which already includes small- (at
the single star or small stellar group level) and large-scale (at the
level of the accretion flow onto the small star-forming sites)
collapse flows. The main difference is that the competitive accretion
scenario in general overlooks the evolution of the MCs that occurs
over timescales of order several {megayears}, in particular the global cloud
contraction and the evolution of cloud properties and the SFR, since
the numerical simulations of that process typically only consider
parsec-scale clumps and durations $\lesssim 1$ Myr, because they are
often aimed at studying the details of the stellar products
\citep[multiplicity, IMF, brown dwarf production, etc.; e.g.,] []
{Bate12, Kuznetsova+15}, rather than at the evolution of the GMC-scale
system, over timescales of a few tens of {megayears}, from their
formation to their dispersal, passing through their gravitational
contraction, as done in converging-flow simulations \citep[e.g.]  []
{VS+07, VS+10, HH08, Hennebelle+08, Banerjee+09, Colin+13, Carroll+14,
Koertgen+16, Wareing+17a, ZA+18}. An exception is the study by
\citet{Smith+09}, which did consider the accretion onto the star-forming
clumps and their mass evolution, concluding that the massive stars form
when the clumps have become massive enough themselves, similarly to
\citet{VS+17}. But at the local level, our scenario and simulations
are fully consistent with competitive accretion.

{{\mgta Also, as discussed in \citet{GV14} and Sec.\
\ref{sec:general_consid}, and shown in Fig.\ \ref{fig:GV_fil}, the GHC
scenario is consistent with the ``conveyor belt'' scenario
\citep{Longmore+14}, since the filamentary accretion flows serve as
funnels for the gas to flow from the large (cloud and clump) scales
where the gas is initially distributed to the small (core and hub)
scales in which it resides once it has been gravitationally compressed
to high densities and forms stars. These filaments accrete {\it from} the
cloud scale, and {\it onto} the cores and hubs. Moreover, secondary star
formation is observed to occur in the simulations when they have
accreted enough mass that they become locally gravitationally
supercritical \citep[see Figs.\ 2 and 3 of] [] {VS+17}.}}

Finally, the scenario outlined here is fully consistent with the
``dynamical model for SF'' presented by \citet{Elm15}, in which the ISM
is in a dynamical cycle between gravitational contraction followed by
feedback-induced dispersal. In that model, the dispersal occurs on the
free-fall time of the star-forming gas. In our scenario, this time is
longer than the free-fall time due to the anisotropy of the infalling
flow, which adopts a filamentary shape, and thus collapses on timescales
of the order of the free-fall time times the aspect ratio of the
structures \citep{Toala+12, Pon+12}. But the overall dynamical state is
similar to that outlined by \citet{Elm15}, with the infalling gas having
virial parameters $\alpha \sim 1$, while the dispersing gas has $\alpha
\gg 1$ \citep[e.g., Fig.\ 11 of] [] {Colin+13}. Observations 
showing that  much of the molecular gas  mass has large
values of the virial parameter  \citep[e.g.,] [] {Heyer+01, Kauffmann+13,
Leroy+15, HD15, Traficante+18a} are thus consistent with the numerical
result that most of the gas in an initially contracting MC is dispersed
by stellar feedback before it makes it into stars. This feature thus
allows the GHC model to avoid the \citet{ZP74} conundrum that
free-falling MCs should have a SFR much larger than observed.

\subsection{Deconstructing recent criticisms to GHC}
\label{sec:deconstr_critic}

The GHC scenario has recently encountered criticisms, which mostly
originate from the original SFR and line-shift-absence conundrums (cf.\
Sec.\ \ref{sec:deconstr}). The modern version of the SFR conundrum is
{that the} star formation rate per free-fall time,
$\eff$, is very low, of the order of $\sim 1\%$, in all
molecular structures, from GMCs to massive star-forming clumps
\citep[e.g.,] [] {KM05, KT07}. The modern version of the
line-shift-absence conundrum is the 
statement that traditional
infall signatures, such as blue-excess or inverse P-Cygni line
profiles are not observed at large scales in MCs \citep[see, e.g., Sec.\
4.6 of] [] {MC15}. In addition, \citet[] [hereafter K19l see also
Krumholz \& McKee 2019] {Krumholz+18}
have recently argued that the kinematic signatures of clusters do not
correspond to what would be expected from the GHC scenario. All of these
criticisms arise from an incomplete consideration of the full
phenomenology of GHC, in particular its evolutionary and extremely
non-isotropic nature, as well as the ubiquitous presence of accretion at
all scales.

\subsubsection{The efficiency per free-fall time}

The quantity $\eff$, which is actually an efficiency rather than a
rate, is defined as the fraction of a cloud's mass that is converted
into stars over a free-fall time \citep{KM05}; that is,
\beq
\eff \equiv \frac{\dot \Mstar} {\Mgas} \tauff,
\label{eq:sfrff}
\eeq
where $\dot \Mstar$ is the SFR, $\Mgas$ is the molecular gas mass, and
$\tauff$ is the free-fall time. In practice, $\dot \Mstar$ is sometimes
measured as $\dot \Mstar \approx \Mstar/\Delta t_{\rm SF}$, where
$\Mstar$ is the instantaneous stellar mass in an embedded cluster and
$\Delta t_{\rm SF}$ is the duration of the star-formation epoch, often
taken as the stellar age spread, typically a few {megayears} \citep[e.g.,] []
{Evans+09, Lada+10, Povich+16, Retes+17}. In this case,
\beq
\eff \approx \frac{\Mstar} {\Delta t_{\rm SF} \Mgas} \tauff.
\label{eq:sfrff_measd}
\eeq
Other studies estimate the SFR from the IR luminosity \citep[e.g.,] []
{Vutisal+16, Liu+16a}. 

In either case, eq.\ (\ref{eq:sfrff}) may be
suitable over large spatial regions or long timescales that involve a
large number of clouds, over which the SFR can be averaged. Indeed,
\citet{ZV14} have performed temporal and ensemble averages of their
accelerating-star formation model (discussed here in Sec.\
\ref{sec:SFR_increase}), and shown that these averages provide good fits
to the observed correlation between SFR and dense gas mass, both at the
individual cloud scale (when a single-cloud model is temporally
averaged), and at the cloud-ensemble scale (when the model is averaged
over time and over a cloud-mass spectrum; see Fig.\ 5 of that paper).

On the other hand, {the} definition {of $\eff$ in eq.} 
(\ref{eq:sfrff}) will give a serious
underestimation of the efficiency if the present free-fall time is used
to estimate it in an individual, evolving star-forming region, since in
this case the collapse rate and the star formation rate are increasing
strongly, and the free-fall time is decreasing strongly. This can be
clearly seen from eq.\ (\ref{eq:rho_of_t}), from which one can write an
evolution equation for the ratio of the instantaneous to the initial
free-fall time of a collapsing object, which reads
\beq
\frac{\tauff(\tau)} {\tauffo} = \left(1 - \tau^2\right)^{a/2},
\label{eq:tff_of_t}
\eeq
where, as in Sec.\ \ref{sec:seq_coll}, $\tau$ is the time in units of
the initial free-fall time ($\tauffo$), and $\tauff(\tau)$ is the
instantaneous value of the free-fall time at time $\tau$. From this
expression, the mean free-fall time up to time $\tau$ is given by 
\beq
\left\langle \frac{\tauff} {\tauffo} \right\rangle_{\tau} = \int_0^\tau (1 - \tau'^2)^{a/2} d\tau'.
\label{eq:avg_tff}
\eeq
For example, toward the end of the collapse, $\tau \rightarrow 1$ and
$\tauff/ \tauffo \rightarrow 0$. However, $\langle \tauff/
\tauffo\rangle_{\tau=1} \approx 0.68$, meaning that the characteristic
collapse timescale is not shorter than 2/3 of the {\it initial}
free-fall time. Using the {\it final} free-fall time severely
underestimates the mean timescale, spuriously producing very low values
of $\eff$ when it is estimated using eq.\ (\ref{eq:sfrff}).

As an example, if a star-forming clump is $\sim 100 \times$ denser than
the cloud it formed from, its free-fall time will be shorter than the
initial one by a factor $\sim 10$, and shorter than the mean one by a
factor $\sim 6.8$. Interestingly, this {factor is of} the order of
magnitude of the factor by which {K19 suggest that} the duration of
the star-formation activity is longer than the measured free-fall time
({a factor $\sim 10$}). The interpretation in the GHC scenario is
that the present-day free-fall time is up to $10\times$ or more shorter
than the mean one over the development of the cluster up to its present
stage. The clouds have been evolving at the free-fall rate, but this
rate was much lower over most of the past history of a forming cluster
than it is at present. This is illustrated, for example, by the
evolutionary track in Fig.\ \ref{fig:KS_ZA}, which shows that the
bursting stage of the model cloud lasts only a small fraction of the
whole evolutionary process (less than 1 Myr in that specific example).

It is worth pointing out that this bias is analogous to that produced in
the estimate of the expected fragmentation level in massive clumps, as
discussed in Sec.\ \ref{sec:time_nth_lack_frag}. In that case, it was a
consequence of the usage of the present-day Jeans mass, rather than the
initial one. Here, it is a consequence of using the present-day $\tauff$
rather than the characteristic one. 

{It is important to note that the above discussion applies to
individual MCs that are observed at random stages of their evolutionary
process.  At the scale of hundreds of parsecs or more, of the type of
regions observed in nearby external galaxy surveys \citep[e.g.,] []
{Bigiel+08, Leroy+08, Onodera+10, Schruba+10, Liu+11}, which contain
ensembles of GMCs as well as more diffuse gas, the GHC interpretation of
the low observed $\eff$ is different. In this type of studies, the SFR
and the gas mass are estimated by means of the region brightness in gas
or star-formation tracers, such as CO and H$_\alpha$ emission,
respectively. In this case, the reason for the low observed values of
$\eff$ is the same as the solution to the SFR conundrum: since most of
the mass in initially contracting MCs is prevented from actually forming
stars, and is instead dispersed or evaporated, large regions are
expected to contain much of this ``molecular debris'' from a previous SF
episode. Note, however, that this requires that an important fraction of
the molecular mass is dispersed rather than evaporated. This can be
considered a prediction of the scenario, which should be tested in
simulations of the full life cycle of MCs, in the presence of various
feedback mechanisms. Existing simulations using photoionising radiation
feedback \citep[e.g.,] [] {Colin+13, ZA+18, Haid+19} suggest that indeed
a substantial fraction of dense gas survives.}

\subsubsection{The absence of infall signatures}

The argument is often made that MCs cannot be in global collapse because
characteristic infall profiles are generally not observed at the cloud
scales \citep[$\gtrsim 10$ pc; e.g.,] [{hereafter MC15}] {MC15}. The
GHC explanation for this is that the collapse flow is far from
isotropic, and instead proceeds mostly along filamentary
structures. This collapse flow {\it is indeed observed}, {except not
so much} as a
velocity gradient along the line of sight which produces an infall line
profile, but rather as longitudinal flows along the filaments, which
feed the central clumps
\citep[e.g.,] [] {Sugitani+11, Kirk+13, FL+14, Motte+14, Peretto+14,
Tackenberg+14, JS+14, Hajigholi+16, Wyrowski+16, Juarez+17, Rayner+17,
Lu+18, Baug+18, Gong+18, Ryabukhina+18, Dutta+18, Chen+19}. Away from
the filaments, the flow is directed mainly towards the filaments
themselves rather than towards the central hubs
\citep[see the left panel of Fig.\ \ref{fig:GV_fil} and the
observational results by] [] {Shimajiri+19}. Therefore, in the GHC
scenario, the flow at cloud scales is highly chaotic and a simple
signature such as typical infall line profiles, which is based on the
assumption of a roughly spherical structure and flow, is not 
generally expected, although, as mentioned in Sec.\ \ref{sec:deconstr}, recent 
studies at the clump \citep{Barnes+18} and GMC \citep{Schneider+15}
scales have indeed found infall signatures by combining optically thick
$^{12}$CO and optically thin $^{13}$CO lines.

\subsubsection{The linewidth-size relation at large and small scales}

MC15 have also argued that the different observed exponents $p$ of
Larson's (1981) linewidth-size relation at the scale of whole GMCs
\citep[$p \sim 0.5$; e.g., ] [] {Solomon+87, HB04} are larger than at the
scale of massive dense cores \citep[$p \sim 0.25$; e.g., ] [] {Plume+97,
CM95}. MC15 arrive to this conclusion because they build a model for the
``adiabatic heating'' of turbulence from gravitatioanl contraction, in
which they find two different regimes: an external region in which $p
\sim 0.5$ and an internal one, dominated by the gravity of a central
stellar object, where $p \sim 0.25$ \citep[see also] []
{Murray+17}. Thus, MC15 suggest that the lower values of $p$ in the
dense cores are indicative of gravitational contraction at the clump
scale, while the higher values at the GMC scale are indicative of
supersonic turbulence rather than gravitational contraction.

However, \citet{BP+11a} have shown that both GMCs and massive clumps do
follow a unique scaling of the form of eq.\ (\ref{eq:vir}). That is, the
value $p \sim 0.5$ applies for roughly-constant-column density objects,
such as molecular clouds, but when the column density varies, the
linewidth-size relation is shifted, as also shown in numerical
simulations by \citet{Camacho+16}. Thus, the shallower linewidth-size
relation for massive clumps can be understood as a consequence that the
larger clumps have lower column densities. That this is so is
demonstrated by the fact that, when the clumps and the GMCs are plotted
in the $\call$--$\Sigma$ diagram, they all fall in the same scaling
relation.

\subsubsection{Radial motions of cluster members}

It has been recenty claimed by K19 that the GHC scenario should produce
radial motions in the members of stellar clusters formed by the
collapsing clouds which, they claim, are not observed. Thus, they
conclude that the GHC scenario does not match the observed young cluster
kinematics. This line of argument {has two problems}:

First, K19 start by stating that the GHC scenario should produce
radial infall motions in the stellar products during the early stages of
the process, and then argue that these inwards radial motions are not
observed, thus concluding that collapse cannot be occurring. This claim
seems to arise from an implicit assumption of roughly spherical
symmetry, which is however not backed up by the very simulations that
follow the collapse process \citep{Kuznetsova+15, Kuznetsova+18a,
VS+17}. In the simulations, the structure of the forming clusters is
hierarchical (or fractal), having been inherited from their parent
clouds. Thus, the clusters consist of groups and sub-groups, and the
motions are a combination of the local plus the global collapse flows,
on a highly filamentary substrate \citep[see, e.g., Figs.\ 2 and 3 of]
[] {VS+17}. Thus, the claim of radial inward motions arises from an
oversimplification of the hierarchical collapse mechanism, rather than
from the actual kinematics observed in the simulations of GHC.

Second, for more advanced stages, K19 argue that, once the gas is
expelled by the stellar feedback, up to 90\% of the stars should be
moving radially away from the dense cluster, and that thus, the GHC
scenario should produce either inward or outward radial motions of the
cluster members. Then they quote \citet{WK18}, \citet{Kuhn+19}, and
\citet{Kounkel+18} as examples that {\it Gaia} shows no such radial
motions in Orion and other complexes, again concluding that the
observations contradict the GHC scenario.

This line of argument is actually confusing, for two reasons. The first
is that the expansion due to gas expulsion should occur independently of
the scenario, since it only involves a shallowing of the gravitational
potential well, caused by the feedback, independently of the mechanism
that assembled the cluster-forming gas. So, the expansion should be
present regardless. The second is that both \citet{Kuhn+19} and
\citet{Kounkel+18} {\it do} report expansion motions, and in fact the
former authors report {\it typical} expansion velocities of $\sim 0.5\,
\kms$, and even a radial expansion velocity gradient, so the reference
to these papers by K19 {\it against} expansion seems
contradictory. In any case, the simulations with feedback presented by
\citet{VS+17} also show expansion (again, see Figs.\ 2 and 3 of that
paper), although a detailed comparison with the observations remains to
be done.

 Concerning \citet{WK18}, these authors do claim that no expansion
is seen in their data. However, they consider much larger regions,
several tens of parsecs across. As mentioned in Sec.\
\ref{sec:clusters}, at size scales this large, the coherence of the
infall motions is lost because the infall timescale is larger than the
cloud-destruction timescale from the feedback, so no clear signature of
the infall is to be expected. This is aided by the multi-scale and
filamentary nature of the infall motions, which contribute to the
non-appearance of a single-focused collapse at those scales. On the
other hand, \citet{WK18} do conclude that their observations support a
hierarchical star formation model.

We thus conclude that the claim by K19, that observed cluster
kinematics do not support the GHC scenario, is unfounded.

\subsection{Caveats} \label{sec:caveats}

In this paper we have presented several calculations of the evolution
of clouds as they undergo gravitational contraction, but of course,
these calculations are only approximate, and are presented mostly as
proof of concept. The main approximations we have used, and which
must be improved upon in more detailed  calculations (such as those
performed semi-analytically in the ZV14 model) are  as follows:

First, we have assumed spherical symmetry and uniform density by using
eq.\ (\ref{eq:r_of_t}) for the evolution of the radius of a collapsing
object. In reality, clouds are expected to be flattened, and thus to
evolve more slowly than a spherical cloud with the same density
\citep{BH04, Toala+12, Pon+12, ZA+12}. Moreover, due to the presence of
fluctuations, the clouds contract non-homologously, as shown by both
similarity studies and numerical simulations \citep[e.g.,] [] {Larson69,
Penston69, Hunter77, Shu77, WS85, FC93, MS13, Keto+15,
Naranjo+15}. Thus, we have substituted the calculation of a single,
non-homologous contraction, by the consideration of a sequence of
uniform spheres of different densities that collapse on different
timescales.

Second, because of this treatment, we have neglected accretion onto an
object at a given scale from its parent object at a larger
scale. Accretion from the next-larger scale has been shown by numerical
simulations to be an essential part of the multi-scale collapse process
\citep[e.g., see Fig. 7 in] [and Sec. 4.3.2 of V\'azquez-Semadeni et
al.\ 2010] {VS+09}. Also, the clumps lose mass  by conversion to
stars and by the effect of stellar feedback. These important aspects of
the evolution remain to be included in a more thorough analytical
treatment. Numerical simulations show that clumps defined by density
thresholds rather than by fixed mass (as assumed in Sec.\
\ref{sec:seq_definitions}) evolve along different trajectories than
those shown in Fig.\ \ref{fig:Lin_Lff}, turning back to lower column
densities as they begin to lose mass (Camacho et al.\ 2019, in
prep.).

Third, we have neglected deviations from spherical or circular
geometries as the collapse proceeds. As proposed by \citet{GV14} and
discussed in Sec.\ \ref{sec:molec_fil_form}, the infall of the large
scales develops filamentary flows, which feed the  faster-evolving,
smaller-scale, more roundish objects. However, this should not introduce
significant errors in the calculation of the collapse timescales since,
as discussed in \citet{Toala+12}, the collapse timescale of a flattened
or filamentary structure is the same as that of a spherical structure
containing the same mass over size scale equal to the largest spatial
extent of the sheet or filament. Since the filaments basically form from
the collapse of the large-scale cloud, their timescale will be the same
of the initial, closer-to-spherical diffuse parent cloud.

Finally, we have neglected any effects of support provided by the
Galactic rotation and shear (i.e., the Toomre criterion). This implies
that our discussion is limited to scales where the shear induced by the
Galactic rotation is negligible, i.e., a few hundred parsecs.

\section{Summary and conclusions}\label{sec:concl}

\subsection{Summary} \label{sec:summary}

\begin{figure*}
\includegraphics[width=\textwidth]{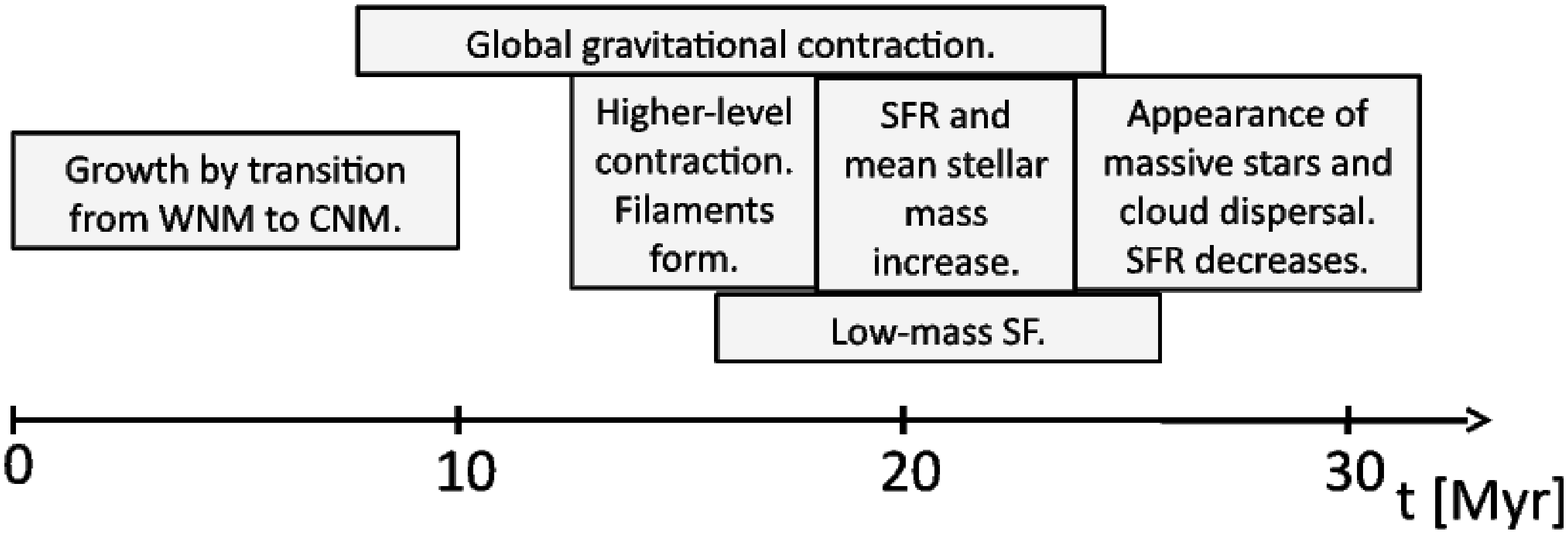}
\caption{ Schematic representation of the evolution of a $10^4$--$10^5
\Msun$ GMC for typical Solar neighborhood conditions, assuming it starts
from WNM gas exclusively. The timescales are approximate, and neglect
the possibility of extended accretion from larger-scale mass
reservoirs. See text for a description of each stage.}
\label{fig:timeline}
\end{figure*}

In this paper we have presented a complete and unified description of
the GHC model, starting from the formation of MCs as thin, cold atomic
sheets, through their growth, evolution of their SFR, until their
destruction by the massive stars they form, preventing most of their
mass {from being} converted to stars, and thus keeping the overall SFE low. The
model describes the phenomenology observed in numerical simulations of
MC formation in the warm atomic medium.  Figure \ref{fig:timeline}
shows a schematic representation of the evolution of a $10^4$--$10^5
\Msun$ GMC for typical Solar neighborhood conditions, assuming it starts
from WNM gas exclusively. The next subsections briefly summarize the
evolutionary sequence.

\subsubsection{Cloud formation and the onset of collapse}

In this scenario, MCs at galactocentric distances where the majority of
the volume of the galactic disk is occupied by diffuse gas \citep[$R
\gtrsim 6.5$ kpc;] [] {Koda+16} initiate their life cycle by
compressions in the predominantly diffuse medium, which nonlinearly
trigger a transition to the cold atomic phase \citep{HP99, KI02}. Clouds
formed by this type of compression tend to be planar, because the
collision of streams or shock fronts are most often two-dimensional
surfaces, which grow by accretion of gas from the warm phase. The clouds
are bounded by a {\it transition front}, where the accreted gas
undergoes the transition from the warm to the cold phase \citep{VS+06,
Banerjee+09}.

Because the cold phase is $\sim 100\times$ denser and colder than
the initial warm phase, the Jeans mass in this material is $\sim 10^4
\times$ smaller than in the warm phase, and thus the cold gas quickly
(in a few to several {megayears}, depending on the coherence scale of the
converging motions)
acquires masses much larger than its Jeans mass \citep{GV14}. The
accumulation of gas produces turbulence, but generally it is only
moderately supersonic \citep[$\Ms \sim$ a few] {Vishniac94, WF00,  KI02,
Heitsch+05, VS+06, Wareing+19}, implying that in general it is insufficient to
prevent the cloud from quickly engaging in global gravitational
collapse, at which point the nonthermal kinetic energy becomes
``locked'' to the gravitational energy, but is never capable of
retarding the collapse \citep{VS+07}.

\subsubsection{Fragmentation and filament formation}

Because the cloud  acquires masses much larger than the Jeans mass,
the collapse becomes nearly pressureless, and thus it proceeds
fastest along the shortest dimension of the cloud, reducing its
dimensionality \citep{Lin+65}, leading to the formation of filaments
\citep{Heitsch+08b, GV14}. Within the filaments, roundish density
fluctuations grow faster than the rest of the filament because the
filaments have longer collapse timescales \citep{Toala+12,
Pon+12}. Thus, these clumps produce a deeper gravitational potential,
and the rest of the filament begins to accrete onto the clump,
developing a longitudinal flow along the filament and onto the clumps
\citep{GV14}. It is important to remark that {\it no strong shock is
produced at the axis of the filament} because a shock only develops in a
gravitationally contracting structure when the density develops a
singularity, but this is avoided by the longitudinal mass flux along the
filament towards the clumps, which apparently leads to the establishment
of a near-stationary state in the filaments (Naranjo-Romero et al.\, in
prep.). The absence of a central shock is indeed observed in
numerical simulations of gravitationally-formed filaments (Fig.\
\ref{fig:GV_fil}), and may constitute a diagnostic to distinguish
between gravity- and turbulence-formed filaments.

As the mean density in the cloud increases, the mean Jeans mass
decreases (eq.\ [\ref{eq:MJrms_of_tau}]). Assuming that the density PDF
{\it of the turbulent fluctuations} (not of the collapsing objects)
retains its lognormal shape and its mean Mach number, the density PDF
evolves simply by shifting to higher densities while retaining its shape
\citep{ZA+12}. Moreover, because the clouds
are sheet-like and the turbulent density fluctuations are nonlinear,
Hoyle-like fragmentation can occur without concern for the objections
raised by \citet{Tohline80}, as discussed in Sec.\
\ref{sec:general_consid}. 

\subsubsection{Increase of the SFR and the mean stellar mass formed}

At the {\it typical} density of the turbulent density fluctuations, less
massive fluctuations become unstable at later times (Fig.\
\ref{fig:ttot_of_mu}). However, the {\it first} fragments that begin to
collapse correspond to the most extreme density fluctuations that have a
mass larger than their corresponding local Jeans mass. Therefore, the
first fragments to collapse have the lowest masses. {Subsequently}, as
the mean density increases, the local collapses can occur for densities
less distant from the mean. Since there is more mass at these densities,
larger-mass fluctuations can collapse, and so {\it the typical mass of
the collapsing objects increases with time}.  Simultaneously, the SFR
increases with time \citep{ZA+12, ZV14, VS+18}, and so the
star-forming clouds evolve towards larger SFRs while increasing the mean
mass of the stars they form (see Fig.\ \ref{fig:cumul_mass_distr} here,
taken from \citet{VS+17}). {This implies that there should be a
correlation between the most massive star in a star-forming core or
clump, and the clump's mass. In turn, this implies the existence of a
correlation between the mass of the most massive star in a cluster and
the cluster mass, which is indeed observed \citep[e.g., ] [] {WK06}}.

\subsubsection{Energy balance evolution in collapsing clumps}
\label{sec:en_bal_evol}

Moreover, as the clump evolves, so does its virial parameter. According
to the discussion in \citet{BP+18} and Sec.\ \ref{sec:seq_definitions},
the virial parameter contains an inertial, possibly compressive,
component\footnote{Recall that this component need not consist of
turbulence that supports the cloud. It can just as well consist of the
external converging flows that produce the clump.} and a
gravitationally-driven infall one ($\vg$). Because the collapse of
fragments starts at an initial radius $R_0$, the infall speed is given
by eq.\ (\ref{eq:vg}), implying that the gravitational component
is expected to be subvirial over a significant fraction of the clump's
evolution. Thus, depending on the initial value of the ratio of the
inertial to the gravitational components, the core may appear super- or
sub-virial.

Moreover, at a given size scale, the more massive (i.e., denser) cores
will have a larger value of $\vg$, and thus their total linewidth will
tend to be dominated by this component, which may nevertheless be
sub-virial. This may explain the often-observed trend for more massive
cores to be more sub-virial \citep[e.g.,] [] {Kauffmann+13, Ohashi+16,
Sanhueza+17, Traficante+18b}. We will explore this scenario in a future
publication.

\subsubsection{Local cloud disruption}

As massive stars begin to form in the region, they begin to feed back
sufficiently strongly on their parent clump. This has the effect of
first disrupting the filaments feeding the central star-forming core or
hub, and later the hub itself (Fig.\ \ref{fig:fil_erosion}) or,
alternatively, allowing the gas in the hub to be exhausted because the
external gas supply has been interrupted. This process leads to a
gradual reduction, and possibly eventual full local supression of the
SFR. Numerical simulations of the process show that the {\it final} SFEs
are $\sim 30\%$ \citep[e.g., ] [] {Colin+13, IH17}. Such values are
reasonable, considering that they refer to the {\it final} value of the
SFE, for which there is no observational determination, because by the
time a cluster appears devoid of gas, it is not possible to
observationally know the total amount of gas that went into its
formation. Currently, the issue that still needs to be properly assesed
is the fact that the effect of the feedback depends strongly on the type
of feedback and the precise location of the injection with respect to
the clouds \citep[e.g.,] [] {Dale15, IH15}.

\subsection{Resolution of the old conundrums} \label{sec:resol_conun}

A crucial question that may be asked of the GHC scenario is whether it
provides an answer to the decades-old conundrums that originally
dispelled the scenario that star-forming molecular clouds may be in
global gravitational collapse \citep{Liszt+74, GK74}. But indeed it
does. The SFR conundrum \citep[] [cf.\ Sec.\ \ref{sec:deconstr} here]
{ZP74}, that the SFR should be much larger than the observed value if
the clouds are in global gravitational collapse, is avoided by the
fact that the star-forming regions are destroyed by the first very
massive stars that form \citep [larger numbers of massive stars may be
necessary for destroying more massive clouds;] [] {Franco+94}, so that
the majority of the cloud mass never manages to be incorporated into
stars, but is rather returned to the ambient medium. Indeed, numerical
simulations of clouds up to $\sim 10^5 \Msun$ including stellar
feedback (ionizing radiation, winds and SNe) show that the clouds are
readily destroyed, with final SFEs that depend on the type and
location of feedback considered \citep[e.g.,] [] {Dale+12, Dale+13,
KT12, KT13, Colin+13, IH15, IH17, Koertgen+16, Wareing+17a,
Wareing+17b, Seifried+18}. Note that the typical quoted efficiencies
for GMCs, of a few percent, refer to the SFE of objects that are still
dominated by the gaseous mass, so by definition they must be small.

The other conundrum, which we referred to as the line-shift conundrum
\citep[] [] {ZE74}, is avoided in the GHC scenario because such line offsets
are not expected to be observed in denser gas, since this gas acquires a
filamentary morphology, and so there is no roughly spherical envelope
around the collapse centers (the hubs). Instead, the gas flows from the
diffuse regions goes onto the filaments, and then longitudinally along
the filaments onto the hubs (see Fig.\ \ref{fig:GV_fil}). In addition,
such shifts are beginning to be marginally observed in more diffuse
molecular gas ($^{12}$CO and $^{13}$CO) \citep{Barnes+18}.

 It should be emphasized that many of the objections against the GHC
scenario arise from a failure to appreciate the self-similar and
non-homologous nature of a realistic gravitational collapse, which
implies that regions farther away from the center can be considered to
be at an earlier evolutionary stage than the more central ones. This is
best illustrated by a similarity description of spherical gravitational
collapse \citep[e.g.,] [] {Larson69, Penston69, Shu77, Hunter77, WS85},
in which the spatial and temporal variables are merged into a single
{\it similarity variable}, defined by $\xi \equiv r/\cs t$; that is, the
radial position is {normalized} by a time-dependent factor amounting to the
distance traveled by a sound wave up to time $t$. Thus, large distances
from the collapse center at a given time behave like shorter distances
at an earlier time, suggesting that a non-homologously-collapsing
spherical core can be thought of as a collection of concentric
collapsing shells, each at a different evolutionary stage.

In the analytic solutions for the spherical problem, the infall speed at
large radii during the prestellar stage approaches a constant \citep
[e.g.,] [] {Larson69, WS85}. However, in numerical simulations, because
the collapse starts from a local fluctuation, the size of the infalling
region is observed to grow over time, similarly to the expanding
rarefaction front of Shu's (1977) inside-out collapse \citep[see, e.g.,
the animations corresponding to Fig.\ 3 of] [noting how the longitudinal
motions along the filaments progressively extend to ever larger
distances] {GV14}. Note, however, that this front lies much further out
than Shu's front, because the infall motions begin much earlier (at the
onset of collapse), rather than at the time of the formation of a
singularity, as he assumed. So, the internal regions of a finite
collapsing prestellar structure exhibit an oustide-in nature (cf.\ Sec.\
\ref{sec:diff_pre_proto}), while the external ones exhibit an inside-out
one. 

Finally, when the feedback from the massive stellar products begins to
disrupt the cloud, regions sufficiently distant that the infall-motion
front has not reached them yet will be effectively decoupled from the
collapse by the feedback, and this material does not participate in the
star formation episode of the region, maintaining the global SFR low.
Thus, even if clouds are dominated by gravity at large scales, the
motions do not necessarily reflect a clear collapse.

\subsection{Implications of multi-scale collapse for the gas flow} 
\label{sec:implic_GHC}

The condition that MCs and their substructures are undergoing  global
hierarchical gravitational contraction has the important implication
that the gas is {\it flowing} from the low- to the high-density
regions. This means that the gas {\it changes location} as it proceeds
to higher densities, and this process {\it takes time}. Simply stated,
before a hydrogen atom reaches the interior of a star, it must traverse
a path where the environmental density increases from the mean density
of the Universe to that in the star's interior.

This apparently trivial consideration has strong implications on notions
such as the one that low-mass dense cores may be Jeans-stable and
hydrostatic, requiring an external pressure to remain confined.
Instead, numerical simulations of spherical collapse on top of a
globally Jeans-unstable nearly uniform background \citep{Naranjo+15}
show that the background is also infalling (accreting) onto the core,
and thus the observational truncation process may artificially discard
part of the infalling material that provides a {\it ram}, rather than 
thermal, pressure, so that the whole system is undergoing gravitational
contraction, rather than in equilibrium.

\subsection{Final considerations}

The GHC scenario provides a unified framework that allows understanding
a wide range of MC properties, such as the apparent virialization at all
scales, the formation of filaments, the acceleration of star formation,
the low overall SFE, and the Bonnor-Ebert-like structure of cores while
allowing them to be contracting structures. It is based on the premise
that, opposite to the standard beliefs, star-forming MCs, rather than
being near-equilibrium structures, are undergoing nearly pressureless
gravitational contraction, albeit in an extremely non-homologous and
hierarchical fashion. This is made possible by means of a Hoyle-like
fragmentation in the nearly isothermal gas that makes up the clouds,
which in turn can occur thanks to the moderate turbulence in the clouds
that allows the formation of nonlinear density fluctuations that have
shorter free-fall times than the whole cloud. Although they do not begin
their collapse until the global collapse has sufficiently reduced the
mean Jeans mass in the cloud, they do terminate their own collapse
earlier than the whole cloud, allowing for its destruction before most
of its mass has made it into stars. 

In this paper we have shown the ability of the GHC scenario to correctly
describe, at both the qualitative (through the analytical approximation
discussed here) and quantitative (through the numerical simulations) the
spatial structure of observed clouds, star-forming regions and young
clusters, and to make predictions about the evolutionary processes they
undergo. {It is noteworthy that the GHC scenario bears a strong
resemblance to the formation of structure in the early Universe, by
means of a cosmic web of filaments in which galaxies form \citep[e.g., ]
[] {Bond+96, Cantalupo+14}, and within which, in turn, the first
clusters form again through filamentary accretion \citep[e.g., ] []
{Safranek+16}.} 

Claims that the scenario does not agree with observations
\citep[e.g.,] [] {Krumholz+18, KM19} appear unfounded, and to originate from a
perceived oversimplification of the mechanisms at play. Observational
works performing detailed comparisons at the structural level {of
clouds and clusters} have also reported qualitative and quantitative
agreement \citep[e.g.,] [] {Motte+18, Getman+18, Getman+19, Chen+19}
with the GHC scenario. In future works we plan to continue testing the
scenario in a variety of situations.

%
%

\section*{acknowledgements}
{We thankfully acknowldge useful and constructive comments from
Peter Barnes, Andi Burkert, Bruce Elmegreen, Neal Evans, Roberto
Galv\'an-Madrid, Lee Hartmann, Pavel Kroupa, Mark Krumholz, Milo\u{s}
Milosavljevi\'c, Nicola Schneider, Alessio Traficante, Chris Wareing,
William Wall, and Long Wang.} This work has received financial support
from CONACYT grants 102488 and 255295 to EV-S, from UNAM-PAPIIT grant
IN113119 to A.P., and UNAM-PAPIIT grant IN100916 to G.G.


\bsp	
\label{lastpage}
\end{document}